\documentclass[epj,nopacs]{svjour}

\usepackage{graphics}
\usepackage{graphicx} 
\usepackage{atlasphysics} 
\usepackage{textcomp}

\usepackage{ifthen}
\usepackage{mathptmx}
\usepackage{helvet}
\usepackage{lineno}
\usepackage{epstopdf}
\usepackage{stfloats} 
\usepackage{hyperref}

\begin{document}

\title{\vspace{-3.0cm}\flushleft{\normalsize\normalfont{CERN-PH-EP-2011-114}} \flushright{\vspace{-0.86cm}\normalsize\normalfont{Submitted to Eur. Phys. J. C}} \\[3.0cm] \flushleft{Performance of Missing Transverse Momentum Reconstruction in Proton-Proton Collisions at $\sqrt{s}$ = 7 TeV with ATLAS}}
\titlerunning{Performance of Missing Transverse Momentum Reconstruction at $\sqrt{s}$ = 7 TeV}

\author{The ATLAS Collaboration
}                    
\institute{CERN, 1211 Geneva 23, Switzerland}
\date{December 6, 2011}

\authorrunning{The ATLAS Collaboration}

\def\ET{$E_{\rm T}$}
\def\metrel{$E_{\rm T}^{\rm miss,rel}$}

\def\pT{$p_{\mathrm{T}}$}
\def\pTe{$p_{\mathrm{T}}^e$}
\def\pTn{$p_{\mathrm{T}}^{\nu}$}
\def\pTell{$p_{\mathrm{T}}^{\ell}$}
\def\pTW{$p_{\mathrm{T}}^W$}
\def\pTZ{$p_{\mathrm{T}}^Z$}
\def\ET{$E_{\mathrm{T}}$}
\def\mT{$m_{\mathrm{T}}$}

\def\etmissvec{$\ensuremath{\mathbf{E_{\mathrm{T}}^{\mathrm{miss}}}}$}
\def\etmissmag{$E_{\mathrm{T}}^{\mathrm{miss}}$}
\def\etmissv{$\textbf{\textit{E}}_{\rm T}^{\rm miss}$}
\def\etmiss{$\textbf{\textit{E}}_{\rm T}^{\rm miss}$}

\def\px {$E_{x}^{\mathrm{miss}}$}
\def\py {$E_{y}^{\mathrm{miss}}$}\def\phimiss{$\phi^{\rm miss}$}
\def\sumet{$\sum E_{\mathrm{T}}$}
\def\et{$E_{\mathrm{T}}$}
\newcommand{\emiss}[1]{\ensuremath{E_{#1}^{\mathrm{miss}}}}
\newcommand{\etreso}{\ensuremath{\sigma(\emiss{x},\emiss{y})}}
\def\metfake{$E_{\mathrm{T}}^{\mathrm{miss,Fake}}$}

\def\met{$E_{\rm T}^{\rm miss}$}
\def\metx{$E_{\rm x}^{\rm miss}$}
\def\mety{$E_{\rm y}^{\rm miss}$}
\def\abseta{\ensuremath{|\eta|}}

\def\metxTruth{\mbox{\ensuremath{\, \slash\kern-.6emE_{x}^\mathrm{True}}}}
\def\metyTruth{\mbox{\ensuremath{\, \slash\kern-.6emE_{y}^\mathrm{True}}}}
\def\metTruth{\mbox{\ensuremath{\, \slash\kern-.6emE_\mathrm{T}^\mathrm{True}}}}

\def\me{\mbox{\ensuremath{\, \slash\kern-.6emE}}}
\def\meP{\mbox{\ensuremath{\, \slash\kern-.6emE_\mathrm{P}}}}
\def\meL{\mbox{\ensuremath{\, \slash\kern-.6emE_\mathrm{L}}}}
\def\mety{$E_{\rm y}^{\rm miss}$}
\def\mexy{E_{x,\,y}^{\rm miss}}

\def\mexycalo{E_{x,\,y}^{\rm Calo,miss}}
\def\metcalo{E_{T}^{\rm Calo,miss}}
\def\mexymuon{E_{x,\,y}^{\rm Muon,miss}}
\def\metmuon{E_{T}^{\rm Muon,miss}}

\def\mexyRefFinal{E_{x,\,y}^{\rm RefFinal,miss}}
\def\metxRefFinal{E_{x}^{\rm RefFinal,miss}}
\def\metyRefFinal{E_{y}^{\rm RefFinal,miss}}
\def\metRefFinal{E_{T}^{\rm RefFinal,miss}}
\def\mexyRefEle{E_{x,\,y}^{\rm RefEle,miss}}
\def\mexyRefGamma{E_{x,\,y}^{\rm RefGamma,miss}}
\def\mexyRefTau{E_{x,\,y}^{\rm RefTau,miss}}
\def\mexyRefJets{E_{x,\,y}^{\rm RefJets,miss}}
\def\mexyRefMuo{E_{x,\,y}^{\rm RefMuo,miss}}
\def\mexyCellOut{E_{x,\,y}^{\rm Cellout,miss}}

\def\metxyTrue{E_{x,\,y}^{\rm miss,True}}
\def\metxTrue{E_{x}^{\rm miss,True}}
\def\metyTrue{E_{y}^{\rm miss,True}}
\def\metTrue{E_{T}^{\rm miss,True}}

\def\mtt{\ensuremath{m_{\tau \tau}}}
\def\nut{\ensuremath{\nu_{\tau}}}
\def\anut{\ensuremath{\overline{\nu_{\tau}}}}
\def\anul{\ensuremath{\overline{\nu_{l}}}}

\def\myFigSize{5.7cm}
\renewcommand{\labelitemi}{$\bullet$}

\abstract{
The measurement of missing transverse momentum in the ATLAS detector, described in this paper, makes use of the full event reconstruction and a calibration based on reconstructed physics objects.
The performance of the missing transverse momentum reconstruction is evaluated using data collected in $pp$
collisions at a centre-of-mass energy of 7 TeV in 2010.
Minimum bias events and 
events with jets of hadrons are used
 from data samples corresponding to an integrated luminosity of about 0.3 \inb~ and 600 \inb~ respectively, together with
events containing a $Z$ boson decaying to two leptons (electrons or muons) or a $W$ boson decaying to a lepton (electron or muon)  and a neutrino, 
 from a data sample corresponding to an integrated luminosity of about 36 \ipb. 
An estimate of the systematic uncertainty on the missing transverse momentum scale is presented.
} 
\maketitle
\section{Introduction}
\label{intro}

In a collider event the missing transverse momentum is defined as the momentum imbalance in the 
plane transverse to the beam axis, where  momentum conservation is expected. Such an imbalance
may signal the presence of unseen particles,  such as neutrinos or stable, weakly-interacting supersymmetric (SUSY) particles.
The vector momentum imbalance in the transverse plane is obtained from  the negative vector sum of the momenta of all particles detected in a $pp$ collison and  is denoted as missing transverse momentum, \etmiss.
The symbol \etmissmag~ is used for its magnitude.

A precise measurement of the missing transverse momentum, \etmiss~, is 
essential for physics at the LHC.  
A large \etmissmag\
is a key signature for searches for new physics processes such as
SUSY and extra dimensions. 
The measurement of \etmiss~ also has a direct impact on the quality of 
a number of measurements of Standard Model (SM) physics, such as the reconstruction 
of the top-quark mass in  \ttbar\ events.
Furthermore, it is crucial in the search for  the Higgs boson in the decay channels $H \rightarrow WW$ and $H \rightarrow \tau \tau$, where a good \etmiss~ measurement improves the reconstruction of the Higgs boson mass \cite{ATLAS_PHYS_PAP}.

This paper describes 
an optimized reconstruction and calibration of  \etmiss~ developed by the ATLAS Collaboration. The performance achieved represents a significant improvement
 $~$ compared to earlier results \cite{ATLAS_PERF_first} presented by ATLAS.
The optimal reconstruction of \etmiss~ in the ATLAS detector is
complex and validation with data, in terms of resolution, scale and tails,  is essential.
A number of data samples encompassing a variety of event
topologies are used. Specifically, the event samples used to assess the quality of the \etmiss~ reconstruction are:  minimum bias events, 
events where jets at high transverse momentum are produced via strong interactions described by Quantum Chromodynamics (QCD)
and events  with leptonically decaying $W$ and $Z$ bosons.
This allows  the full exploitation of the detector capability in the reconstruction and calibration of different physics objects and optimization of the \etmiss\ calculation.
Moreover, in events with \Wln~, where $\ell$ is an electron or muon,
the \etmiss~  performance can be studied in events where genuine \etmissmag~ is present due to the neutrino, thus allowing a validation of the \etmissmag~ scale.
In simulated events, the  genuine \etmissmag,  $E_{\rm T}^{\rm miss,True}$, is  calculated from all generated non-interacting particles in the event and it is also referred to as true \etmissmag~  in the following.

 An important requirement on the measurement of \etmiss\ is the minimization
of the impact of limited detector coverage, finite detector resolution,
the presence of dead regions and different sources of noise that can produce
fake \etmissmag. The ATLAS calori\-meter coverage extends to large
pseudorapidities
\footnote{ATLAS uses a right-handed coordinate system with its origin at
  the nominal interaction point (IP) in the centre of the detector and the
  $z$-axis coinciding with the axis of the beam pipe. The $x$-axis points
  from the IP to the centre of the LHC ring, and the $y$ axis points
  upward. Cylindrical coordinates $(r,\phi)$ are used in the transverse
  plane, $\phi$ being the azimuthal angle around the beam pipe. The
  pseudorapidity is defined in terms of the polar angle $\theta$ as
  $\eta=-\ln\tan(\theta/2)$.}
 to minimize the impact of high energy particles escaping in the very forward direction. Even so, there are inactive
transition regions between different calorimeters that produce fake
\etmissmag. Dead and noisy readout channels in the detector, if
present,  as well as cosmic-ray and beam-halo muons crossing the detector,
 will produce fake  \etmissmag. Such sources can
significantly enhance the back\-ground from multi-jet events in SUSY searches with 
large \etmissmag~ or the background from \Zll\ events accompanied by jets of high transverse momentum (\pT)  in Higgs boson searches in final states with two leptons and \etmissmag .
Cuts are applied to clean the data against all these sources  (see Section \ref{sec:Data}), and more severe cuts to suppress fake  \etmissmag~ are applied in analyses for SUSY searches, after which, for selected events with high-\pT~ jets, the tails of the \etmissmag~ distributions are well described by MC simulation \cite{SUSY_PAPER}.

This paper is organised as follows. 
Section~\ref{sec:detector} gives a brief introduction to  the ATLAS detector.
Section~\ref{sec:Data} and Section~\ref{sec:MC} 
describe the data and Monte Carlo samples used and the event selections applied.
Section~\ref{sec:Rec} outlines how \etmiss\ is reconstructed and
calibrated. 
Section~\ref{sec:Perf} presents the \etmiss~performance for data 
and Monte Carlo simulation, first in  minimum bias and 
 jet events and then in $Z$ and $W$ events.
The systematic uncertainty on the \etmissmag~ absolute scale is discussed in 
Section~ \ref{sec:syst}.
Section \ref{sec:W_in-situ} describes the determination of the \etmissmag~ scale in-situ using \Wln~ events. Finally, the conclusions are given in Section~\ref{sec:conclu}.

\section{The ATLAS Detector} \label{sec:detector}

The ATLAS detector \cite{ATLAS_PHYS_PAP} is a multipurpose particle
physics apparatus with a forward-backward symmetric
cylindrical geometry and near 4$\pi$ coverage in solid angle.
The inner tracking detector (ID) covers the
pseudorapidity range $|\eta| <$ 2.5, and consists of a silicon
pixel detector, a silicon microstrip detector (SCT), and,
for $|\eta| <$ 2.0, a transition radiation tracker (TRT). The
ID is surrounded by a thin superconducting solenoid providing 
a 2 T magnetic field.
 A high-granularity lead/liquid-argon (LAr) sampling electromagnetic
 calorimeter covers the region $|\eta|<$~3.2.
An iron/scintillator-tile calorimeter provides hadronic coverage in the range $|\eta| <$ 1.7. 
LAr technology is also used for the hadronic calorimeters in the end-cap region 1.5~$<|\eta|<$~3.2 and for  both electromagnetic and hadronic
measurements in the forward region up to $|\eta|<$~4.9.
The muon spectrometer surrounds
the calorimeters. It consists of three large air-core superconducting toroid systems, precision tracking chambers providing accurate muon tracking out to $|\eta|$~=~2.7, and additional detectors for triggering in the region $|\eta| <$~2.4.

\section{Data samples and event selection}
\label{sec:Data}

During  2010 a large number of proton-proton collisions, at a centre-of-mass energy
of 7 TeV, were recorded with stable proton beams as well as nominal magnetic field conditions. 
Only data with a fully functioning calorimeter, inner detector and muon spectrometer are analysed. 

Cuts are applied to clean the data sample against instrumental noise and out-of-time 
energy deposits in the calorimeter (from cosmic-rays or beam-induced background).  
Topological clusters reconstructed in the calorimeters  (see Section \ref{calorec}) 
at the electromagnetic energy (EM) scale\footnote{ 
 The EM scale is the basic calorimeter signal scale for the ATLAS calorimeters. It provides the correct scale for energy deposited by electromagnetic showers. It does not correct for the lower energy hadron shower response nor for energy losses in the dead material.}
 are used as the inputs to the jet finder  \cite{CONF_NOTE_JET}. 
In this paper the anti-$k_t$  algorithm \cite{ANTI_KT}, with distance parameter $R$~=~0.6, is
used for jet reconstruction.
The reconstructed jets are required to pass basic jet-quality selection criteria.
In particular events are rejected if any jet in the event with transverse momentum \pT$>$20 GeV is 
caused by sporadic noise bursts in the end-cap region, coherent noise in the electromagnetic calorimeter or reconstructed  from large
out-of-time energy deposits in the calorimeter.
These cuts largely suppress the residual sources of fake \etmissmag~ due to those instrumental effects  which remain after the data-quality requirements.

The 2010 data sets used in this paper  correspond to a total integrated 
luminosity \cite{Lumi,Lumi1} of  approximately 600 \inb~ for 
 jet events and to 0.3 \inb~ for minimum bias events. Trigger and selection criteria for these events are described in Section \ref{sec:QCD_selection}.
For  the \Zll~ and \Wln~ channels,  the samples analysed correspond to an integrated luminosity  of  approximately 36 \ipb. Trigger and selection criteria, similar to those developed for the $W$/$Z$ cross-section measurement~\cite{WZXS}, are applied. These criteria are described in Sections \ref{sec:zll_event_selection} and \ref{sec:wl_event_selection}.

\subsection{Minimum bias and di-jet event selection}
\label{sec:QCD_selection}

For the minimum bias events, only the early period of data taking, with a minimal pile-up contribution,  is studied.
Selected minimum bias events were triggered by the minimum bias
trigger scintillators (MBTS),
mounted at each end of the detector in front of the LAr 
end-cap calorimeter cryostats \cite{MINBIAS_PAPER}.

Events in the QCD jet sample  are required to have passed 
the first-level calorimeter trigger,  
which indicates a significant energy deposit in a certain region of the calorimeter, with the most inclusive trigger with a nominal \pT~ threshold at 15 GeV.
The event sample used in this analysis consists of two subsets of 300 \inb~ each, corresponding to two periods with different pile-up and trigger conditions\footnote{Pile-up in the following refers to the contribution of additional $pp$ collisions superimposed on the hard physics process.} .
One subset corresponds to the periods with lower pile-up conditions
 with, on average,  1 to 1.6 reconstructed vertices per event.
 The other subset corresponds to the periods with higher pileup conditions, 
 where the peak number of visible inelastic interactions per bunch crossing 
 goes up to 3.
In the following, di-jet events are used, selected requiring the presence of exactly two jets with \pT~$ >$ 25 GeV and $|\eta| < 4.5$. 
Jets are calibrated with the local hadronic calibration (see Section \ref{calorec}). 

For each event, at least one good primary vertex is required 
 with a $z$ displacement from the nominal $pp$ interaction point less than 200 mm and 
with at least five associated tracks.
After selection, the samples used in the analysis presented here correspond to 14 million minimum bias events and 13 million di-jet events. 

 \subsection{$\textit{{\textbf{Z}}} \  \rightarrow \ell \ell$~ event
selection}
\label{sec:zll_event_selection}

Candidate \Zll~ events, where $\ell$ is an electron or a muon,  are required to pass an $e$/$\gamma$ or muon trigger with a  \pT~ threshold between 10 and 15 GeV, where the exact trigger selection varies depending on the data period analysed. 
For each event, at least one good primary vertex, as defined above, 
is required.\\
\indent The selection of  $\Zmm$ events requires the presence of exactly two good muons.
A good muon is defined to be a muon reconstructed  in the muon spectrometer with a matched track in the inner detector with transverse momentum above 20 GeV and $|\eta|<$ 2.5 \cite{Muon_Perf}. 
Additional requirements on the number of hits used to reconstruct the track in the inner detector are applied. The $z$ displacement of the muon track from the primary vertex is required to be less than 10 mm.
 Isolation cuts are applied around the muon track.\\
\indent  The selection of $\Zee$ events requires the presence of exactly two identified electrons 
with   $|\eta|<$ 2.47, which pass the
``medium"  identification criteria \cite{WZXS,egamma} and have transverse momenta above 20 GeV. Electron candidates in 
the electromagnetic calorimeter transition region, 1.37 $< |\eta| <$ 1.52, are not considered for this study. 
Additional cuts are applied to remove electrons falling into 
regions where the readout of the calorimeter was not fully operational. \\
\indent In both the \Zee~ and the \Zmm~ selections, the two leptons are required to have opposite charge and  the reconstructed invariant mass of the di-lepton system,  $m_{\ell\ell}$,  is required to be consistent with the $Z$ mass, $66 < m_{\ell\ell} < 116$ GeV.\\ 
\indent With these selection criteria, about 9000 \Zee~ and 13000 \Zmm~ 
events are selected. 
The estimated background contribution to these samples is less than 2$\%$ in
both channels \cite{WZXS}.

\subsection{$\textit{{\textbf{W}}} \  \rightarrow \ell \nu$~ event selection}
\label{sec:wl_event_selection}

Lepton candidates are selected with lepton identification criteria similar to those used for the $Z$ analysis. The differences for the selection of \Wen~ events are that the  ``tight" electron identification
criteria  \cite{egamma,WZXS} are used and an isolation cut is applied on the electron cluster in the calorimeter to reduce contamination from QCD jet background.
 The event is rejected if it contains more than one reconstructed lepton.
The \etmissmag, calculated as described in Section \ref{sec:Rec}, is required to be greater than 25 GeV,  and
the reconstructed lepton-\etmiss~  transverse mass, $m_T$, is required to be greater than 50 GeV.

With these selection criteria, about 8.5$\times10^4$ \Wen~ and 1.05$\times 10^5$ \Wmun~ events are selected. 
The background contribution to these samples is estimated to be about  5$\%$ in
both channels \cite{WZXS}.

\section{Monte Carlo simulation samples}
\label{sec:MC}

Monte Carlo (MC) events are generated using the {\sc Pythia6} program \cite{PYTHIA} with 
the ATLAS minimum bias tune (AMBT1) of the {\sc Pythia} fragmentation and hadronisation parameters \cite{MONTE-CARLO}. The generated events are 
processed with the detailed  {\sc Geant4}  \cite{GEANT4} simulation of the ATLAS 
 detector.

The minimum bias MC event samples are generated using non-diffractive as well as single- and double-diffractive processes, where the different components are weighted according to the cross-sections given by the event generators.

The jet MC samples, generated using a 2-to-2 QCD matrix element and subsequent parton shower development, are used for comparison 
with the two subsets of data taken with different pile-up conditions. In the earlier sample the fraction of events with at least 
two observed interactions is at most  of the order of 8 -- 10 \%, while in the sample taken later in 2010 this fraction ranges from 10 \% to more than 
50 \%. These samples are generated in the \pT~ range  8 -- 560 GeV, in separated parton \pT~ bins to provide a larger statistics also in  the high-\pT~ bins. Each sample is weighted according to its cross-section.

MC  events for the study of SM backgrounds in \Zll~ and \Wln~ analyses are also generated using  {\sc Pythia6}. The only exceptions are the \ttbar~  background and the \Wen~ samples used in Section \ref{sec:Wscale_fromPT}, which are generated with the {\sc MC@NLO} program \cite{MCAtNLO}. 
For  the study of  the total transverse energy of the events, samples produced with  {\sc Pythia8}  
\cite{PYTHIA8} are used as well.

MC samples were produced with different levels of pile-up in order
to reflect the conditions in different data-taking periods. In particular, two event samples were used for jets: one was simulated with a pile-up model where only pile-up collisions originating from the primary bunch crossing are considered (in-time pile-up) and a second one was simulated with a realistic configuration of the LHC bunch group structure, where pile-up collisions from successive bunch crossings are also included in the simulation.
In the case of events containing  \Zll~ or \Wln, MC samples with in-time pile-up configuration are used, because these data correspond to periods where the contribution of out-of-time pileup is small.

The trigger and event selection criteria used for the data are also applied to the MC simulation.

\section {\etmiss\ reconstruction and calibration}
\label{sec:Rec}

The \etmiss\ reconstruction includes contributions from energy deposits in the calorimeters 
and muons reconstructed in the muon spectrometer. The two \etmiss~ components are calculated as:
\begin{eqnarray}
        E_{x(y)}^{\mathrm{miss}} = E_{x(y)}^{\mathrm{miss,calo}}
                                   + E_{x(y)}^{\mathrm{miss,\mu}}.
\label{eq1} 
\end{eqnarray}
Low-\pT~ tracks are used to recover low \pT~ particles which are missed in the calorimeters (see Section  \ref{sec:eflow}), and muons reconstructed from the inner detector are used to recover muons in regions not covered by the muon spectrometer (see Section  \ref{muorec}).
The two terms in the above equation
are referred to
as the calorimeter and muon terms, and will be described in more detail in the following sections.
The values of  \etmissmag~ and its azimuthal coordinate (\phimiss) are then calculated as:
\begin{eqnarray}
 E_{\mathrm{T}}^{\mathrm{miss}}=\sqrt{\left(E_{x}^{\mathrm{miss}}\right)^{2} +\left(E_{y}^{\mathrm{miss}}\right)^{2}} ~~~ , \nonumber \\ 
\phi^{\rm miss}=\textrm{arctan}(E_{y}^{\mathrm{miss}} , E_{x}^{\mathrm{miss}}).
\label{eq12} 
\end{eqnarray}

\subsection{Calculation of the \etmiss\ calorimeter term}
\label{calorec}

In this paper, the \etmiss\ reconstruction uses calorimeter cells calibrated according to the 
reconstructed physics object to which they are associated.
 Calorimeter cells are associated with a
 reconstructed and identified high-\pT\  parent object in a chosen order:
electrons, photons, hadronically decaying $\tau$-leptons,
jets and muons. 
 Cells not associated with any such objects are also taken into account in the \etmiss~ calculation. Their contribution, named  $E_{T}^{\mathrm{miss,CellOut}}$ hereafter,  is important for the \etmiss~ resolution \cite{JET_ETMISS}. 

Once the cells are associated with objects as described
above, 
the \etmiss \ calorimeter term is calculated as follows
 (note that the $E_{x(y)}^{\mathrm{miss,calo},\mu}$  term is not always 
 added, as explained in Section \ref{muorec}, and for that reason it is written between parentheses):
\begin{eqnarray}
       E_{x(y)}^{\mathrm{miss,calo}} =
              E_{x(y)}^{\mathrm{miss},e}         +
              E_{x(y)}^{\mathrm{miss},\gamma}    +
              E_{x(y)}^{\mathrm{miss},\tau}      +
              E_{x(y)}^{\mathrm{miss,jets}}     \nonumber \\ 
 +
              E_{x(y)}^{\mathrm{miss,softjets}}   + 
              (E_{x(y)}^{\mathrm{miss,calo},\mu}) +
              E_{x(y)}^{\mathrm{miss,CellOut}}
\label{eq7} 
\end{eqnarray}
where each term
is calculated from the negative sum of calibrated cell energies inside the corresponding objects, as:
 \begin{eqnarray}
 E_{x}^{\mathrm{miss,term}}=-\sum_{i=1}^{N_{\rm cell}^{\rm term}}E_i\sin\theta_i\cos\phi_i\ ~~~ , \nonumber \\ 
 E_{y}^{\mathrm{miss,term}}=-\sum_{i=1}^{N_{\rm cell}^{\rm term}}E_i\sin\theta_i \sin\phi_i\ 
 \label{eq2}  
          \end{eqnarray} 
where $E_i$, $\theta_i$ and $\phi_i$ are the energy, the polar angle
and the azimuthal angle, respectively.
The summations are over all cells associated with specified objects in the pseudorapidity range 
\footnote{This $\eta$ cut is chosen because the MC simulation does not describe data well  in the very forward region.}
$|\eta|<4.5$.

Because of the high granularity of the calorimeter, it is 
crucial to suppress noise contributions and to limit  the cells used in the \etmiss~ sum to those containing a significant signal.
This is achieved by using only cells belonging to three-dimensional topological
clusters, referred as  topoclusters  hereafter~\cite{clusters}, with the exception of electrons and
photons for which a different clustering algorithm is used \cite{egamma}. 
The topoclusters are seeded by cells with deposited energy~\footnote{$\sigma_{\rm noise}$ is the 
Gaussian width of the EM cell energy distribution measured in randomly triggered events far from 
collision bunches.} $|E_i|>4\sigma_{\rm noise}$,
and are built by iteratively adding neighbouring cells with $|E_i|>2\sigma_{\rm noise}$ and, finally, 
by adding all neighbours of the accumulated cells. 

The various terms in Equation \ref{eq7} are described in the following:      
\begin{itemize}
\item
              $E_{x(y)}^{\mathrm{miss},e}$, 
              $E_{x(y)}^{\mathrm{miss},\gamma}$,
              $E_{x(y)}^{\mathrm{miss},\tau}$   are reconstructed from cells in clusters associated to  electrons, photons and $\tau$-jets from hadronically decaying $\tau$-leptons, respectively;
\item
 $E_{x(y)}^{\mathrm{miss,jets}}$ is reconstructed from cells in clusters associated to 
jets with calibrated \pT~$>$ 20 GeV;
\item
 $E_{x(y)}^{\mathrm{miss,softjets}}$ is reconstructed from cells in clusters associated to jets with 
 7 GeV $<$~ \pT~$<$ 20 GeV;
 \item
 $E_{x(y)}^{\mathrm{miss,calo},\mu}$ is the contribution to \etmiss\
          originating from the energy lost by muons in the calorimeter
  (see Section  \ref{muorec});
 \item
 the $E_{x(y)}^{\mathrm{miss,CellOut}}$ term is
calculated from the cells in topoclusters which are not included in the
reconstructed objects.
\end{itemize}

All these terms are calibrated independently as described in Section \ref{sec:RefFinalConfig}.
The final $ E_{x(y)}^{\mathrm{miss}}$ is calculated from 
Equation~\ref{eq1}
adding the  $E_{x(y)}^{\mathrm{miss},\mu}$ term, described in Section \ref{muorec}.

\subsection{Calculation of the \etmiss\ muon term}
\label{muorec}
The \etmiss\ muon term is calculated from the momenta of muon tracks reconstructed
with  $\abseta<2.7$:  
\begin{eqnarray}
       E_{x(y)}^{\mathrm{miss},\mu} =
               - \sum_{\mathrm{muons}} p_{x(y)}^{\mu}
\label{eq40} 
\end{eqnarray}
where the summation is over selected muons.
In the region $\abseta<2.5$, only well-reconstructed 
muons in the muon
spectrometer with a matched track in the inner detector are
considered (combined muons). The matching requirement considerably reduces  
contributions from fake muons (reconstructed muons not corresponding to true muons). 
These fake muons can sometimes be created from high hit
multiplicities in the muon spectrometer in events 
where some particles from very energetic jets punch through the calorimeter into the
muon system.

In order to deal appropriately with the energy deposited by the muon in the calorimeters, 
$E_{x(y)}^{\mathrm{miss,calo},\mu}$, 
the muon term is calculated differently for isolated and
non-isolated muons, with 
non-isolated muons defined as 
those within a distance 
$\Delta R=\sqrt{(\Delta\eta)^{2}+(\Delta\phi)^{2}} < 0.3$ of  
a reconstructed jet in the event:
\begin{itemize}
\item
The \pT\ of an isolated muon is determined  
from the combined measurement of the inner detector and muon  
spectrometer, taking into account the energy deposited in the calori\-meters. 
In this case the energy  
lost by the muon in the calorimeters ($E_{x(y)}^{\mathrm{miss,calo},\mu}$)
  is not added to the calorimeter  term (Equation \ref{eq7}) to avoid double counting of energy.
\item
  For a non-isolated muon, the energy deposited in the calori\-meter cannot 
  be resolved from the calorimetric energy depositions of the particles in the jet.
  The muon spectrometer measurement of the muon momentum after energy loss in  
the calorimeter is therefore used, so the $E_{x(y)}^{\mathrm{miss,calo},\mu}$ term is
added to the calorimeter term (Equation \ref{eq7}). 
Only in cases in which there is a significant 
mis-match between the spectrometer and the combined measurement,
the combined measurement is used and a parameterized estimation of the 
muon energy loss in the calorimeter \cite{Muon_Perf} is subtracted.
\end{itemize}
 For higher values of pseudorapidity  ($2.5 < \abseta < 2.7$), 
outside the fiducial volume of the inner detector,  
there is no matched track requirement and the muon spectrometer \pT~ alone  is  
used for both isolated and non-isolated muons.

 Aside from the loss of muons outside the acceptance of the muon
spectrometer ($\abseta>2.7$),  muons can be lost in other small inactive regions 
(around $|\eta| =0$ and $|\eta| \sim 1.2$) 
 of the muon spectrometer.
The muons which are reconstructed by segments matched to inner detector tracks extrapolated to the muon spectrometer 
are used to recover their contributions to  \etmiss~ in the   $|\eta| \sim 1.2$ regions \cite{Muon_Perf}.

Although the core of the \etmiss\ resolution is
not much affected by the muon term,
any muons which are not  re\-con\-struc\-ted, badly measured, or fake, can be a source of  fake \etmissmag.

\subsection{Calibration of \etmiss}
\label{sec:RefFinalConfig}

The calibration of \etmiss~ is performed using the  scheme described below, where the cells are calibrated separately according to their parent object:

\begin{itemize} 
\item
The $E_{\mathrm{T}}^{\mathrm{miss},e}$ term is calculated from reconstructed electrons passing the ``medium" electron identification requirements, with \pT~$>$~10 GeV  and calibrated with the default electron calibration \cite{WZXS}.
\item
 The $E_{\mathrm{T}}^{\mathrm{miss},\gamma}$ term is calculated from photons reconstructed  with the ``tight" photon identification requirements \cite{egamma}, with \pT~$>$~10 GeV at the EM scale. Due to the low photon purity, the default photon calibration is not applied. 
 \item 
 The $E_{\mathrm{T}}^{\mathrm{miss},\tau}$  term is calculated from $\tau$-jets reconstructed with the  ``tight" $\tau$-identification requirements \cite{TauPerf},  with \pT~$>$~10 GeV, calibrated with the local hadronic calibration (LCW) scheme~\cite{LCW}. The LCW scheme uses properties of clusters to calibrate them individually.
It first classifies calorimeter clusters as electromagnetic or hadronic,
according to the cluster topology,
and then weights each calorimeter cell in clusters according to the cluster energy
and the cell energy density.
Additional corrections are applied to the cluster energy for the average energy deposited
in the non-active material before and between the calorimeters and for unclustered calorimeter energy.
 \item
 The $E_{\mathrm{T}}^{\mathrm{miss,softjets}}$  term is  calculated from jets (reconstructed using the anti-$k_t$ algorithm with R=0.6) with  $7~<~ $\pT $~<~20$ GeV
  calibrated with the LCW calibration.
\item
The $E_{\mathrm{T}}^{\mathrm{miss,jets}}$ term is calculated from jets   with  \pT $~>~20$ GeV
 calibrated with the LCW calibration and the jet energy scale (JES) factor \cite{JES}  applied.
 The JES factor corrects the energy of jets, either at the EM-scale or after cluster calibration,
 back to particle level. The JES is derived as a function of reconstructed jet $\eta$ and \pT~using the generator-level information in MC simulation.
\item
 The $E_{\mathrm{T}}^{\mathrm{miss,CellOut}}$ term is calculated from topoclusters outside reconstructed objects with the LCW calibration and from reconstructed tracks as  described in Section \ref{sec:eflow}. 
\end{itemize}
Note that  object classification criteria and calibration can be chosen according to specific analysis criteria, if needed.

\subsubsection{Calculation of the $E_{\mathrm{T}}^{\mathrm{miss,CellOut}}$ term with a track-cluster matching algorithm }
\label{sec:eflow}

In  events with $W$ and $Z$~ boson production, the calibration 
of the $E_{\mathrm{T}}^{\mathrm{miss,CellOut}}$ term is of particular importance
because, due to the low  particle multiplicity in these events,  this \etmiss\ contribution balances the $W/Z$ boson \pT\ to a large extent \cite{JET_ETMISS}.   An energy-flow algorithm 
 is used to improve the calculation of the low-\pT~ contribution to \etmiss\ ($E_{\mathrm{T}}^{\mathrm{miss,CellOut}}$). 
Tracks are added to recover the contribution from low-\pT~ particles which do not reach the calorimeter or do not seed a topocluster.
Furthermore the track momentum is used instead of the topocluster energy for tracks associated to topoclusters,  thus exploiting the better calibration and resolution of tracks at 
  low momentum compared to topoclusters.
  
   Reconstructed tracks with \pT~$>$ 400 MeV, passing track quality selection criteria such as the number of hits and $\chi^2$ of the track fit, are used for the calculation of the 
   $E_{\mathrm{T}}^{\mathrm{miss,CellOut}}$ term. 
     All selected tracks are extrapolated to the second layer of the
  electromagnetic calorimeter and very loose criteria are used  for association
  to reconstructed objects or topoclusters, to avoid double counting.
    If a track is neither associated to a topocluster nor a reconstructed object, its
transverse momentum is added to the calculation of $E_{\mathrm{T}}^{\mathrm{miss,CellOut}}$.
 In the case where
the track is associated to a topocluster, its transverse momentum is used for the calculation of
 the $E_{\mathrm{T}}^{\mathrm{miss,CellOut}}$ and the topocluster energy is discarded, 
  assuming that the topocluster energy corresponds to the charged particle giving the track.
  It has to be noticed that there is a strong correlation between the 
number of particles and topoclusters, so, in general no neutral energy is
lost replacing the topocluster by a track, and the neutral topoclusters are kept in most of the cases.
If more than one topocluster is associated to a track, only  the topocluster with the largest energy 
is excluded  from  the  \etmiss\ calculation, assuming that this energy corresponds to the track.

\section{Study of \etmiss~ performance}
\label{sec:Perf}

In this section the distributions of  \etmiss~ in minimum bias, di-jet,
  \Zll~ and \Wln~ events from data are compared with the expected distributions from the MC samples.
The performance of \etmiss~  in terms of resolution and scale is also derived.

Minimum bias, di-jet events and $\Zll$ events are used to investigate
the \etmiss~ performance without relying on MC detector simulation. 
In general, apart from a small contribution from the semi-leptonic decay of
heavy-flavour hadrons in jets, no genuine \etmissmag~ is expected in these
events.
Thus most of the \etmiss~ reconstructed in these events is a direct
result of imperfections in the reconstruction process or in the detector response.

\subsection{\etmiss~ performance in minimum bias and di-jet events}
\label{sec:met_rec}

\begin{figure*}
\begin{center}
\begin{tabular}{lr}
\hfill \includegraphics[width=.49\linewidth,height=\myFigSize]{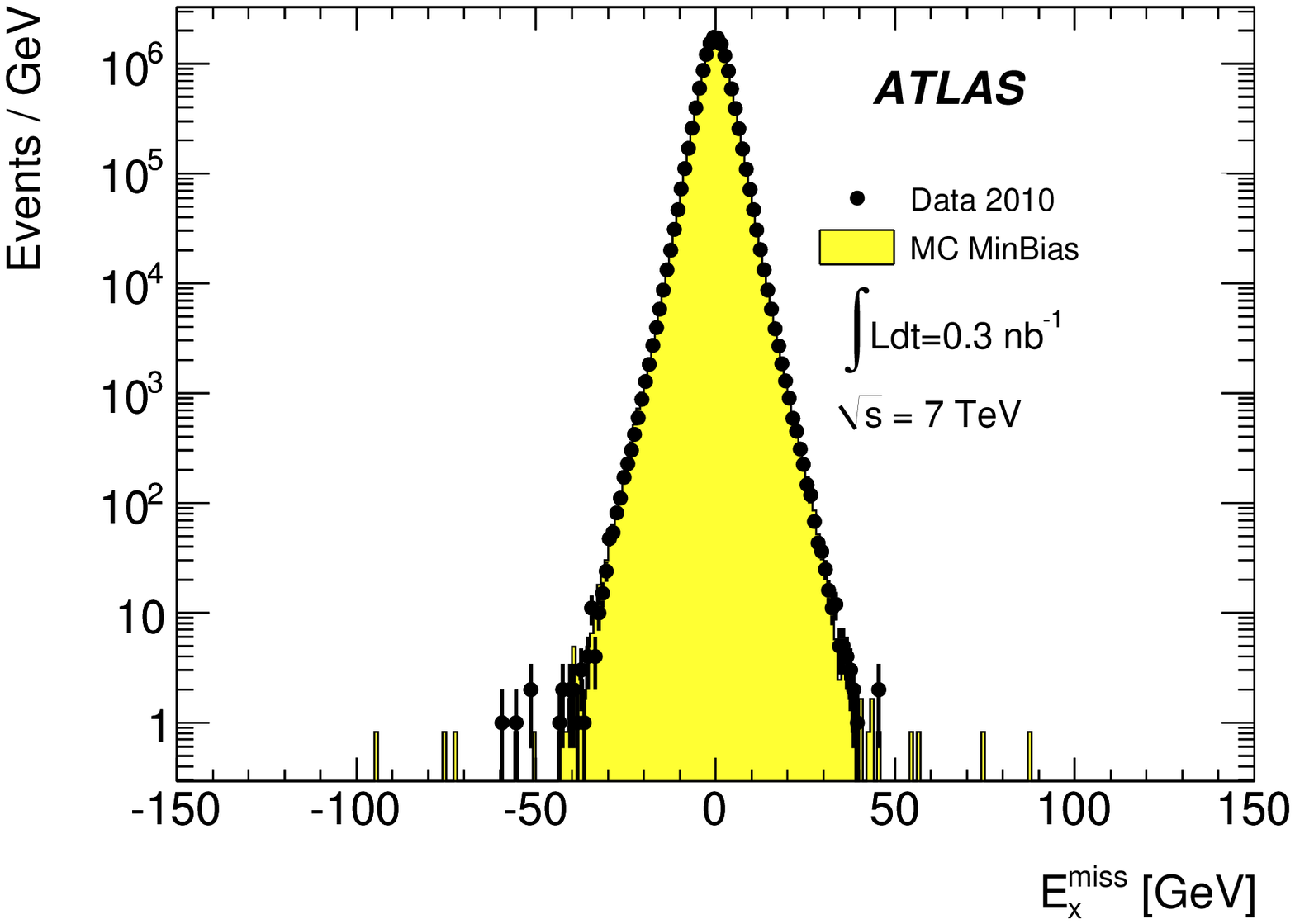} 
\hfill \includegraphics[width=.49\linewidth,height=\myFigSize]{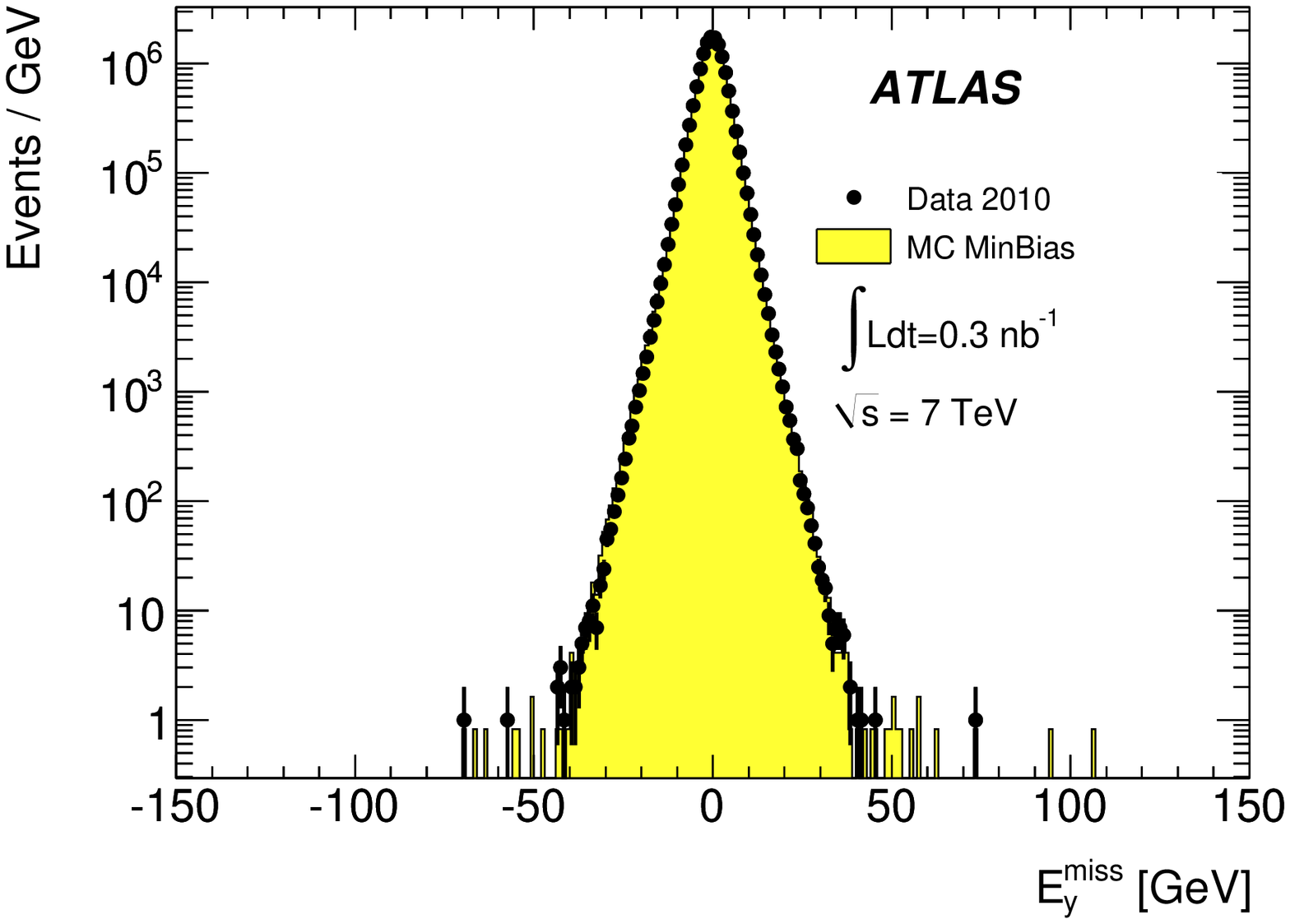} \\
\hfill \includegraphics[width=.49\linewidth,height=\myFigSize]{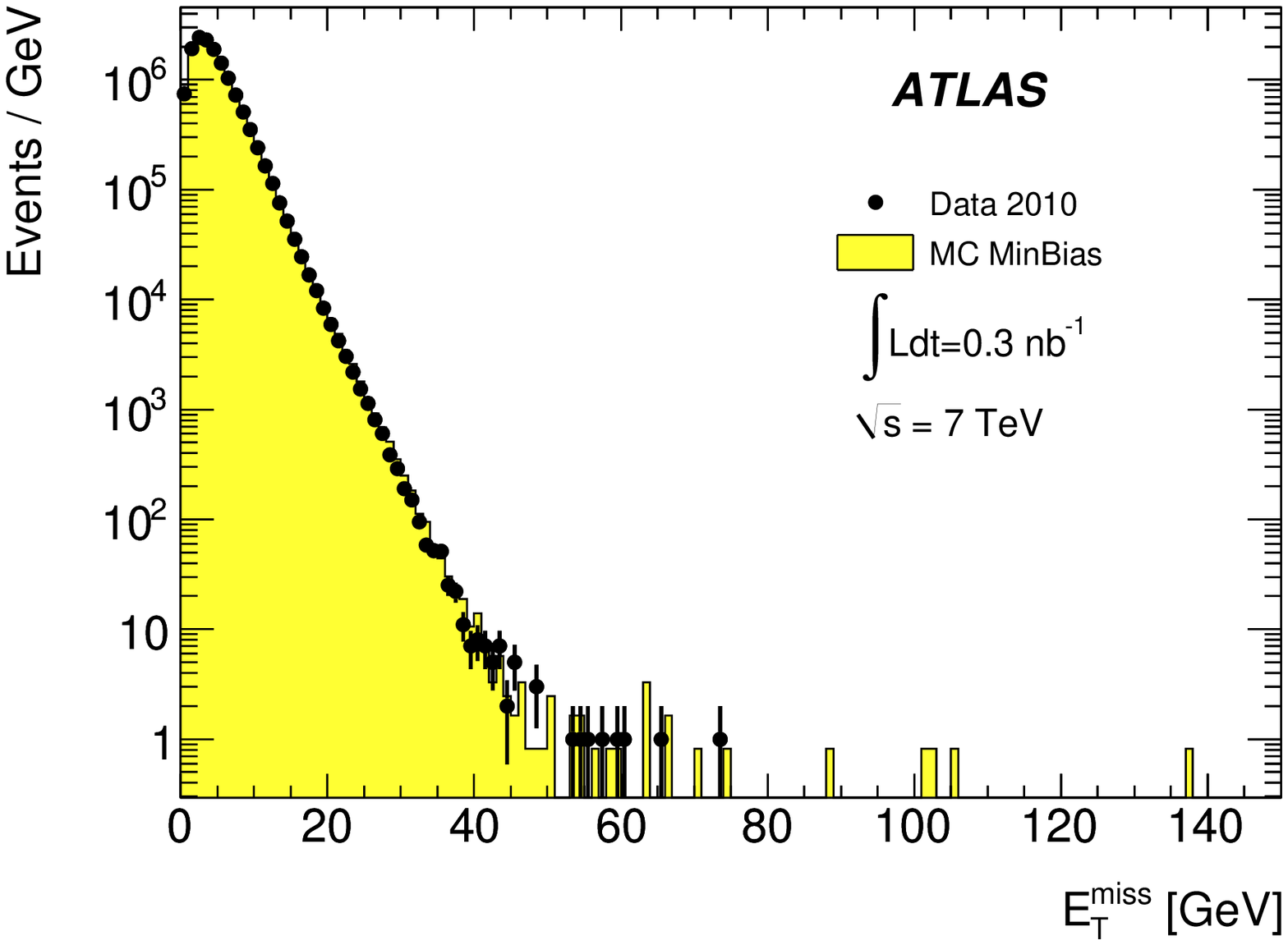}  
\hfill \includegraphics[width=.49\linewidth,height=\myFigSize]{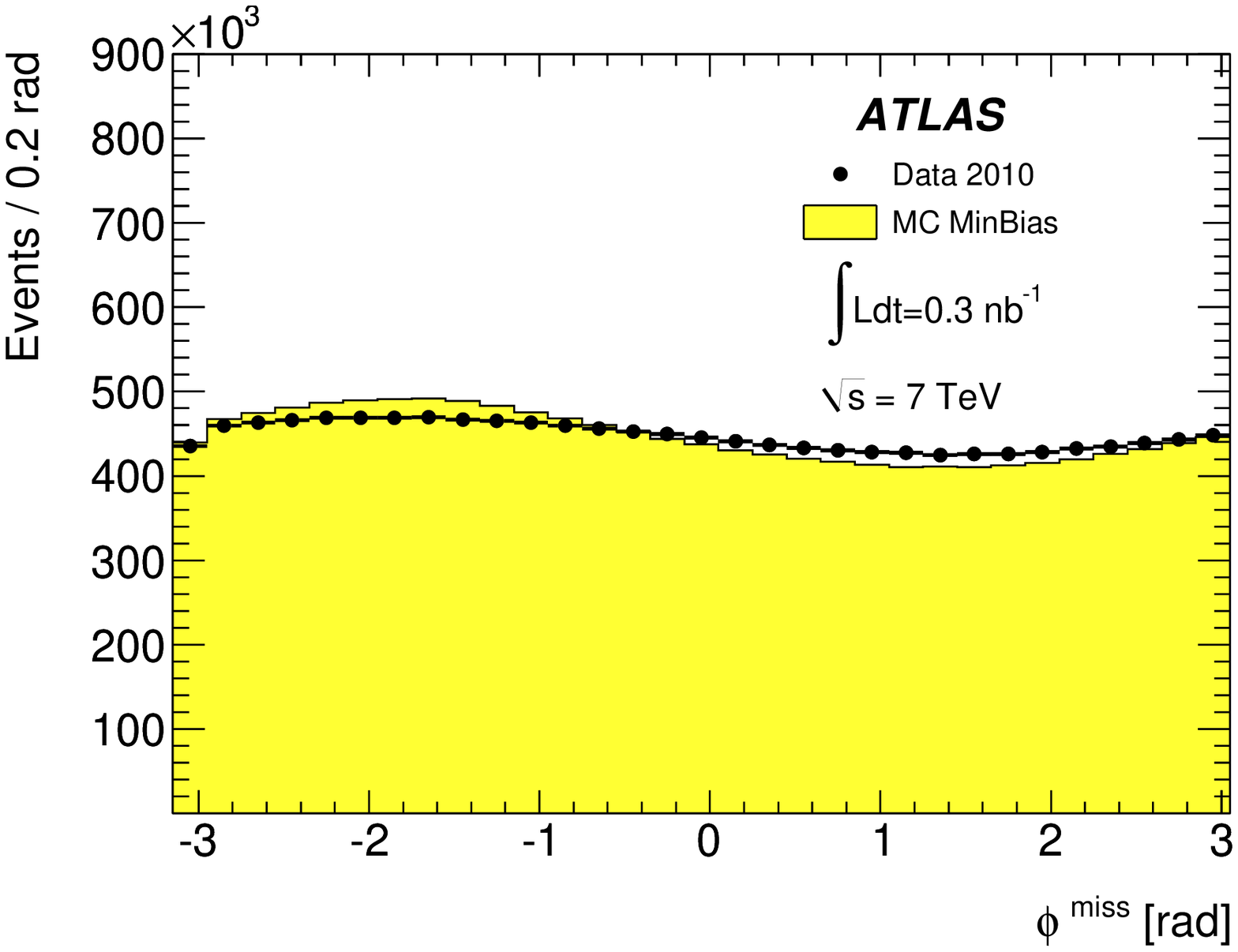}  
\end{tabular}
\end{center}
\caption{\it Distribution of \px~ (top left), \py~ (top right),   \etmissmag~ (bottom left), 
\phimiss~ (bottom right) as measured in a 
data sample of minimum bias events.
The expectation from MC simulation, 
normalized to the number of events in data, 
is superimposed.}
\label{fig:basic_MB}
\end{figure*}

\begin{figure*}
\begin{center}
\begin{tabular}{lr}
\hfill \includegraphics[width=.49\linewidth,height=\myFigSize]{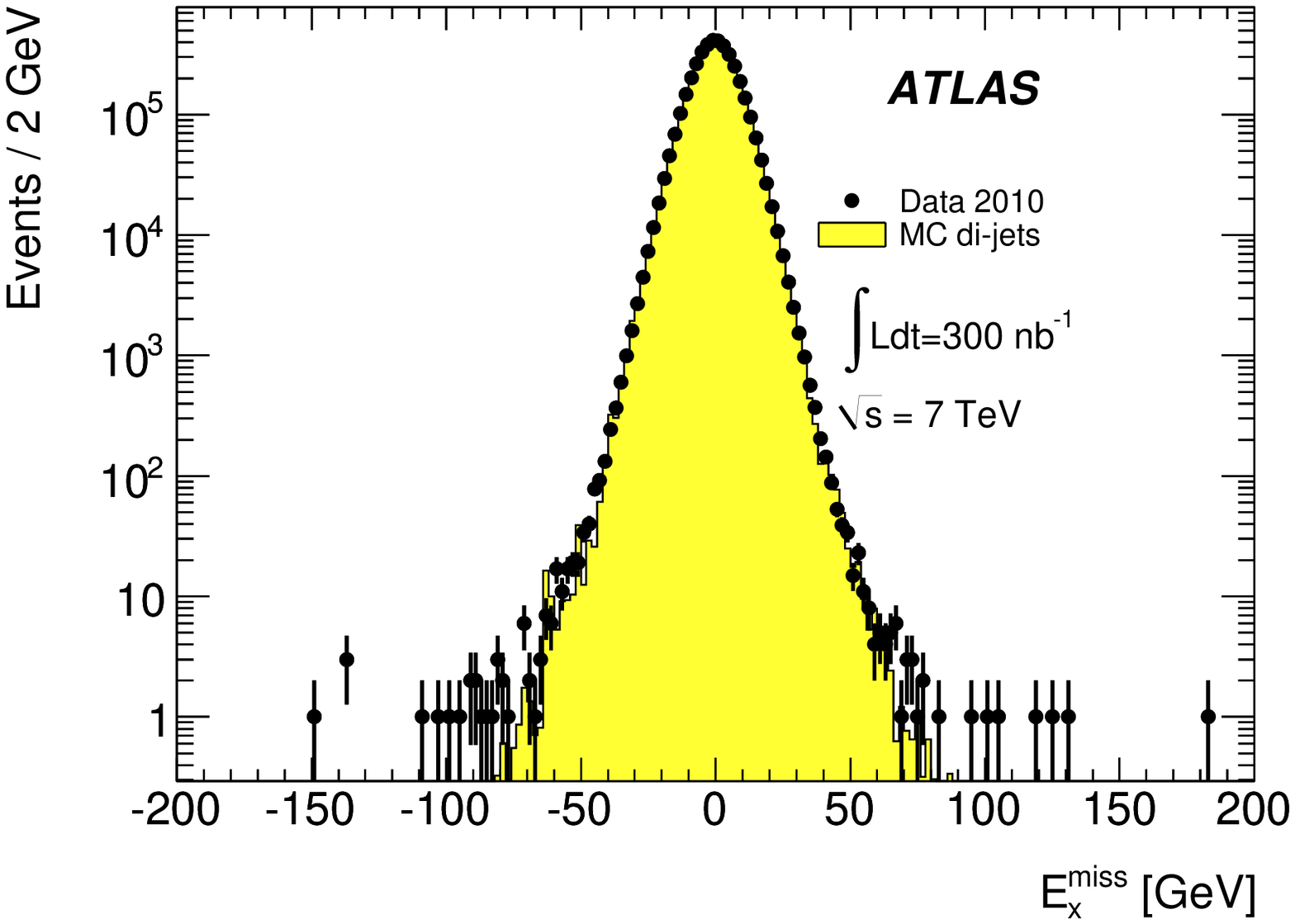}  
\hfill \includegraphics[width=.49\linewidth,height=\myFigSize]{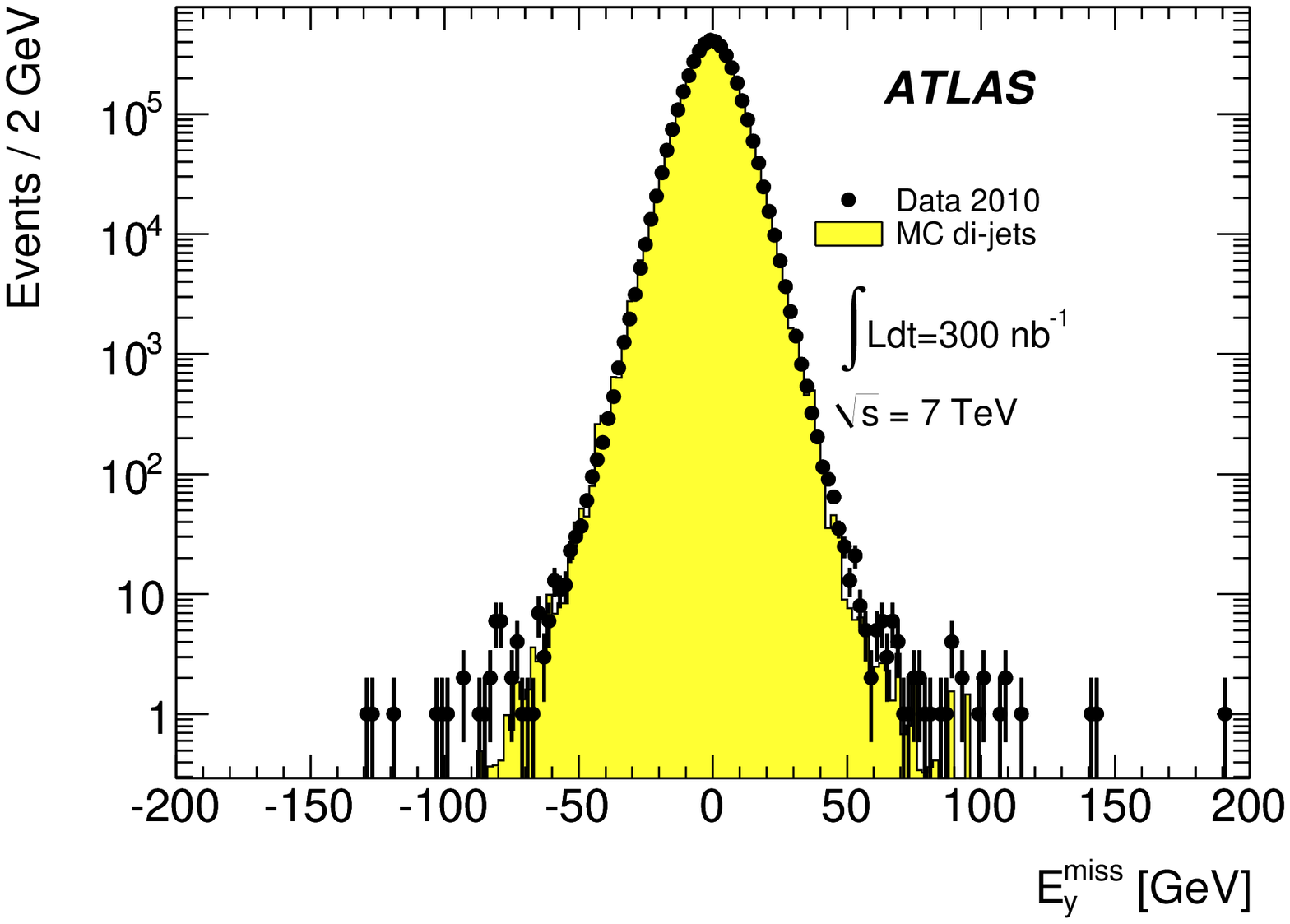}  \\
\hfill \includegraphics[width=.49\linewidth,height=\myFigSize]{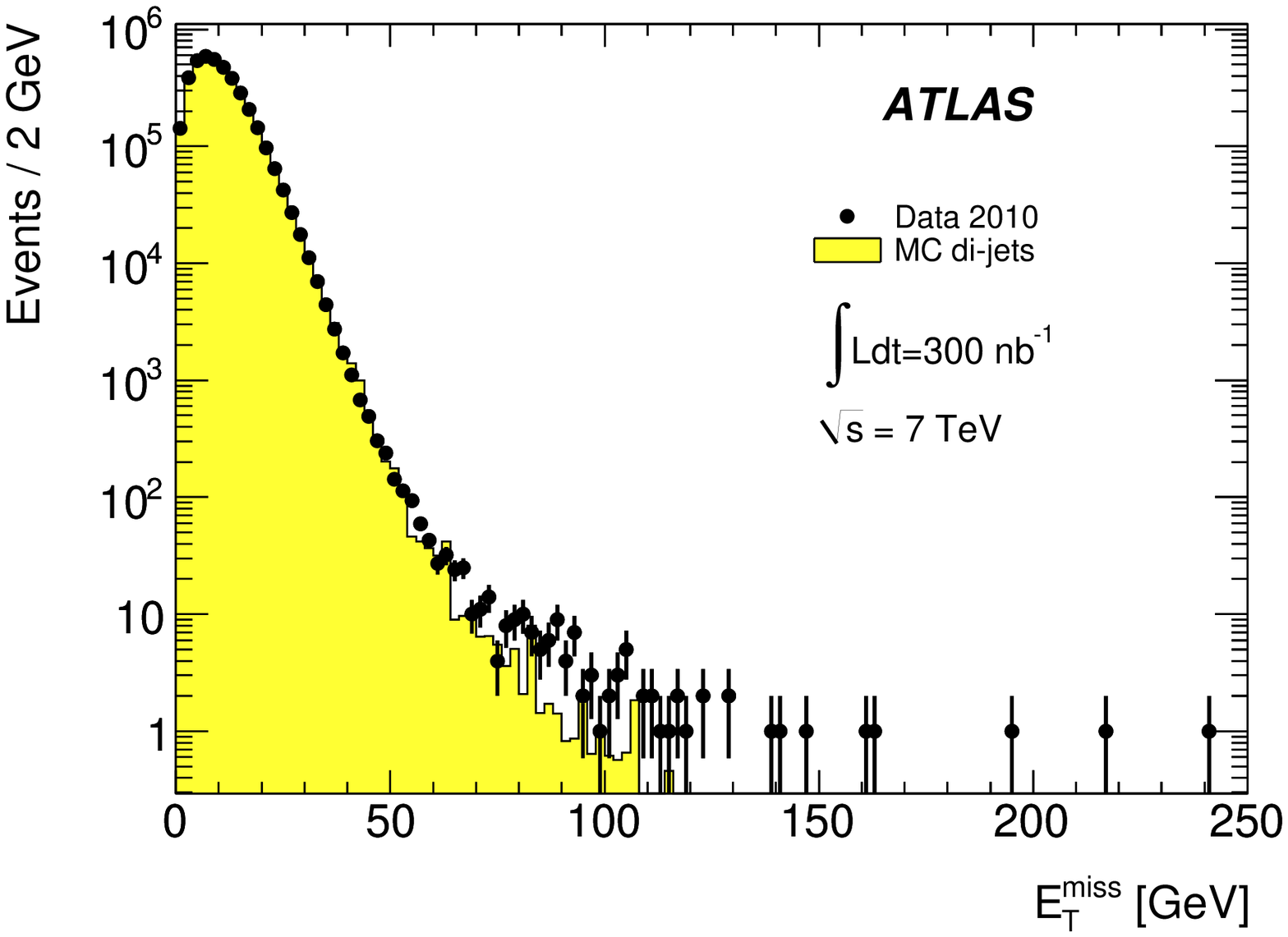}  
\hfill \includegraphics[width=.49\linewidth,height=\myFigSize]{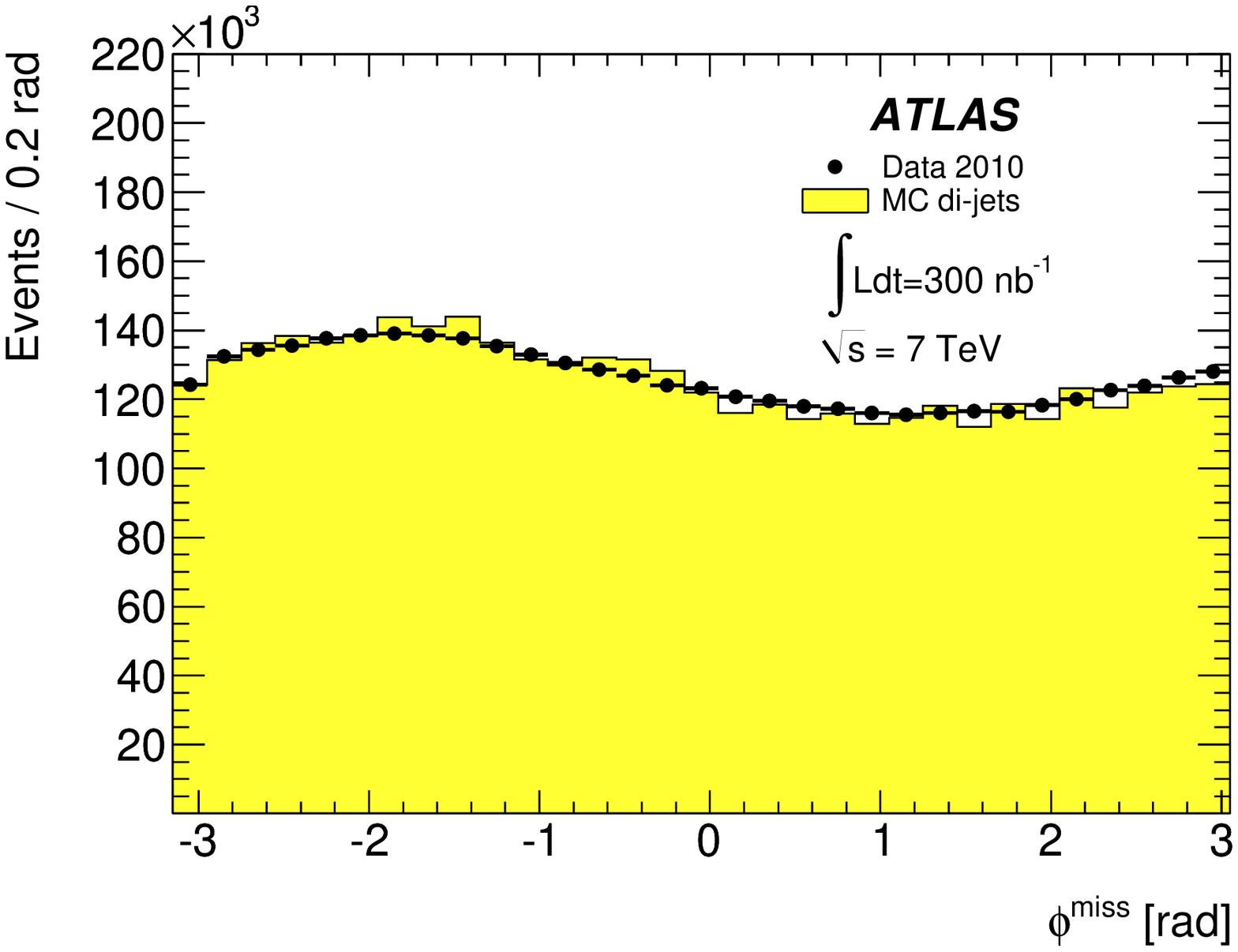}  
\end{tabular}
\end{center}
\caption{\it Distribution of \px~ (top left), \py~ (top right),   \etmissmag~ (bottom left), 
\phimiss~ (bottom right) as measured in the data sample of di-jet events.
The expectation from MC simulation, 
normalized to the number of events in data, 
is superimposed. The events in the tails are discussed in the text.}
\label{fig:basic_QCD}
\end{figure*}

The distributions  of  \px~, \py~,   \etmissmag~ and  \phimiss~  for data and MC simulation  are shown in  Figure~\ref{fig:basic_MB}  for minimum bias events.
The distributions are shown only for events with total transverse energy (see definition at the end of this section)  greater than 20 GeV in order to reduce the contamination
of fake triggers from the MBTS.
Figure~\ref{fig:basic_QCD} shows the distributions  of the same variables for the 
di-jet sample. The di-jet sample corresponding to the periods with higher pileup conditions (see Section \ref{sec:QCD_selection}) is used.
The MC simulation expectations are superimposed, normalized to the number of events in the data. 

In di-jet events a reasonable agreement is found between data and simulation for all basic quantities, while there is some disagreement in minimum bias events, attributed  to
imperfect modelling of soft particle activity in the MC simulation.
The better agreement 
between data and MC simulation in the \phimiss
\par\noindent
 distribution for the di-jet sample can be partly explained by the fact that the \etmiss~ is not corrected for the primary vertex position;
the primary vertex position in data is better reproduced by the MC simulation
for the di-jet sample than in the case of the minimum bias sample.

Events in the tails of the \etmissmag~ distributions  have been carefully checked, 
in order to understand the origin of the large measured \etmissmag. 
The tails are not completely well described by MC simulation, but, 
both in data and in MC simulation they are in general due to mis-measured jets. 
In minimum bias events there are more events in the tail in MC simulation
and this can be due to the fact that the MC statistics is larger than in data. 
 In di-jet events, there are more events in the tail in data.
 More MC events would be desirable.
In di-jet events there are 19 events with \etmissmag~ $>$ 110 GeV in the data.
The majority of them (13 events) are due to mis-measured jets,
where in most of the  
cases at least one jet points to a transition region between calorimeters.
Two events are due to a combination of mis-measured jets with an overlapping muon, and one event is due to a fake high-\pT~ muon. 
Finally two events look like good \bbbar~ candidates, and 
one event has one reconstructed jet and no activity in the other  
hemisphere.

The events with fake \etmissmag~  due to mis-measured jets and jets containing leptonic decays of heavy hadrons can be rejected by a cut based on the azimuthal angle between the jet and \etmiss, $\Delta\phi$({\bf jet},\etmiss).
Since the requirement of event cleaning depends on the physics analysis,
the minimal cleaning cut is applied and careful evaluation of tail events is performed in this paper.
Analyses that rely on a careful understanding and reduction of the tails of the \etmissmag~ distribution (e.g. SUSY  searches such as Ref. \cite{SUSY_PAPER}) have performed more detailed studies to characterize the residual tail in events containing high-\pT~ jets. These analyses use tighter jet cleaning cuts, track-jet matching, and angular cuts on  $\Delta\phi$({\bf jet},\etmiss) to further reduce the fake \etmissmag~ tail. In Ref. \cite{SUSY_PAPER} a fully data-driven method (described in detail in Ref. \cite{JET_ETMISS}) was then employed to determine the residual fake \etmissmag~ background.

\begin{figure*}
\begin{center}
\begin{tabular}{lr}
\hfill \includegraphics[width=.49\linewidth,height=\myFigSize]{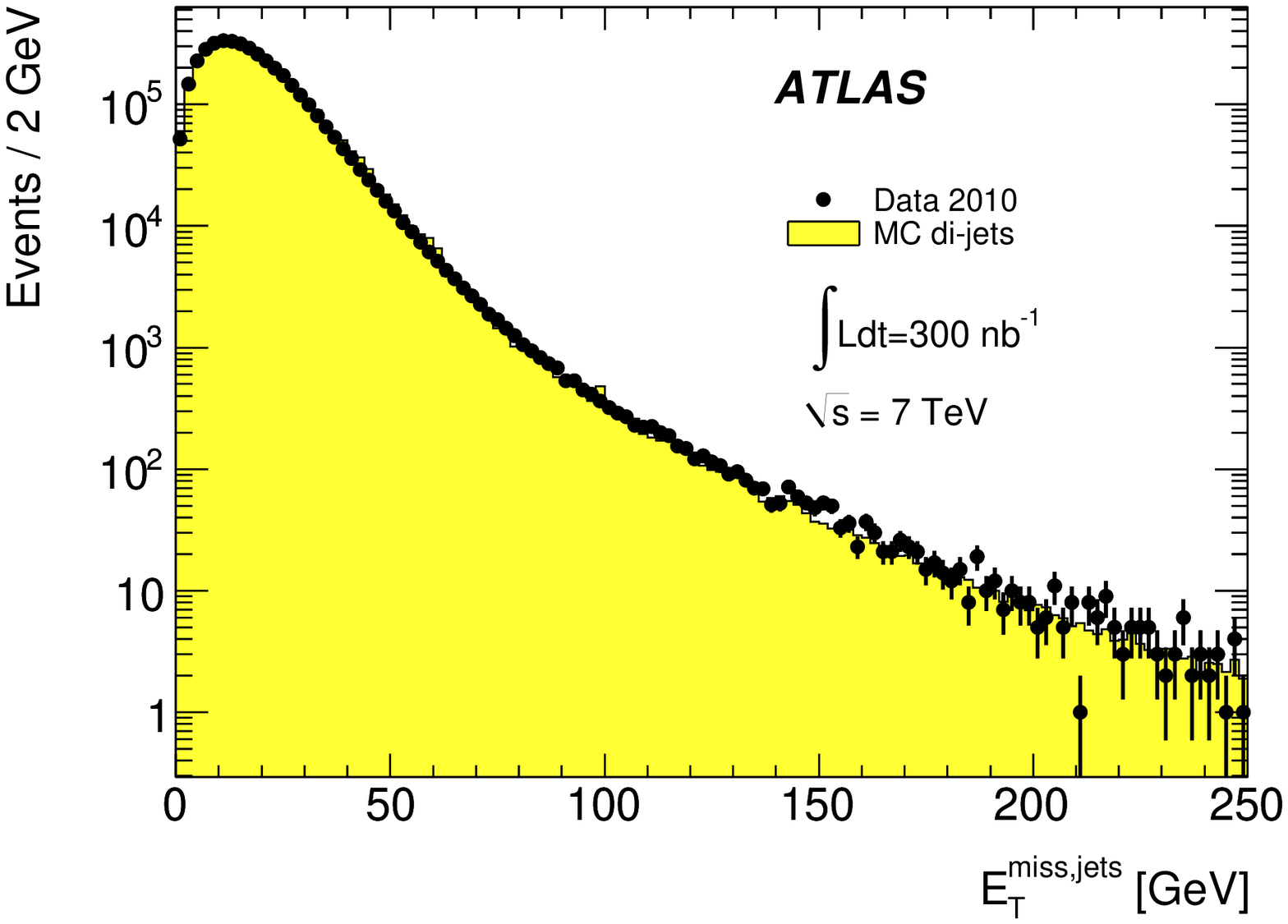}  
\hfill \includegraphics[width=.49\linewidth,height=\myFigSize]{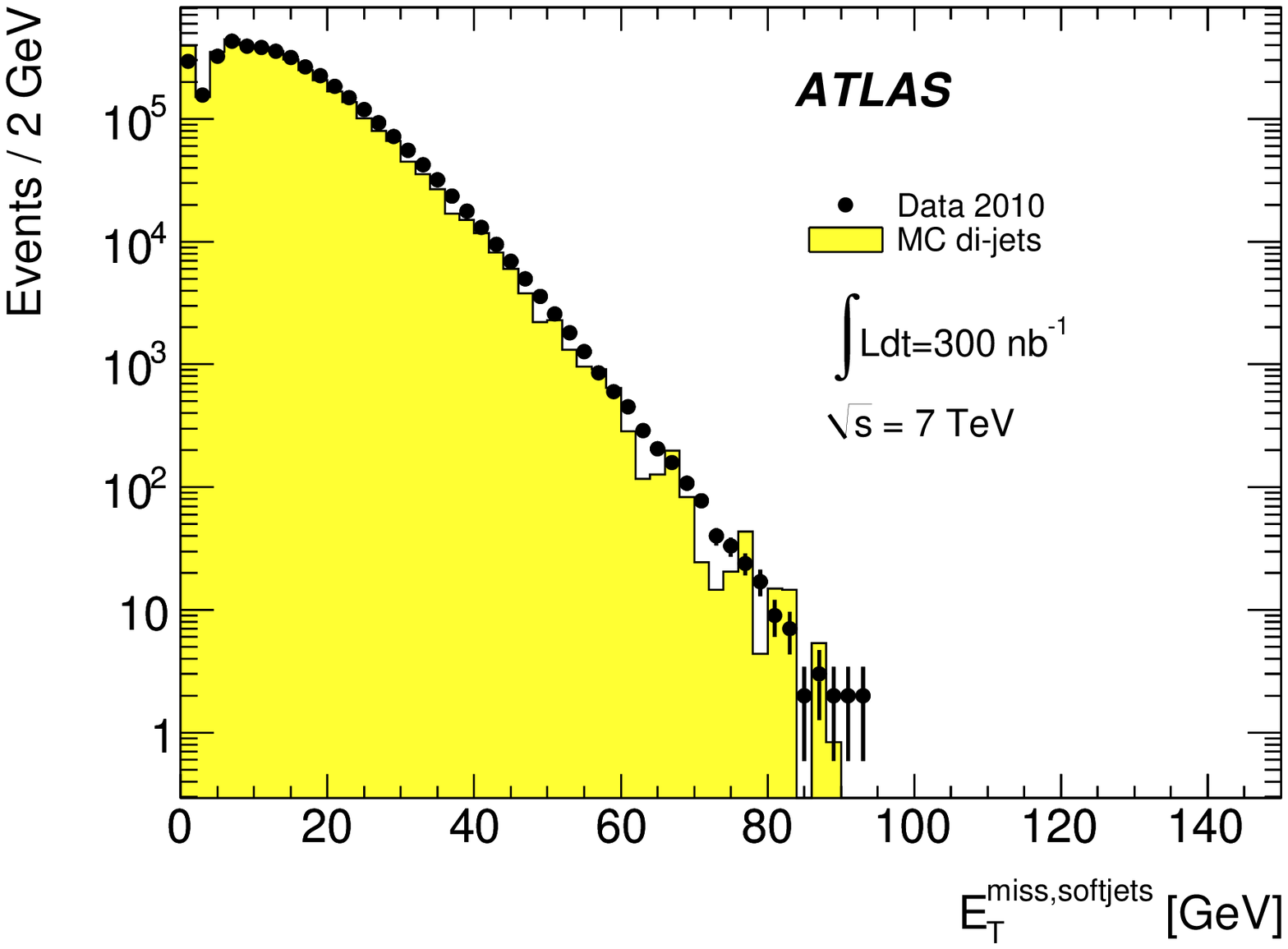}  \\
\hfill \includegraphics[width=.49\linewidth,height=\myFigSize]{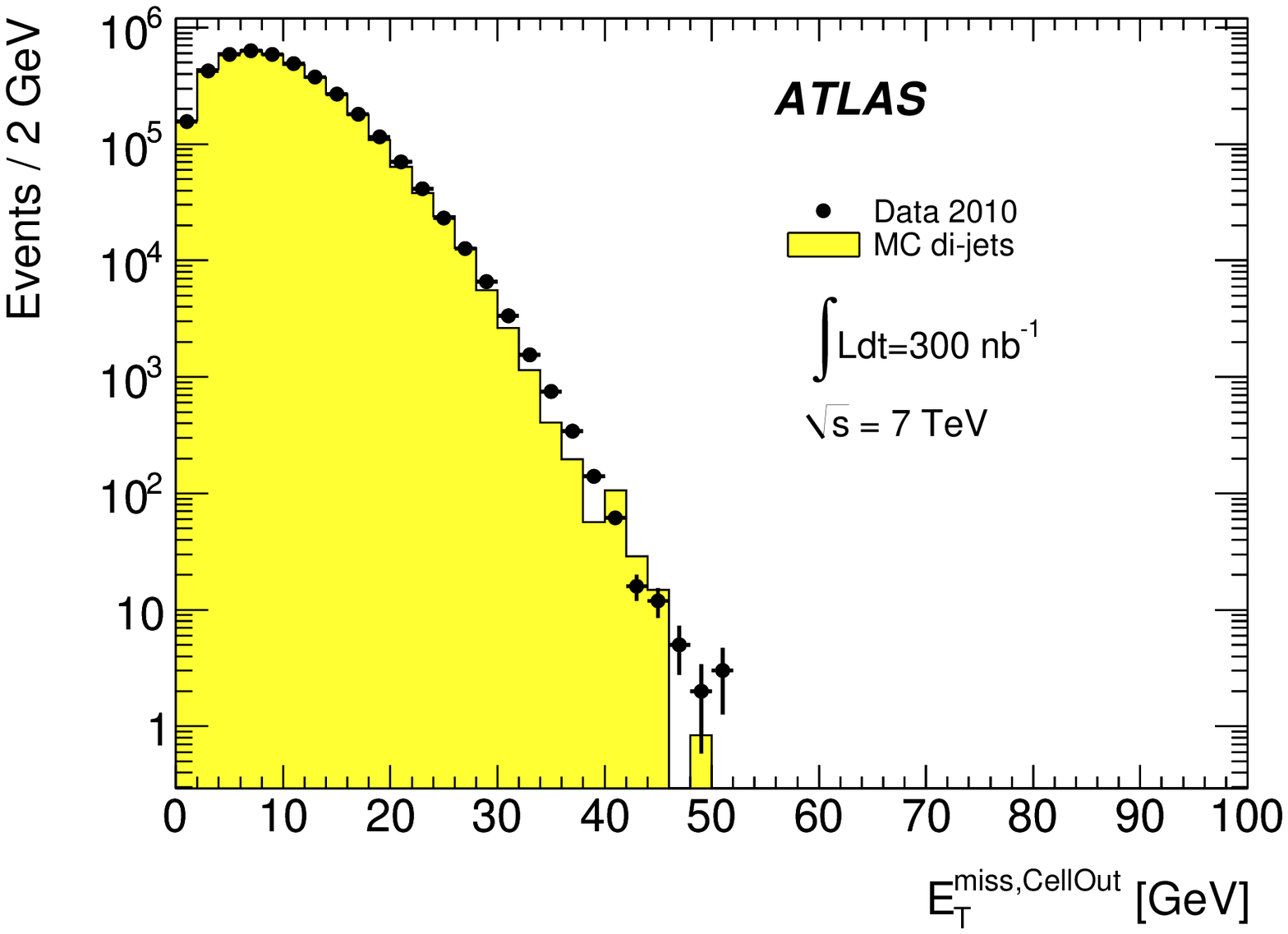}  
\hfill \includegraphics[width=.49\linewidth,height=\myFigSize]{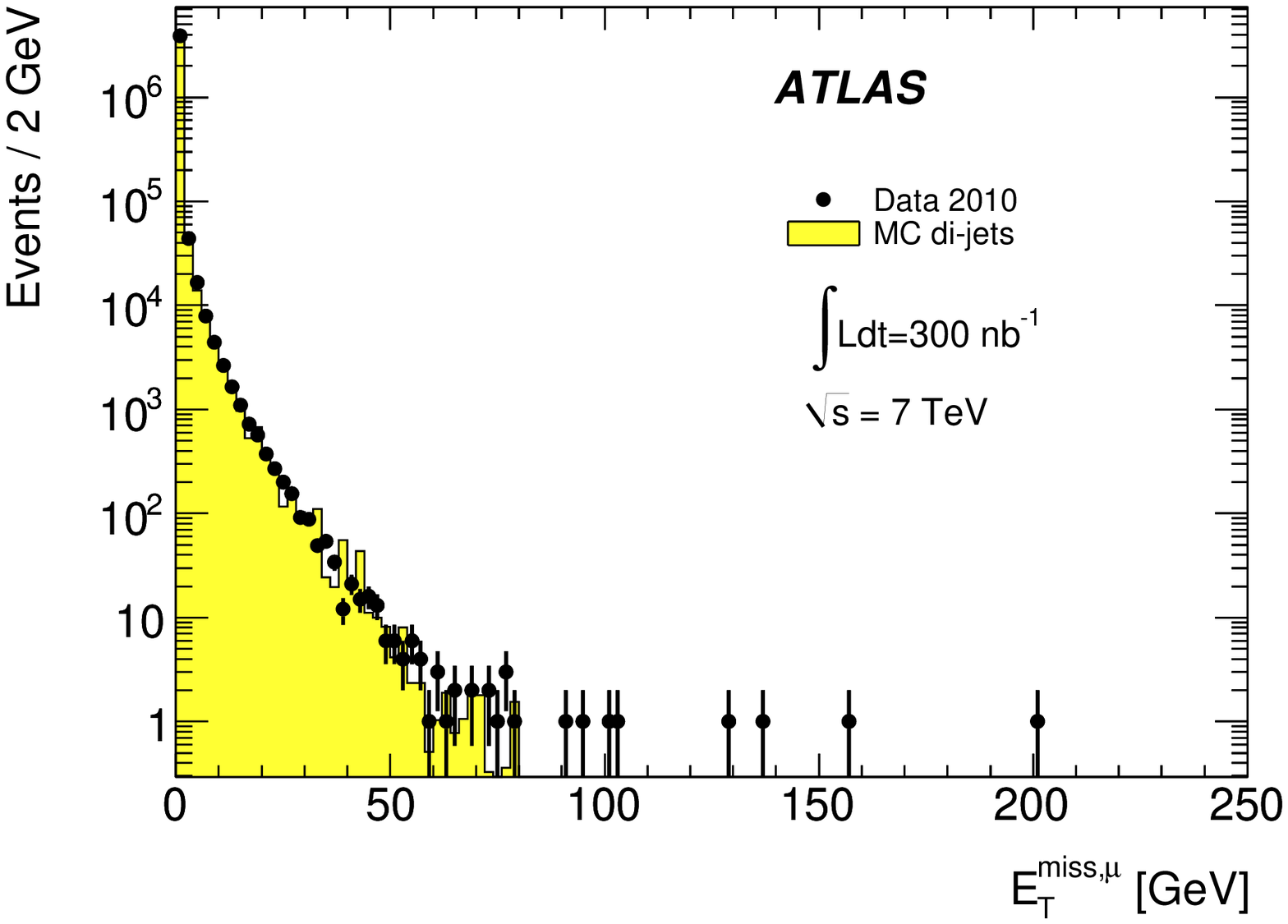}  

\end{tabular}
\end{center}
\caption{\it Distribution of   \etmissmag~ computed with cells from topoclusters in jets (top left), 
in soft jets (top right), from topoclusters outside reconstructed objects (bottom left) and from reconstructed muons (bottom right) for data 
for di-jet events.
The expectation from MC simulation, 
normalized to the number of events in data, 
is superimposed.  The events in the tail of the  $E_{T}^{\mathrm{miss,\mu}}$ distribution are discussed in the text.}
\label{fig:terms_QCD}
\end{figure*}

The contributions from jets, soft jets and topoclusters not associated to
the reconstructed objects and muons are shown in Figure~\ref{fig:terms_QCD} for the 
di-jet events. The data-MC agreement is good for all of the terms contributing to \etmiss.
The tails observed in the muon term are mainly due to reconstructed fake muons and to one 
cosmic-ray muon, which can be rejected by applying a tighter selection for the muons used in the \etmiss~ reconstruction,  
based on $\chi^2$ criteria for the combination, isolation criteria and requirements on the number of hits in muon chambers
used for the muon reconstruction.

In the following some distributions are shown for the total  transverse energy, \sumet, 
which is an important quantity
to parameterise and understand the \etmiss~ performance.
It is defined as:
\begin{eqnarray}
\sum E_{\rm T} &=&\sum_{i=1}^{N_{\rm cell}} E_i \sin\theta_i\
\end{eqnarray}
where $E_{i}$ and $\theta_{i}$ are  the energy and the polar angle, respectively, of calorimeter cells associated to  topoclusters within $\abseta< 4.5$. 
Cell energies are calibrated according to the scheme described in Section \ref{sec:RefFinalConfig} for \etmiss~ .

\begin{figure*}
\begin{center}
\begin{tabular}{lr}
\hfill \includegraphics[width=.49\linewidth,height=\myFigSize]{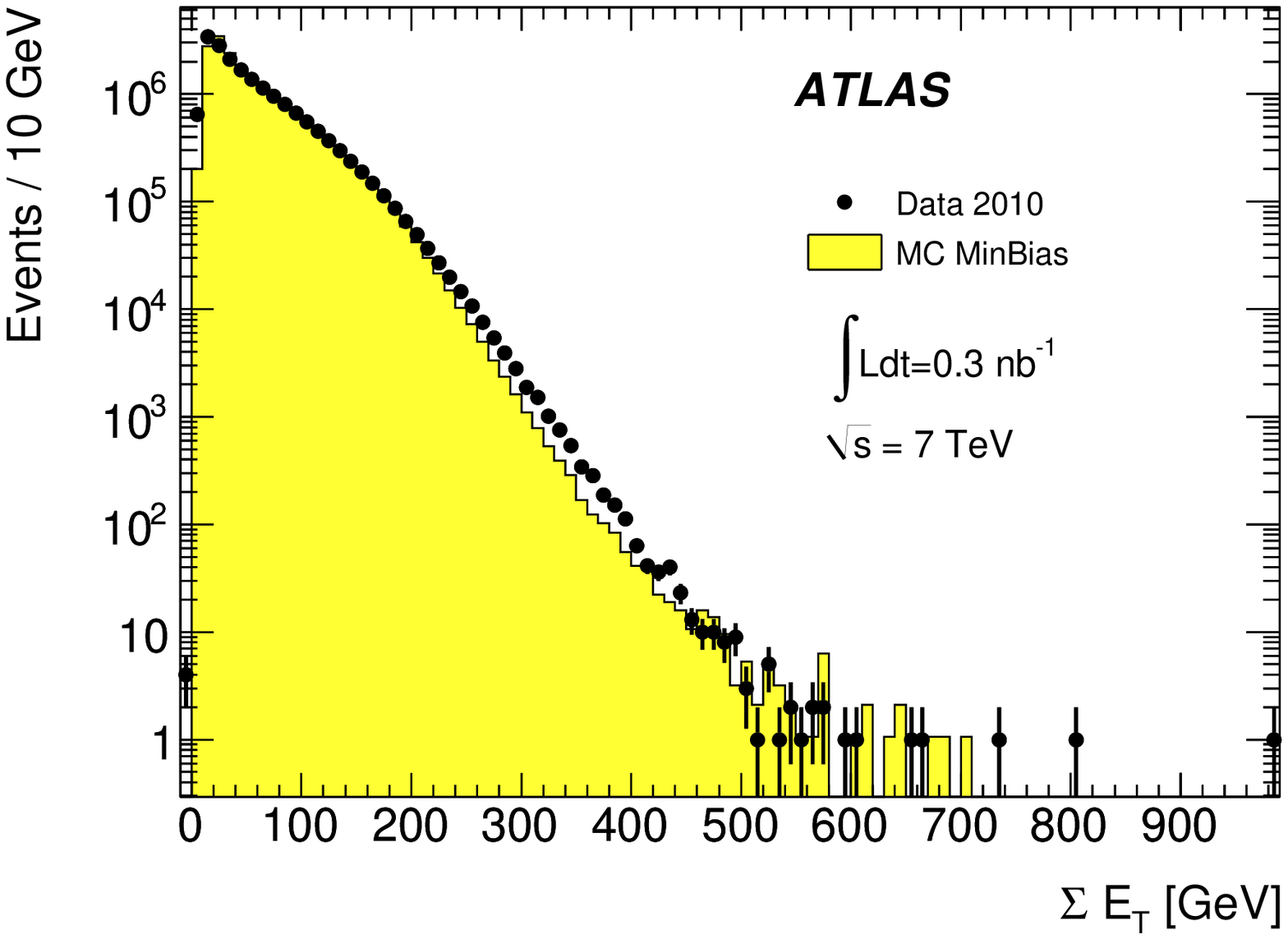}  
\hfill \includegraphics[width=.49\linewidth,height=\myFigSize]{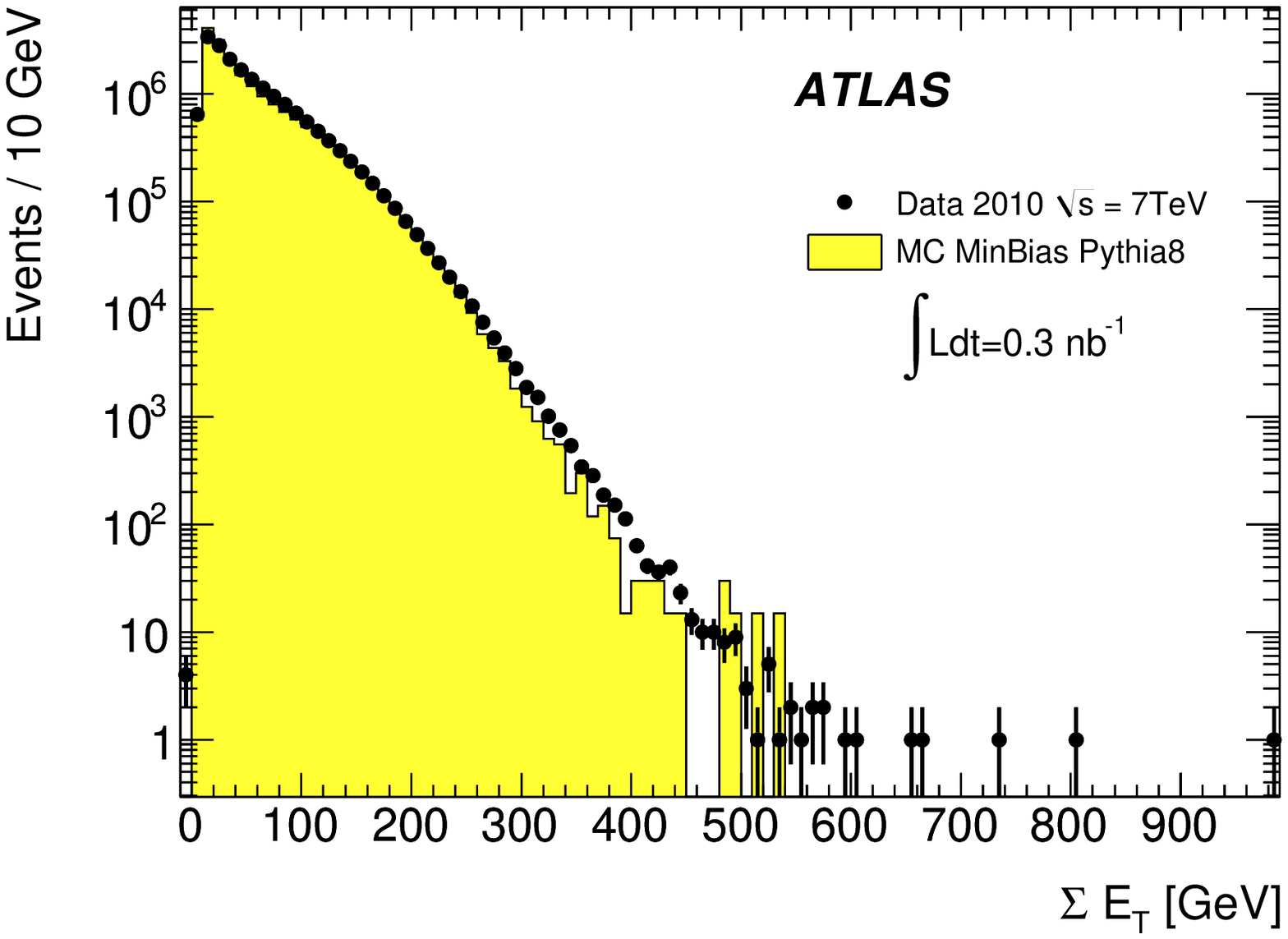} \\ 
\hfill \includegraphics[width=.49\linewidth,height=\myFigSize]{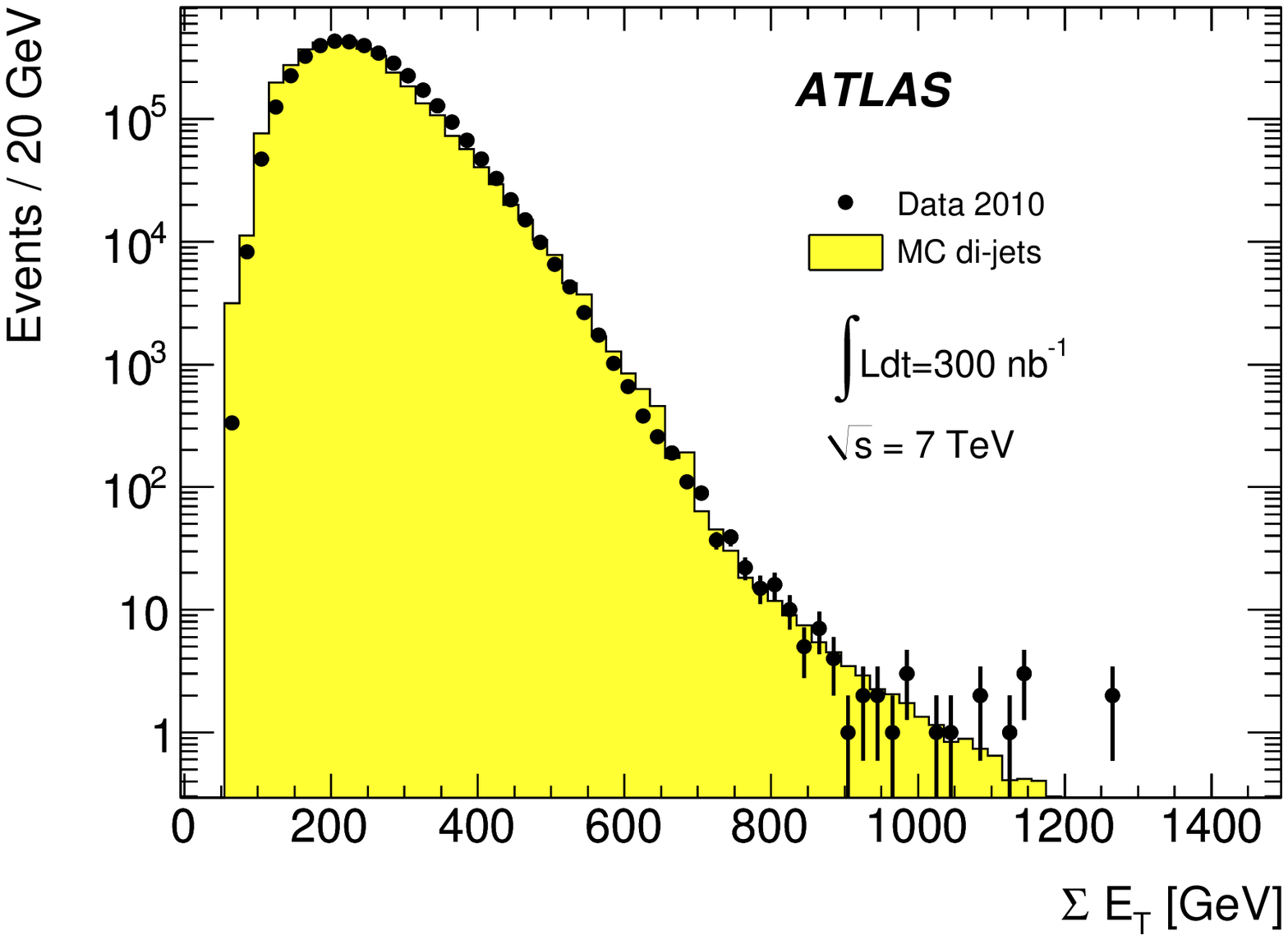}  
\hfill \includegraphics[width=.49\linewidth,height=\myFigSize]{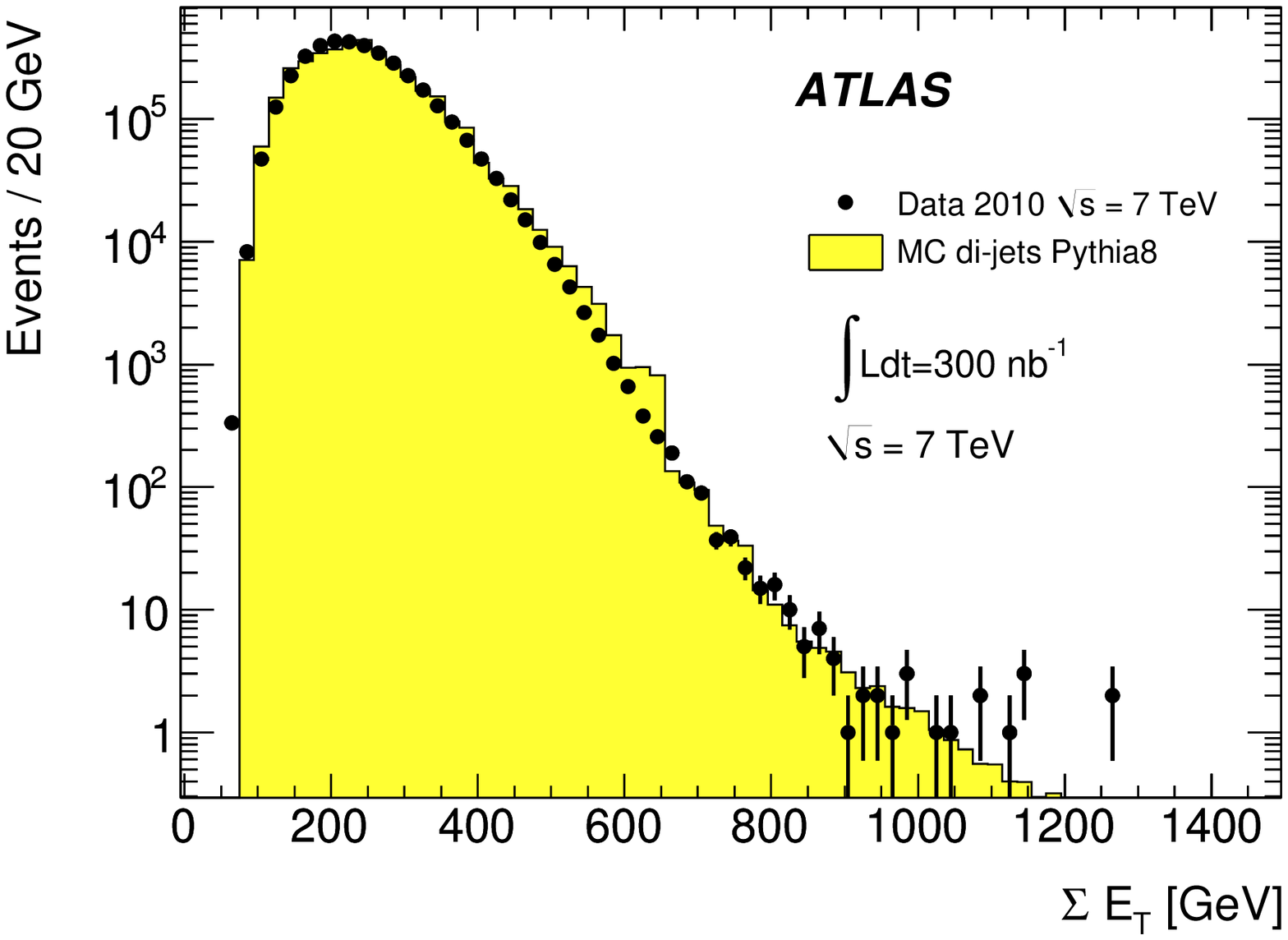}  
\end{tabular}
\end{center}
\caption{\it  Distribution of  \sumet~ as measured in a 
data sample of  minimum bias events (top) and di-jet events  (bottom) 
selecting two jets with \pT~ $>$ 25 GeV.
The expectation from MC simulation, 
normalized to the number of events in data, 
is superimposed.
On the left {\sc Pythia6} (ATLAS tune)  is compared with the data. On the right {\sc Pythia8}  is compared with the
data. }
\label{fig:sumet}
\end{figure*}

The data distributions of \sumet~ for minimum bias and di-jet events from the subset corresponding to lower pileup conditions 
(see Section \ref{sec:QCD_selection}) are compared to MC predictions from two versions of  {\sc Pythia} in Figure ~\ref{fig:sumet}. 
The left-hand distributions show comparisons with the ATLAS tune of {\sc Pythia6}. The right-hand distributions show the comparisons with the default tune of {\sc Pythia8}. 
Due to the limited number of events simulated,
    the distribution for the di-jet {\sc Pythia8} MC sample
    is not smooth, and is zero in the lowest
    \sumet\ bin populated by data.
    This is not understood, also if it can be partly explained
    by the fact that the low \sumet\  region is populated by events from
    the jet MC sample generated in the lowest parton \pT\ bin 
    (17-35 GeV), which is the most suppressed by the di-jet selection
    (a factor about 20 more than other samples) and has a large
    weight, due to cross-section.
    Moreover the {\sc Pythia8} jet MC sample in the 8-17 GeV
    parton \pT\ bin is not available.
 In the case of the minimum bias sample, due to the very 
limited number of events simulated (about a factor 25 less
respect to data), the tails in the  {\sc Pythia8} MC distribution are
strongly depleted.

 The {\sc Pythia8}  MC \cite{PYTHIA8} version used in this paper  has not yet been tuned to the ATLAS data.
  The current tune \cite{PYTHIA82} uses the {\sc CTEQ 6.1} parton distribution functions ({\sc PDF}) instead of the  
{\sc MRST LO$^{**}$} as used in {\sc Pythia6},  and its diffraction model  
differs, including 
higher-$Q^2$ diffractive processes. 
The comparison of the mean values and the shapes of the two different MC distributions with data seems to indicate that a better agreement is obtained with the {\sc Pythia8} but, 
due to the reduced {\sc Pythia8} MC statistics,
no firm conclusion can be drawn.
In the rest of the paper, the {\sc Pythia6} MC samples with the ATLAS tune are used for comparison with data; this version is used as the baseline for  {\sc Pythia} MC samples for 2010 data analyses. 

\subsection{\etmiss~ performance in $\textit{{\textbf{Z}}} \  \rightarrow \ell \ell$ events}
\label{sec:Perf_Z}
The absence of genuine \etmissmag~ in \Zll~ events, coupled with the clean event signature
and the relatively large cross-section, means that it is a good channel to study  \etmiss~ performance. 

\begin{figure*}[htbp]
\begin{center}
\begin{tabular}{lr}
\includegraphics[width=.49\linewidth, height=\myFigSize]{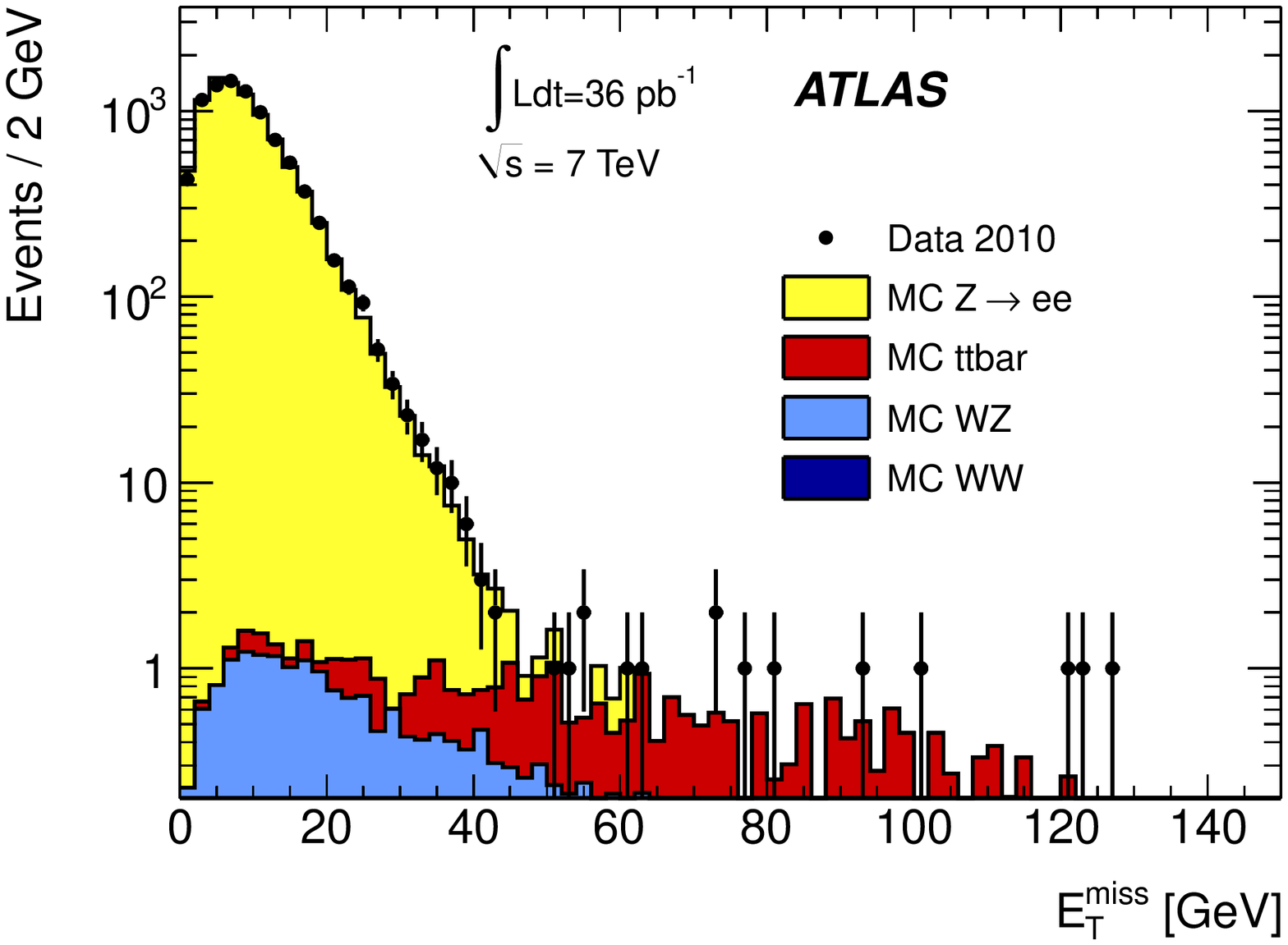} 
\includegraphics[width=.49\linewidth, height=\myFigSize]{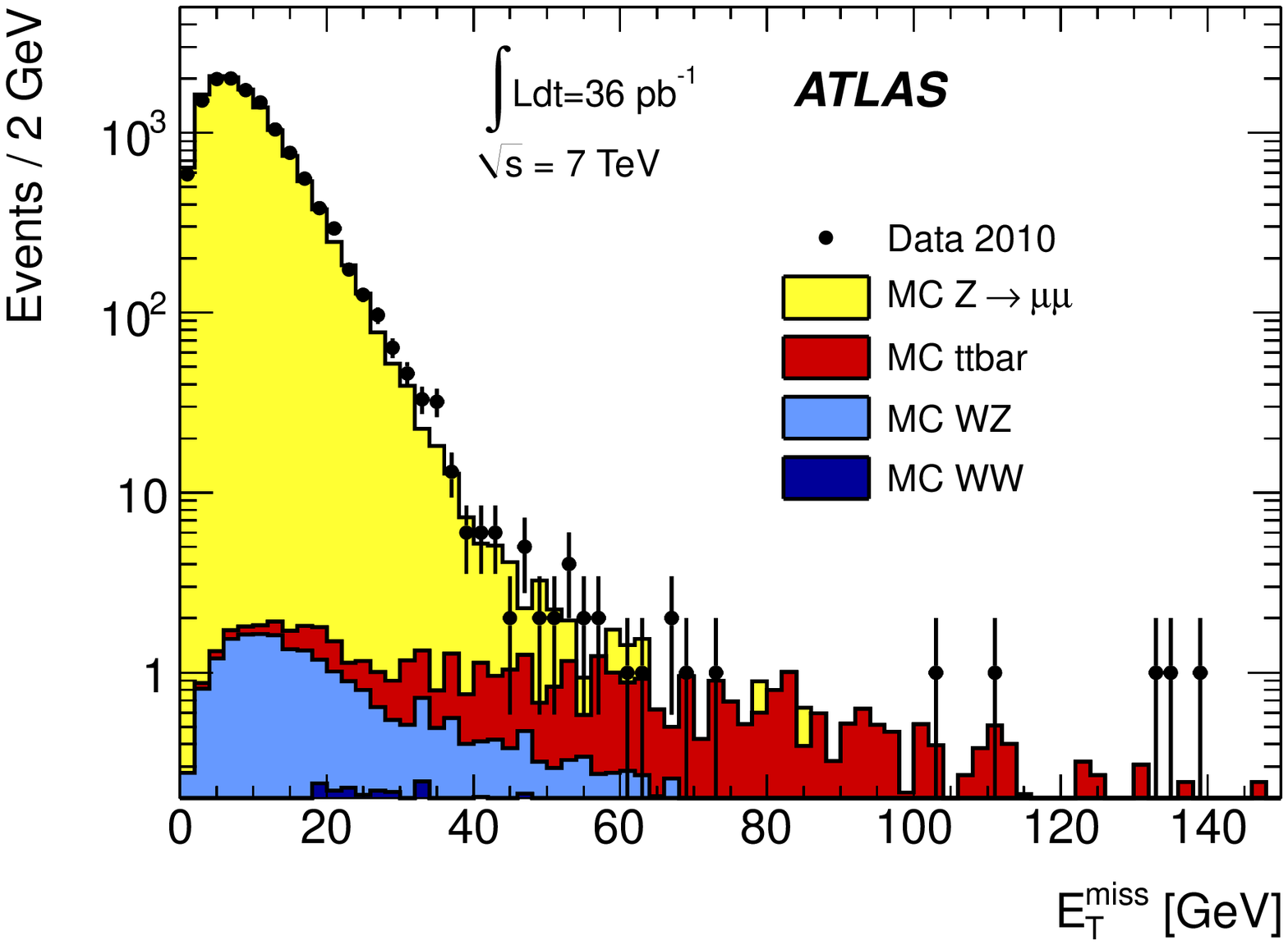} \\
\includegraphics[width=.49\linewidth, height=\myFigSize]{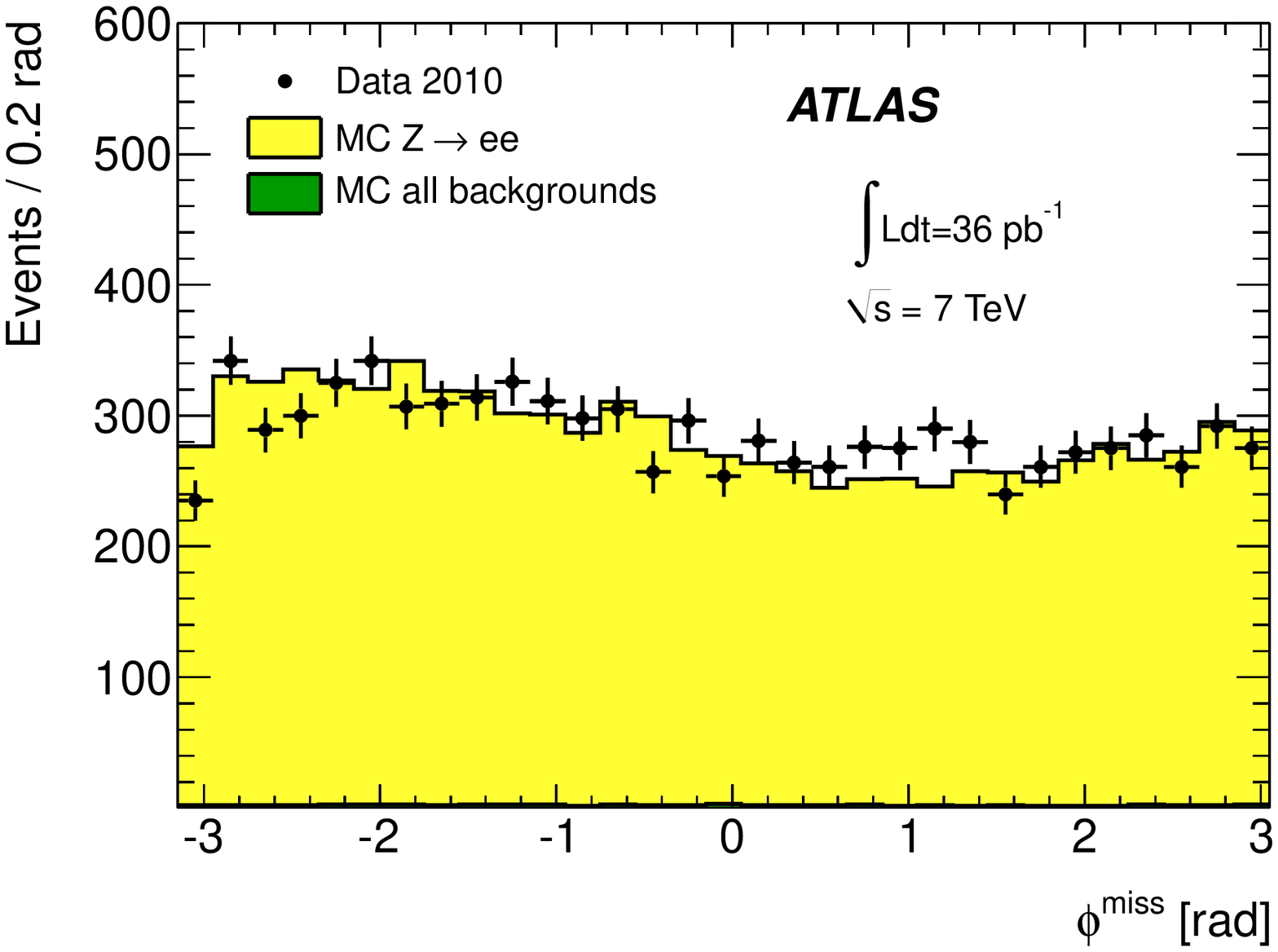} 
\includegraphics[width=.49\linewidth, height=\myFigSize]{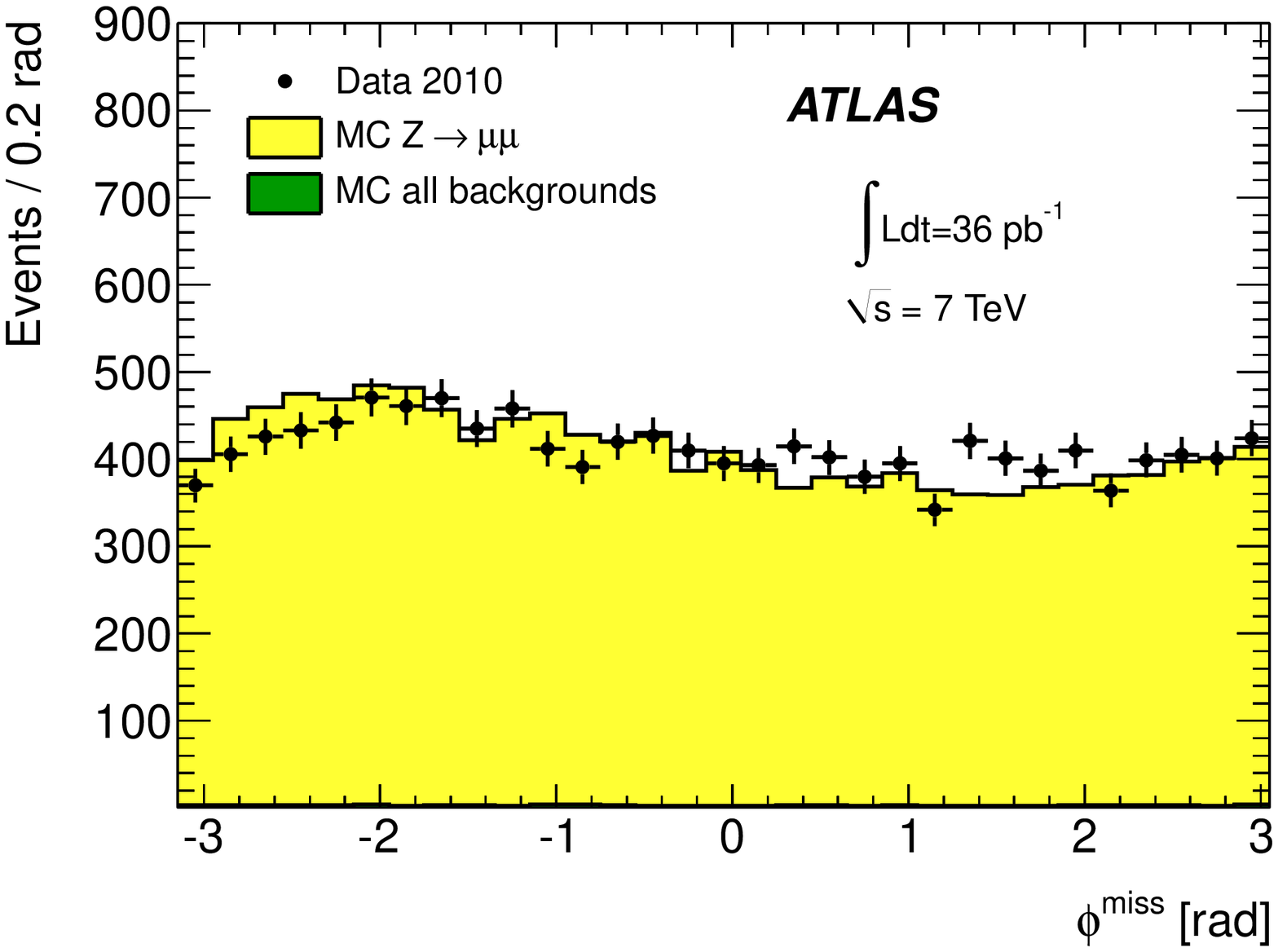}
\end{tabular}
\end{center}
\caption{\it Distribution of \etmissmag~ (top)  and \phimiss (bottom) as measured in a 
data sample of \Zee~(left)  and of \Zmm~ (right).  
The expectation from Monte Carlo simulation is superimposed and normalized to data, after each MC sample is 
weighted with its corresponding cross-section. The sum of all backgrounds is shown in the lower plots.
}
\label{fig:METZ_basic}
\end{figure*}

\begin{figure*}
\begin{center}
\begin{tabular}{lr}
\includegraphics[width=.49\linewidth, height=\myFigSize]{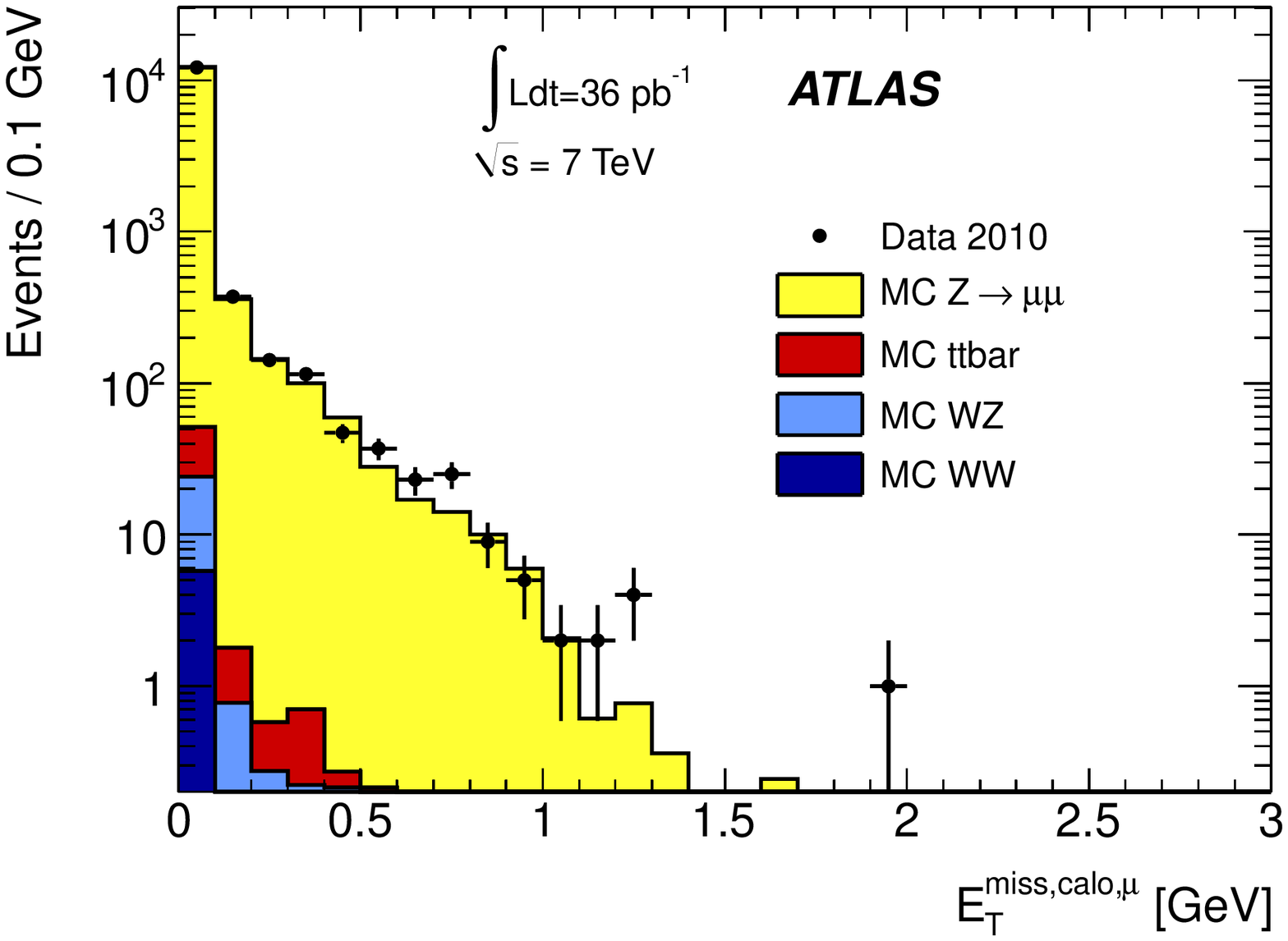}
\includegraphics[width=.49\linewidth, height=\myFigSize]{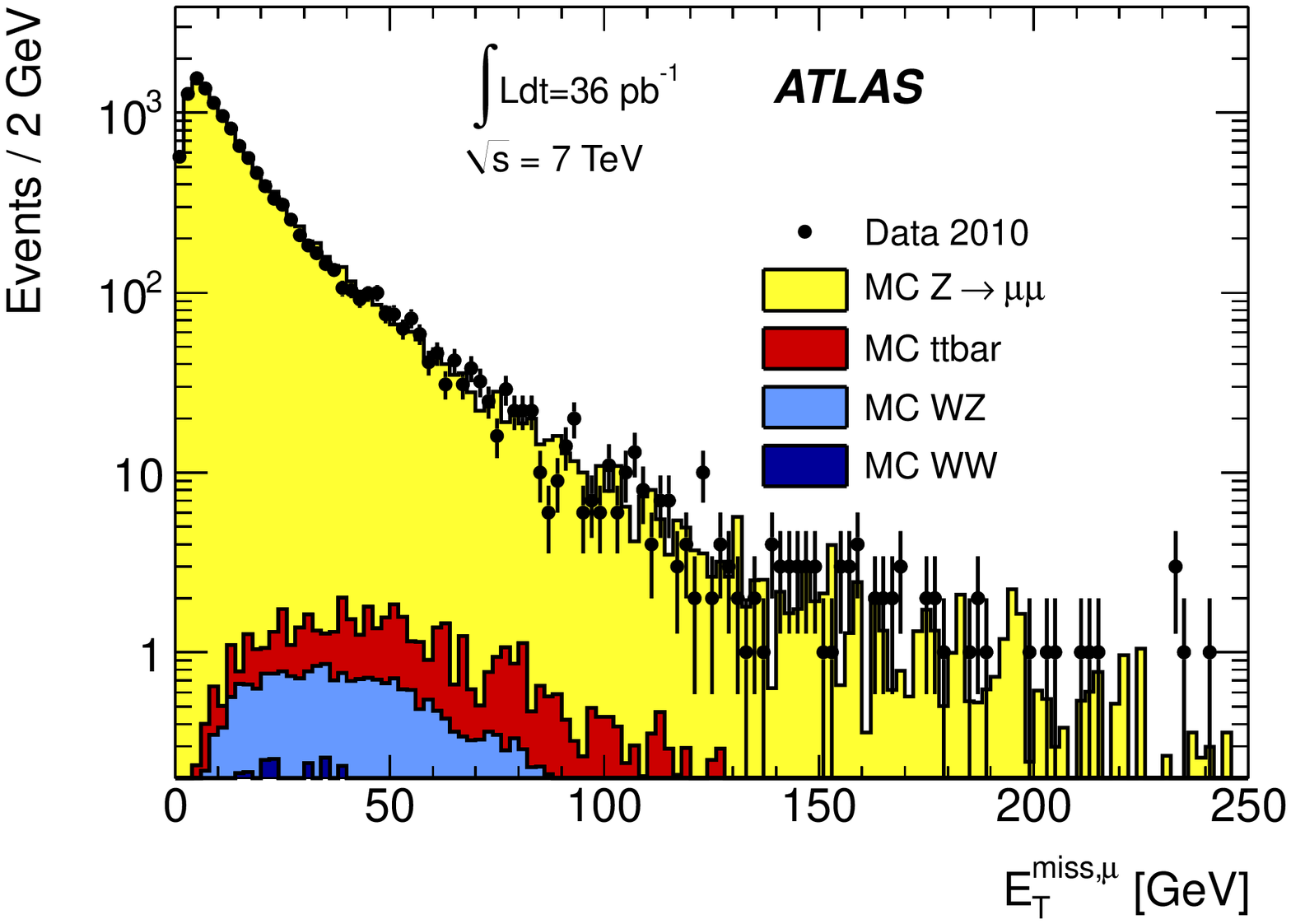}
\end{tabular}
\end{center}
\caption{\it Distribution of \etmissmag~computed with calorimeter cells associated to 
muons  ($E_{\mathrm{T}}^{\mathrm{miss,calo},\mu}$) (left) and computed from reconstructed muons ($E_{\mathrm{T}}^{\mathrm{miss,\mu}}$) (right) for \Zmm~ data.
The expectation from Monte Carlo simulation is superimposed and
normalized to data, after each  MC sample is 
weighted with its corresponding cross-section.
}  
\label{fig:METZ_terms2}
\end{figure*}

\begin{figure*}
\begin{center}
\begin{tabular}{lr}
\includegraphics[width=.49\linewidth, height=\myFigSize]{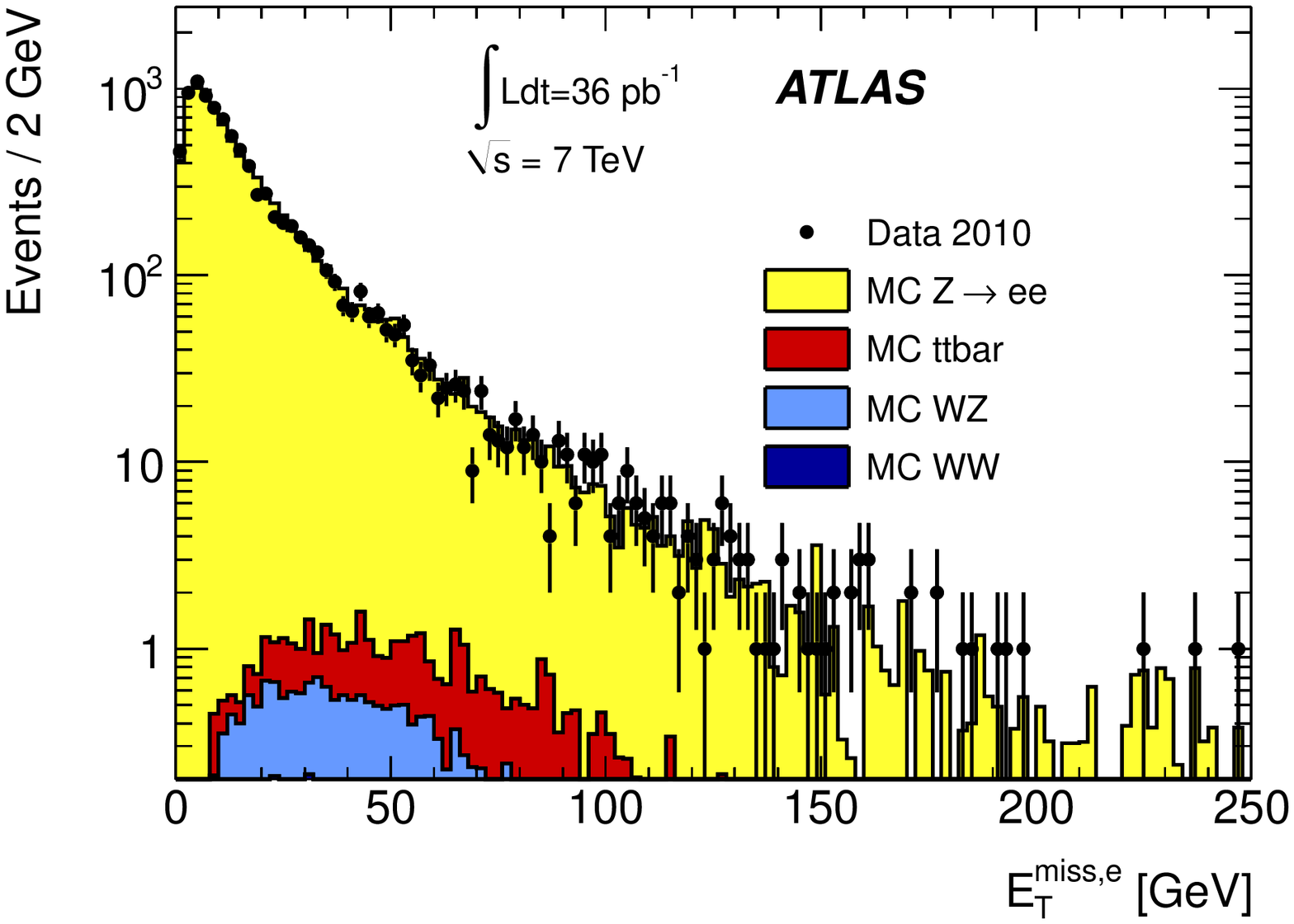}
\includegraphics[width=.49\linewidth, height=\myFigSize]{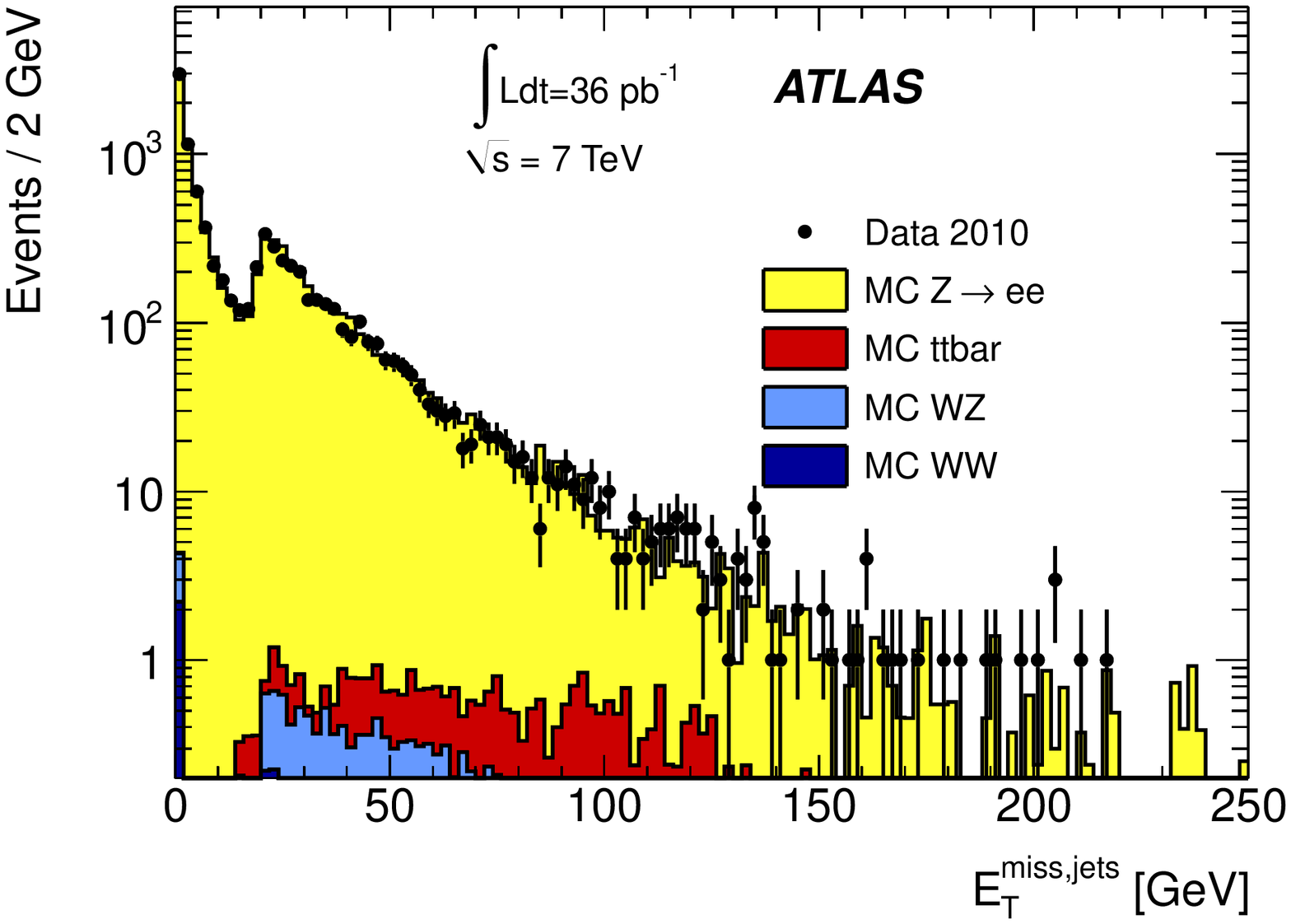}\\
\includegraphics[width=.49\linewidth, height=\myFigSize]{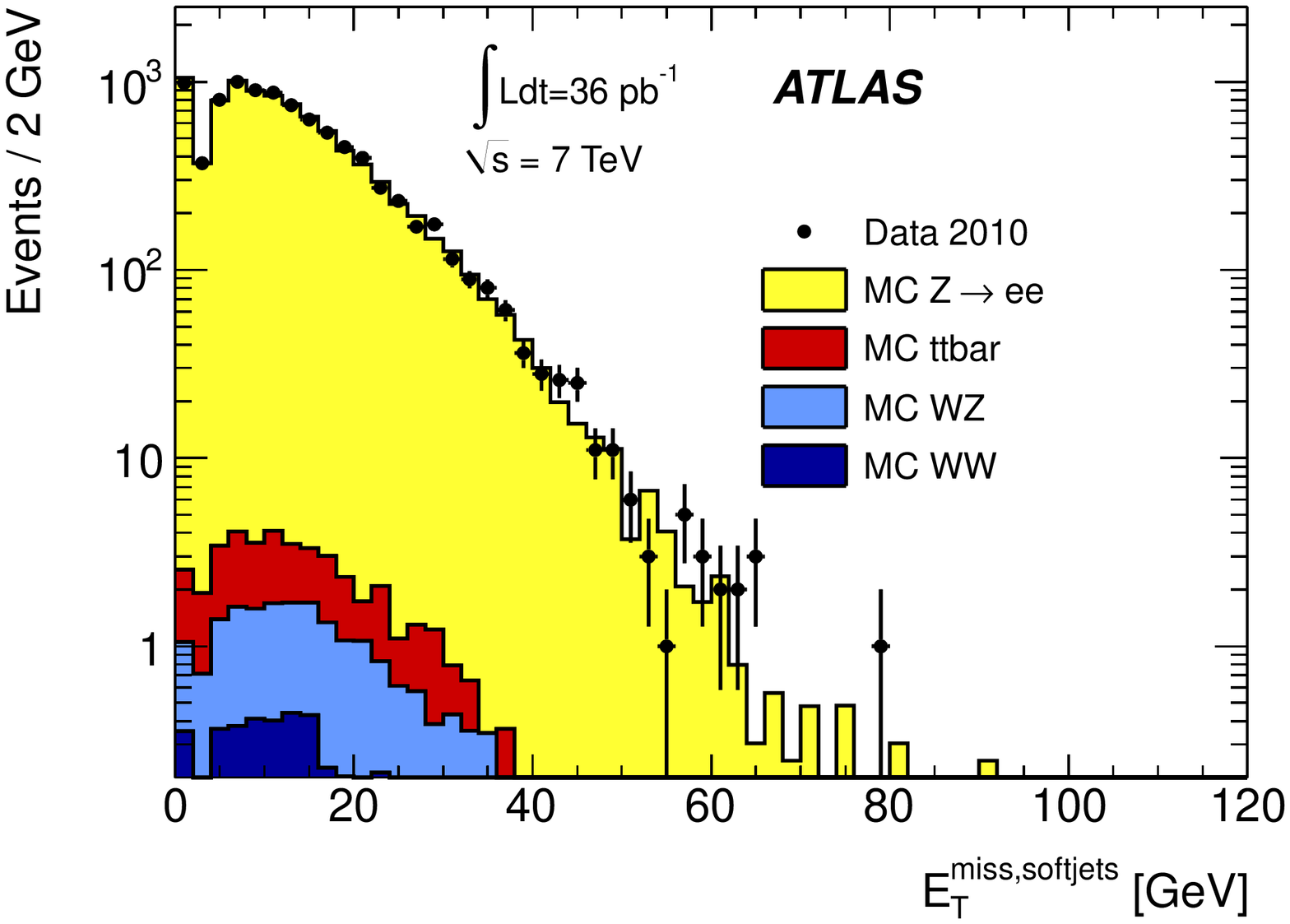}
\includegraphics[width=.49\linewidth, height=\myFigSize]{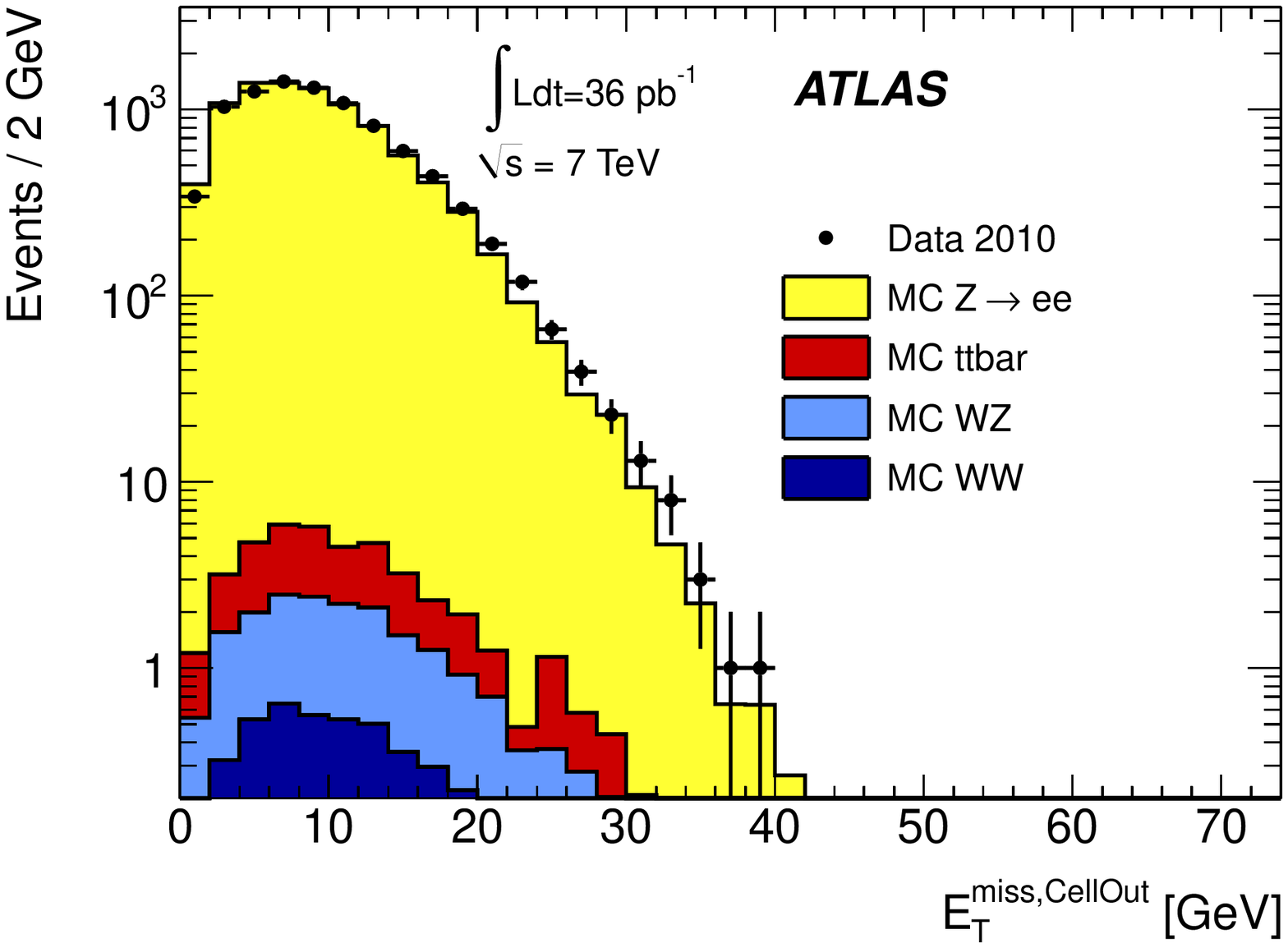}
\end{tabular}
\end{center}
\caption{\it Distribution of \etmissmag~computed with cells associated to 
electrons ($E_{\mathrm{T}}^{\mathrm{miss},e}$)  (top left), jets with 
\pT~ $ >  20$ GeV ($E_{\mathrm{T}}^{\mathrm{miss,jets}}$) (top right), jets with 7 GeV $< $ \pT~ $<$  20 GeV ($E_{\mathrm{T}}^{\mathrm{miss,softjets}}$) (bottom left) and from topoclusters outside 
reconstructed objects ($E_{\mathrm{T}}^{\mathrm{miss,CellOut}}$) (bottom right) for \Zee~ data.
The expectation from Monte Carlo simulation is superimposed and 
normalized to data, after each MC sample is 
weighted with its corresponding cross-section.
}  
\label{fig:METZ_terms3}
\end{figure*}

The distributions of \etmissmag~ and  \phimiss ~  for data and MC simulation 
are shown in Figure~\ref{fig:METZ_basic} for \Zee~ and \Zmm~ events.
The contributions  due to muons
 are shown  for \Zmm~ events in Figure~\ref{fig:METZ_terms2}. Both the contributions from energy deposited in calorimeter cells associated to muons, taken at the EM scale, and the contributions from reconstructed muons are shown. 
For \Zee~ events, the contributions from electrons, jets, soft jets and topoclusters outside
 the reconstructed objects are shown separately in  
 Figure~\ref{fig:METZ_terms3}.
The peak at zero in the distribution of the jet term
corresponds to events where there are no jets with \pT~ above 20 GeV, and the
small values ($<$ 20 GeV) in the distribution are due to events with two jets whose
transverse momenta balance.
The MC simulation expectations, from \Zll~ events and from the dominant SM backgrounds,  are superimposed. 
Each MC sample is  weighted with its corresponding cross-section and then the total MC expectation is normalized to the number of events in data.
Reasonable agreement between data and MC simulation is observed in all distributions.

 Events in the tails of the \etmissmag~ distributions in Figure~\ref{fig:METZ_basic} have been carefully checked.
The 22 events with the highest $E_{\mathrm{T}}^{\mathrm{miss}}$ values, above $60$ GeV, have been examined in detail to check whether they are related to cosmic-ray muon background,  fake muons, badly measured jets or jets pointing to dead calorimeter regions.
 The events in the tails 
  are found to be compatible with either signal candidates, including \ttbar,   $WW$ and $WZ$ di-boson events, all involving real $E_{\mathrm{T}}^{\mathrm{miss}}$, or events in which the \etmiss~vector is close to a jet in the transverse plane.  The latter category of events can arise from mis-measured jets,
and be rejected at the analysis level with cuts on $\Delta\phi$({\bf jet}, \etmiss) (see Section \ref{sec:met_rec}).
 
\subsubsection{Measuring \etmiss~ response in $\Zll$ events}

From the event topology \cite{JET_ETMISS} in events with $\Zll$ decay one can define an axis in the transverse plane 
such that the component of \etmiss~ along this axis is sensitive to detector resolution and biases. 
The direction of this axis,  $\mathbf { \textbf{ \textit {A}}_Z}$, is defined by the reconstructed momenta
 of the leptons:
\begin{equation}
\mathbf {\textbf{ \textit {A}}_Z}=(\mathbf{{ \textbf{ \textit { p}}_T}^{\ell^+}}+\mathbf{{ \textbf{ \textit { p}}_T}^{\ell^-}})/|\mathbf{{ \textbf{ \textit {p}}_T}^{\ell^+}}+\mathbf{{ \textbf{ \textit {p}}_T}^{\ell^-}}|
\end{equation}
where $\mathbf{{ \textbf{ \textit {p}}_T}^{\ell}}$ are the vector transverse momenta of the lepton and anti-lepton.
The direction of $\mathbf { \textbf{ \textit {A}}_Z}$ thus reconstructs the direction of motion of the Z boson.
The perpendicular axis in the transverse plane, $\mathbf { \textbf{ \textit {A}}_{AZ}}$, is a unit vector placed at right angles to
$\mathbf{\textbf{ \textit {A}}_Z}$, with positive direction anticlockwise from the direction of the $Z$ boson.

\begin{figure*}[ht]
\begin{center}
\begin{tabular}{lr}
\includegraphics[width=0.49\linewidth, height=\myFigSize]{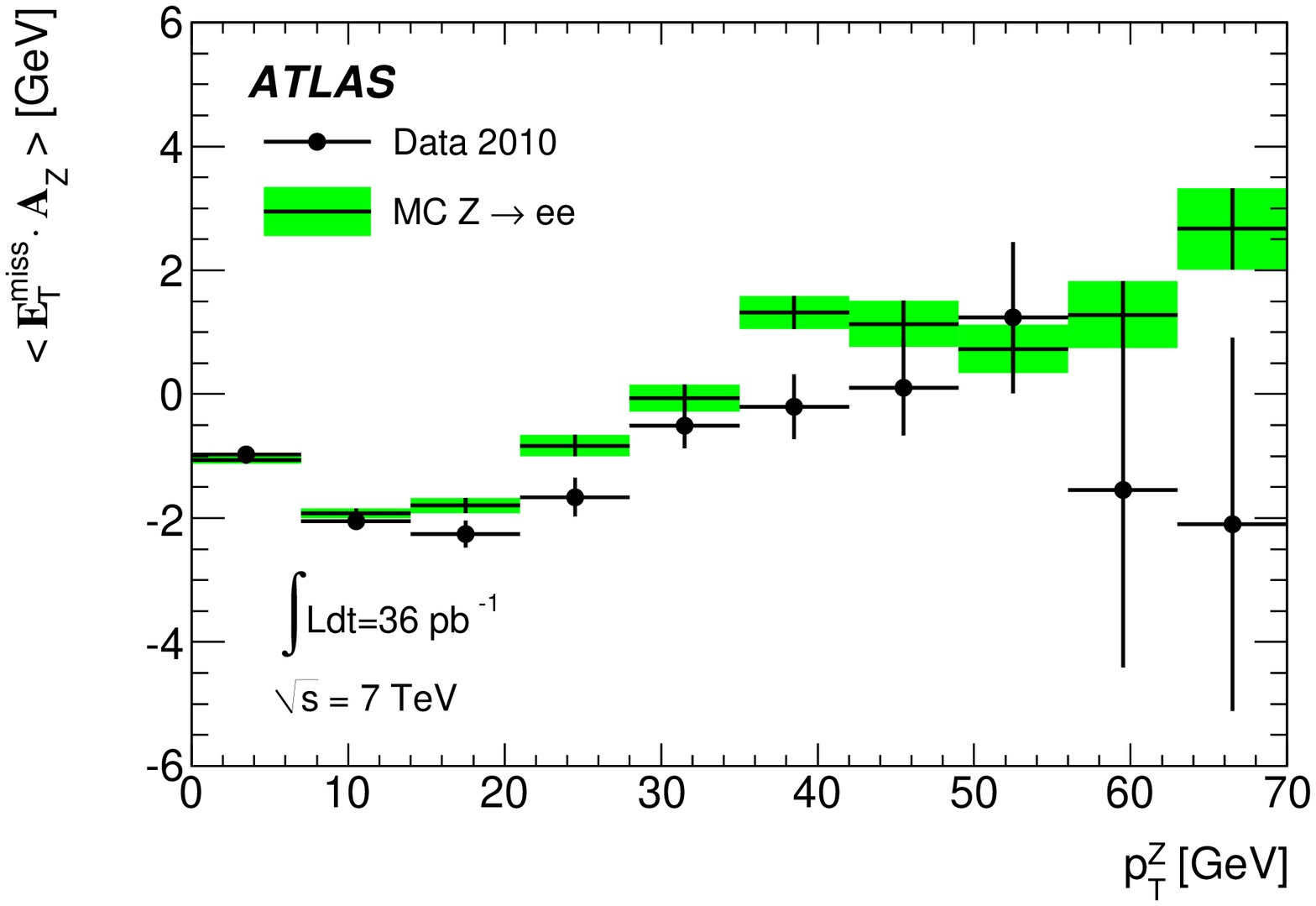} \label{figure_Diagnostic_elecPar}
  \includegraphics[width=0.49\linewidth, height=\myFigSize]{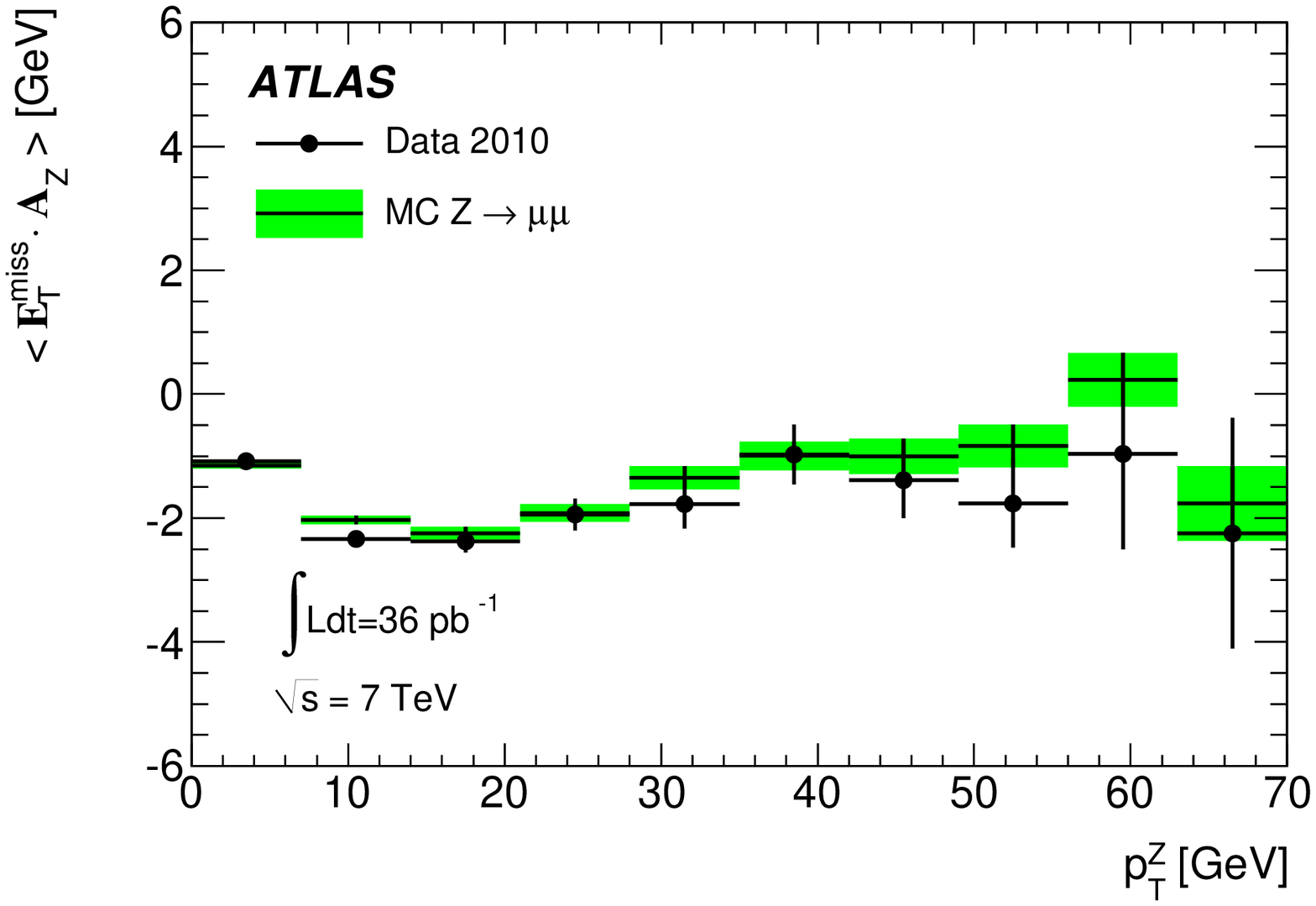} \label{figure_Diagnostic_muonPar} \\
\end{tabular}
\end{center}
\caption{\it Mean values of $\MEtZ$ 
 as a function of \pTZ~ in $\Zee$ (left) and $\Zmm$ (right) events.}
\label{figure_Diagnostic}
\end{figure*}

The mean value of the projection of \etmiss~  onto the longitudinal axis, $\langle\MEtZ\rangle$, 
is a measure of the \etmiss~ scale, as this axis is sensitive to the balance between
the leptons and the hadronic recoil. 
Figure  \ref{figure_Diagnostic}  shows the value of $\langle\MEtZ\rangle$ as a
function of \pTZ. These mean values are used as a diagnostic to validate the \etmiss~ 
reconstruction algorithms. If the leptons perfectly balanced the hadronic recoil,
regardless of the net momentum of the lepton system, then the \MEtZ~ would
be zero, independent of \pTZ. 
Instead, $\langle\MEtZ\rangle$ displays a small bias in both 
the electron and muon channels which is reasonably reproduced by the MC simulation.
The observed bias  is slightly negative for low values of 
\pTZ, suggesting either that the \pT~ of the lepton system is overestimated or that the 
magnitude of the hadronic recoil is underestimated.  The same sign and magnitude of bias is seen in both electron and muon channels, suggesting that 
the hadronic recoil, here dominated by $E_{\mathrm{T}}^{\mathrm{miss,CellOut}}$ and by soft jets, is the source of bias. The component of the \etmiss~ along the perpendicular axis, $\MEtAZ$, 
displays no bias, and, indeed there is no mechanism for generating such a bias.

\begin{figure*}[ht]
\centering
\includegraphics[width=0.49\linewidth, height=\myFigSize]{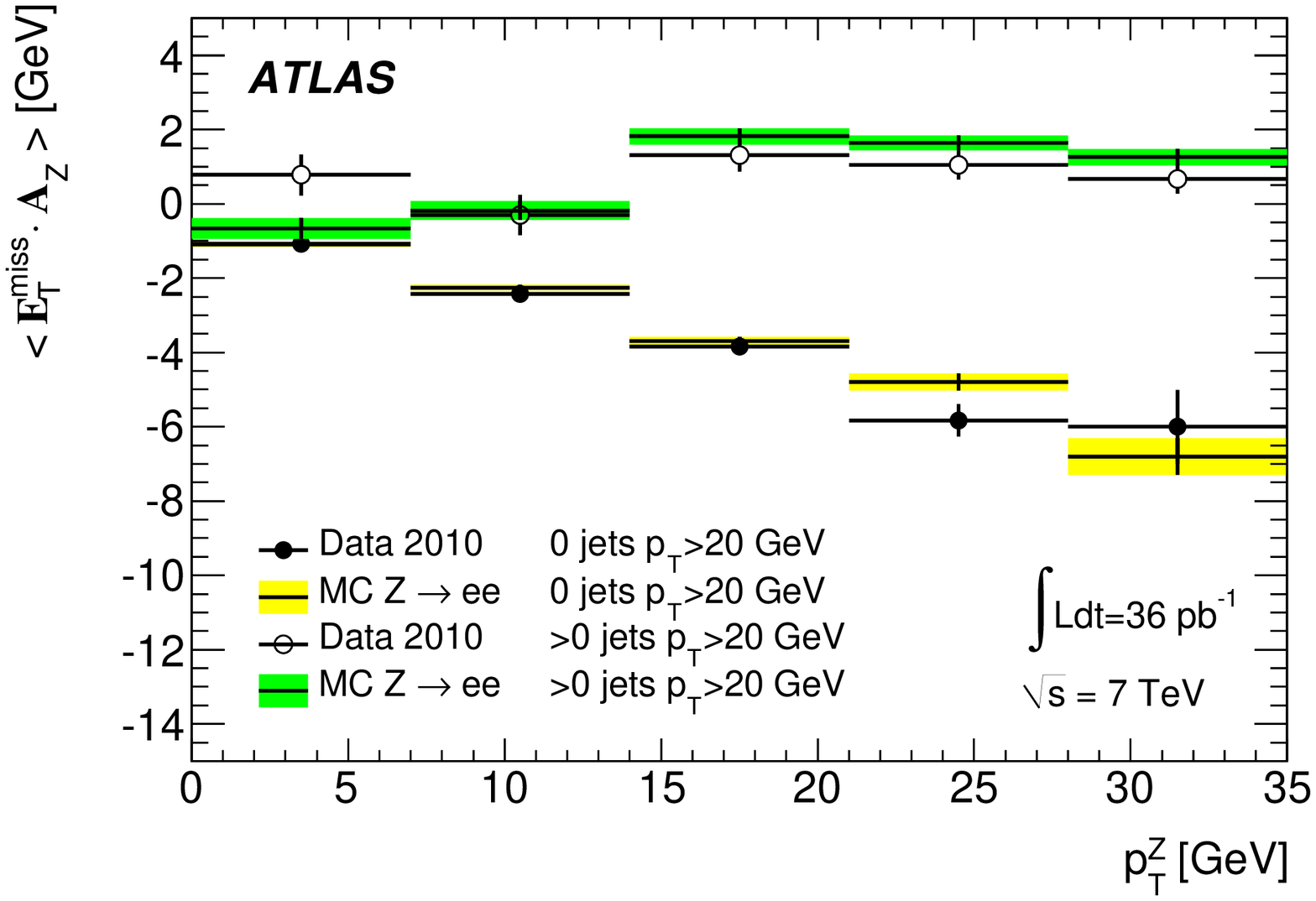}
\includegraphics[width=0.49\linewidth, height=\myFigSize]{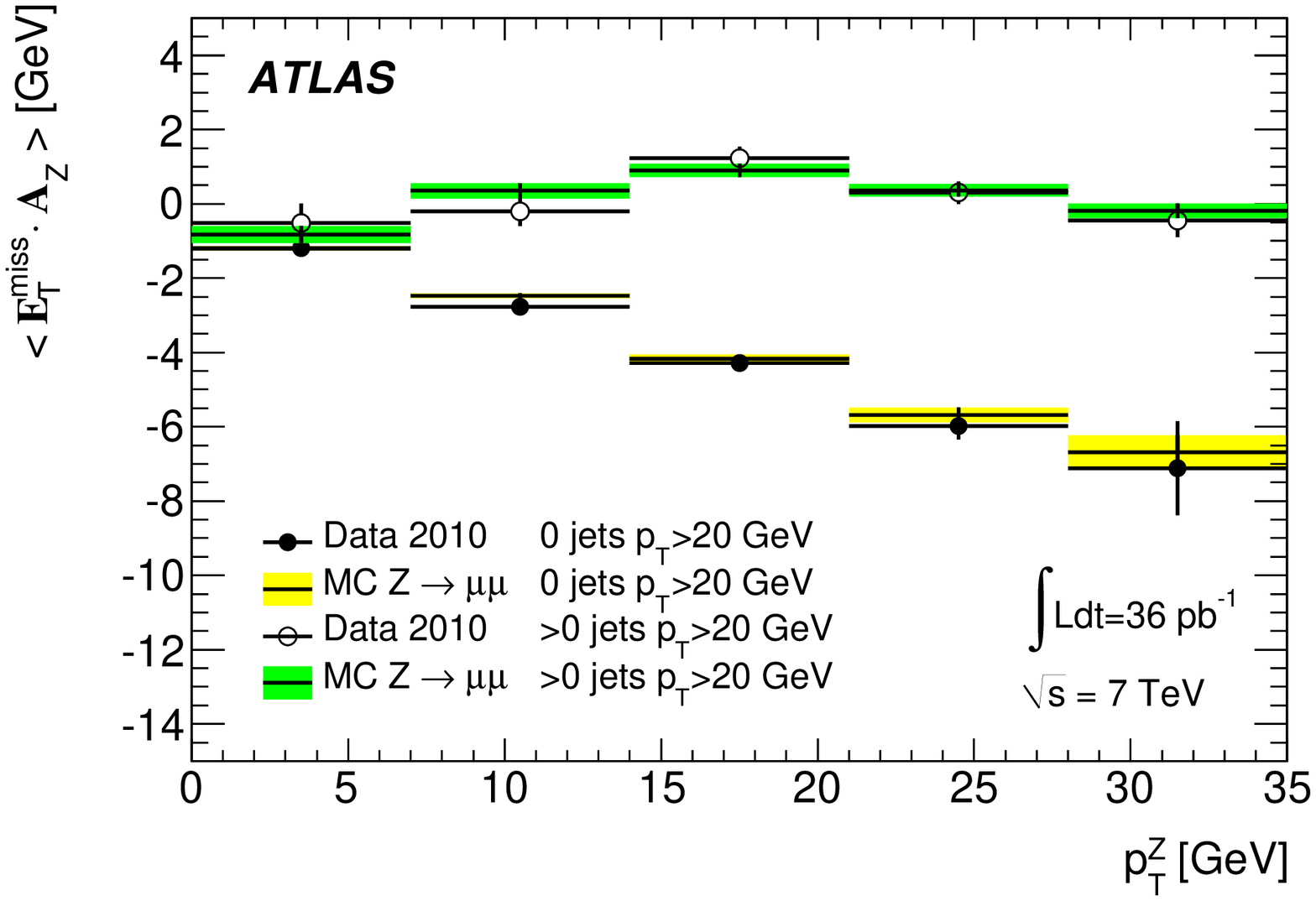} 
\caption{\it Mean value of $\MEtZ$ as a function of \pTZ~ requiring either zero jets with \pT~$>$ 20 GeV or at least 1 jet with \pT~$>$ 20 GeV in the event for \Zee~ (left) and \Zmm~ (right) events.  \label{figure_Diagnostic_byJetNum}}
\end{figure*}

\begin{figure*}[htbp]
\begin{center}
\begin{tabular}{lr}
\includegraphics[width=.49\linewidth, height=\myFigSize]{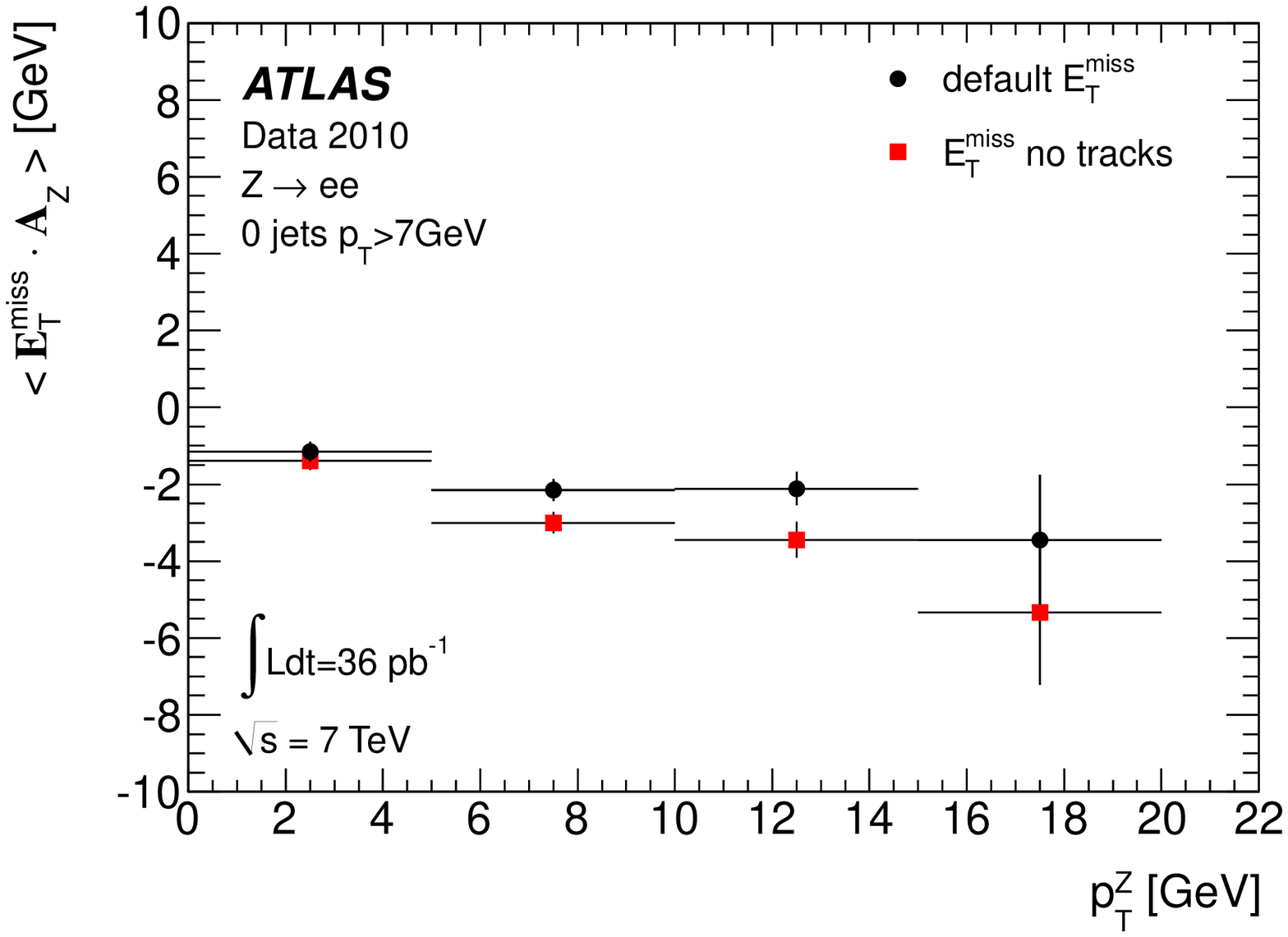}
\includegraphics[width=.49\linewidth, height=\myFigSize]{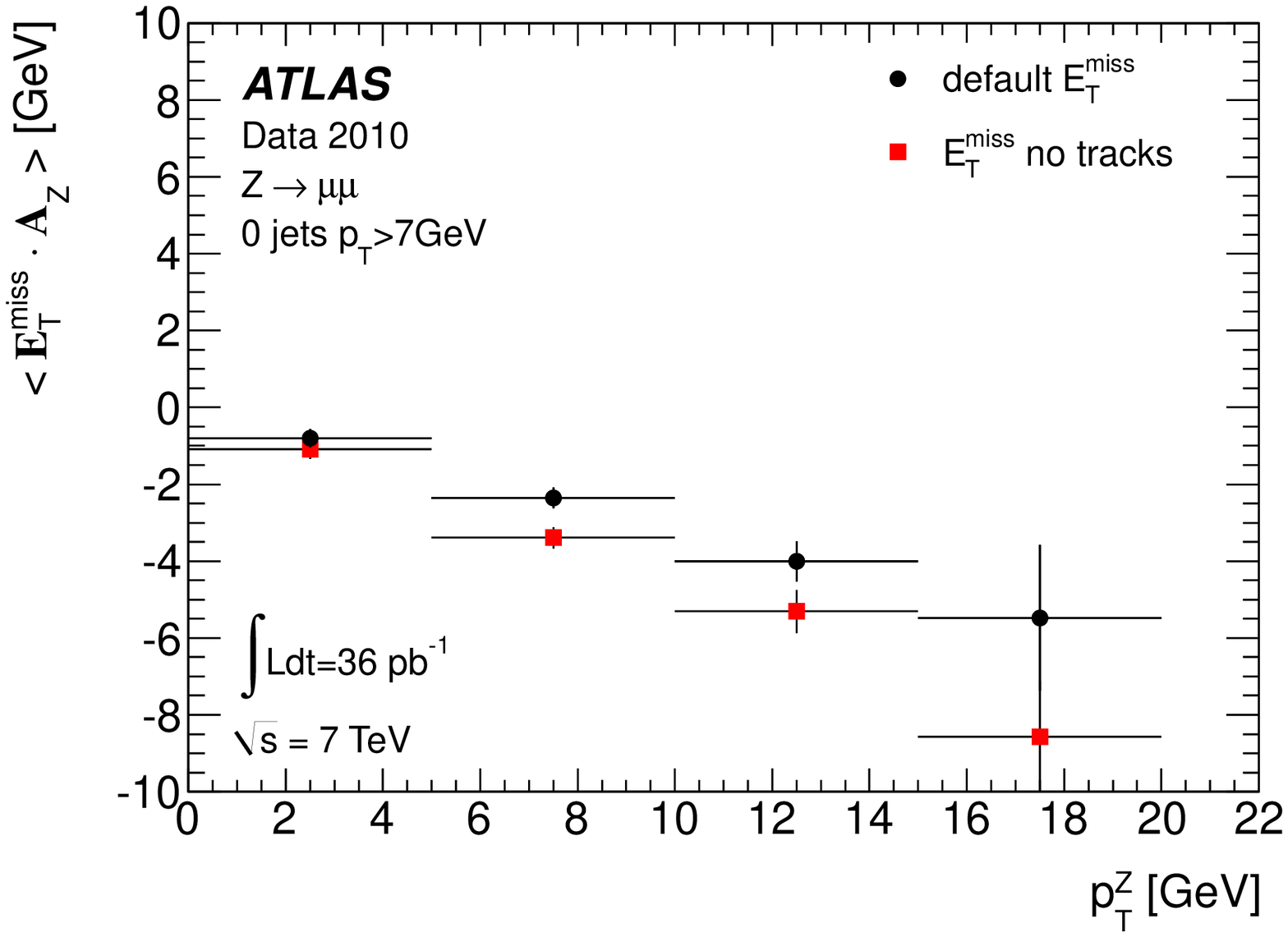}
\end{tabular}
\end{center}
\caption{\it Mean value of $\MEtZ$ as a function of \pTZ~ in \Zee~(left) and \Zmm~(right) 
 for events with no jets with \pT~$>$ 7 GeV.
The default  \etmiss~ is compared with \etmiss~ calculated in the same way with the exception that the track-cluster matching  algorithm is not used for the calculation of $E_{\rm{T}}^{\mathrm{miss,CellOut}}$.}
\label{fig:METZ_diagn}
\end{figure*}
In Figure \ref{figure_Diagnostic_byJetNum} the dependences of  $\langle\MEtZ\rangle$ on \pTZ~ 
are shown separately for events with $\Zll$  produced in association with zero jets or with at least one jet,  with the jet definition as described in Section \ref{sec:QCD_selection}.
The figure demonstrates that there
is a negative bias in $\langle\MEtZ\rangle$ for events with zero jets, which increases with \pTZ~
 up to 6 GeV. 
 A similar bias is observed in both electron and muon channels, hence it is interpreted as coming from imperfections in the calibration of the soft hadronic recoil (the $E_{\mathrm{T}}^{\mathrm{miss,CellOut}}$ and the  $E_{\mathrm{T}}^{\mathrm{miss,softjets}}$ terms).
In events with at least one jet there is a small positive bias
in the electron channel at high \pTZ, which is visible also in the muon channel for \pTZ~ in the region 15-20 GeV.

Figure \ref{fig:METZ_diagn} shows $\langle\MEtZ\rangle$ for  \Zll~ events where there are neither high \pT~ nor soft jets, for two cases of \etmiss~ reconstruction: 
calculating the $E_{\mathrm{T}}^{\mathrm{miss,CellOut}}$ term with  the track-cluster matching algorithm (see Section \ref{sec:eflow})
 or calculating this term from the calorimeter topoclusters only (denoted as \etmissmag~ no tracks).
The plots show a lower bias for the case with the track-cluster matching algorithm, indicating that it improves the reconstruction of the  $E_{\mathrm{T}}^{\mathrm{miss,CellOut}}$ term.

\subsection{\etmiss~ performance in $\textit{{\textbf{W}}} \  \rightarrow \ell \nu$ events}
\label{sec:Perf_W}

In this section the  \etmiss~ performance is studied  in \Wen~ and \Wmun~ events.
 In these events genuine \etmissmag~ is expected due to the
presence of the neutrino, therefore  the \etmissmag~ scale can be checked.

\begin{figure*}[htbp]
\begin{center}
\begin{tabular}{lr}
\includegraphics[width=.49\linewidth,height=\myFigSize]{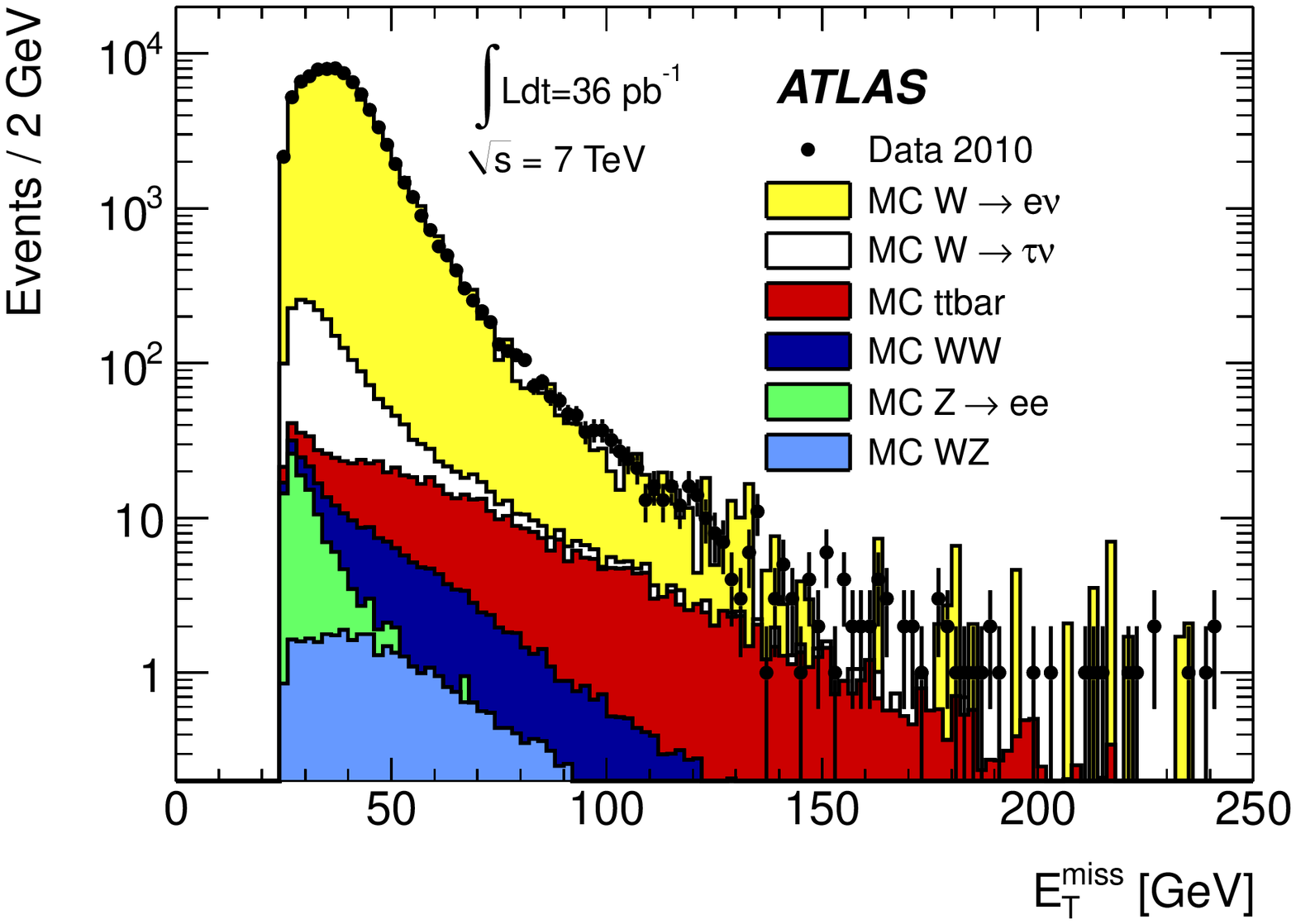} 
\includegraphics[width=.49\linewidth,height=\myFigSize]{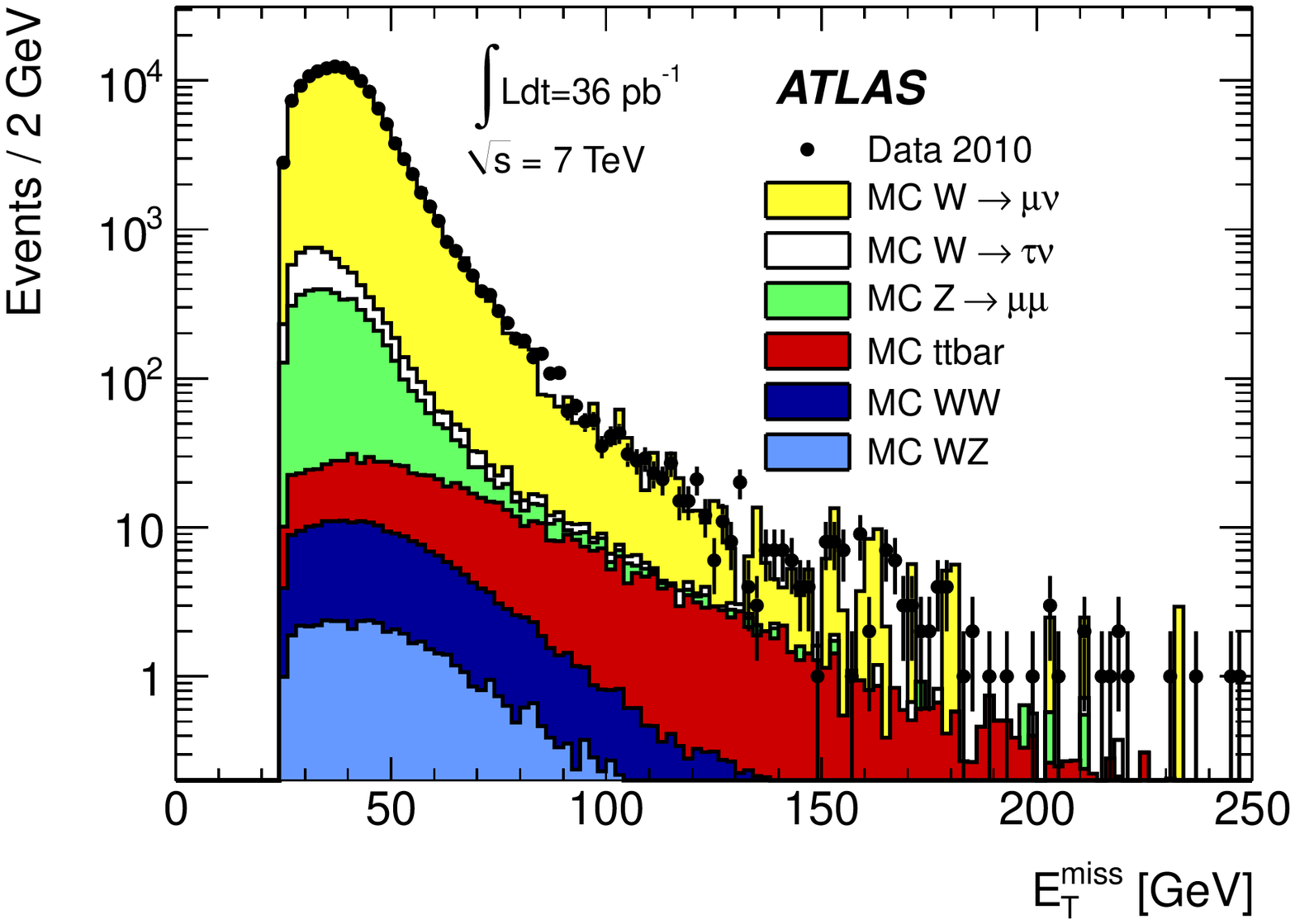} \\
\includegraphics[width=.49\linewidth,height=\myFigSize]{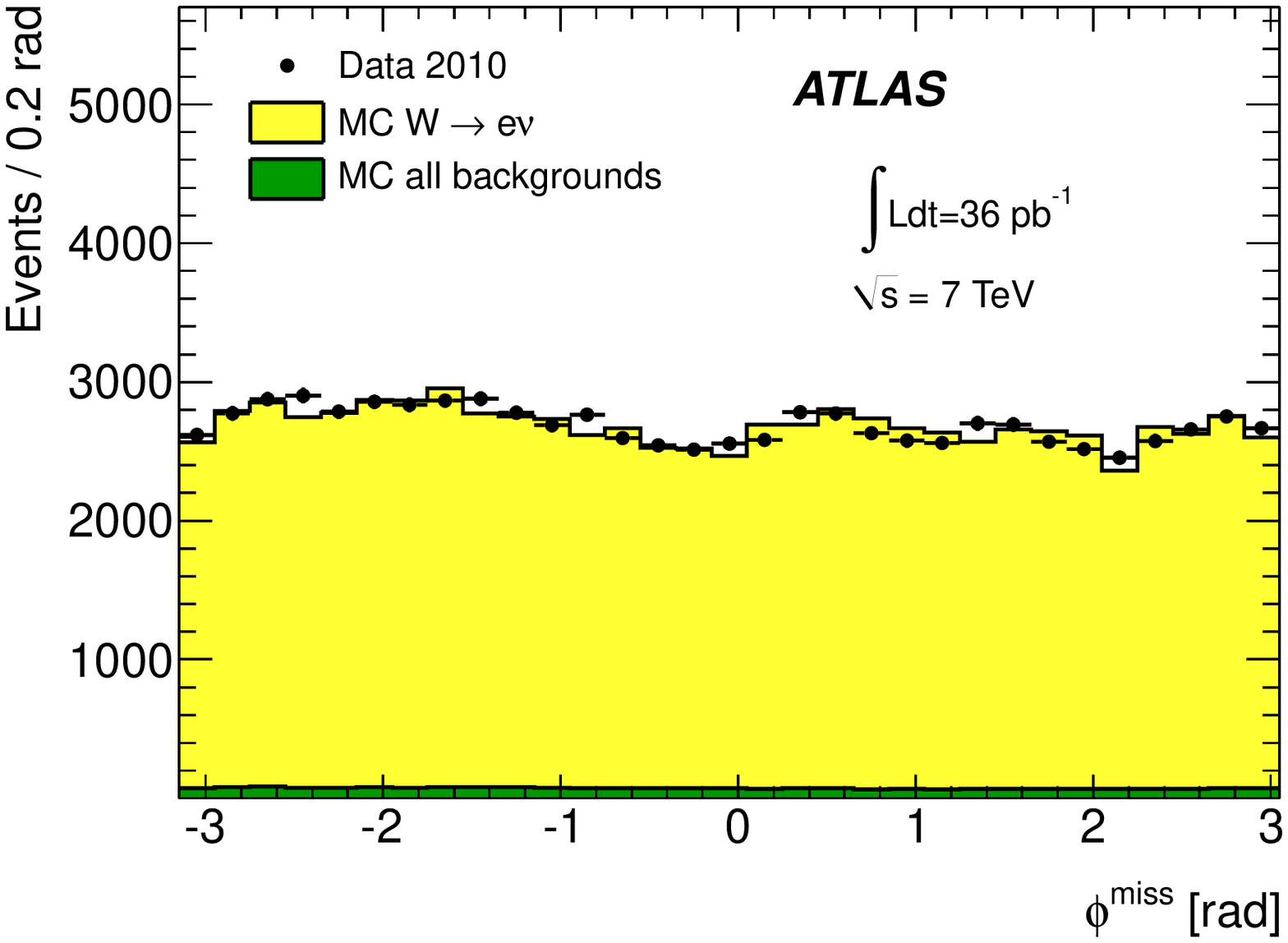}
\includegraphics[width=.49\linewidth, height=\myFigSize]{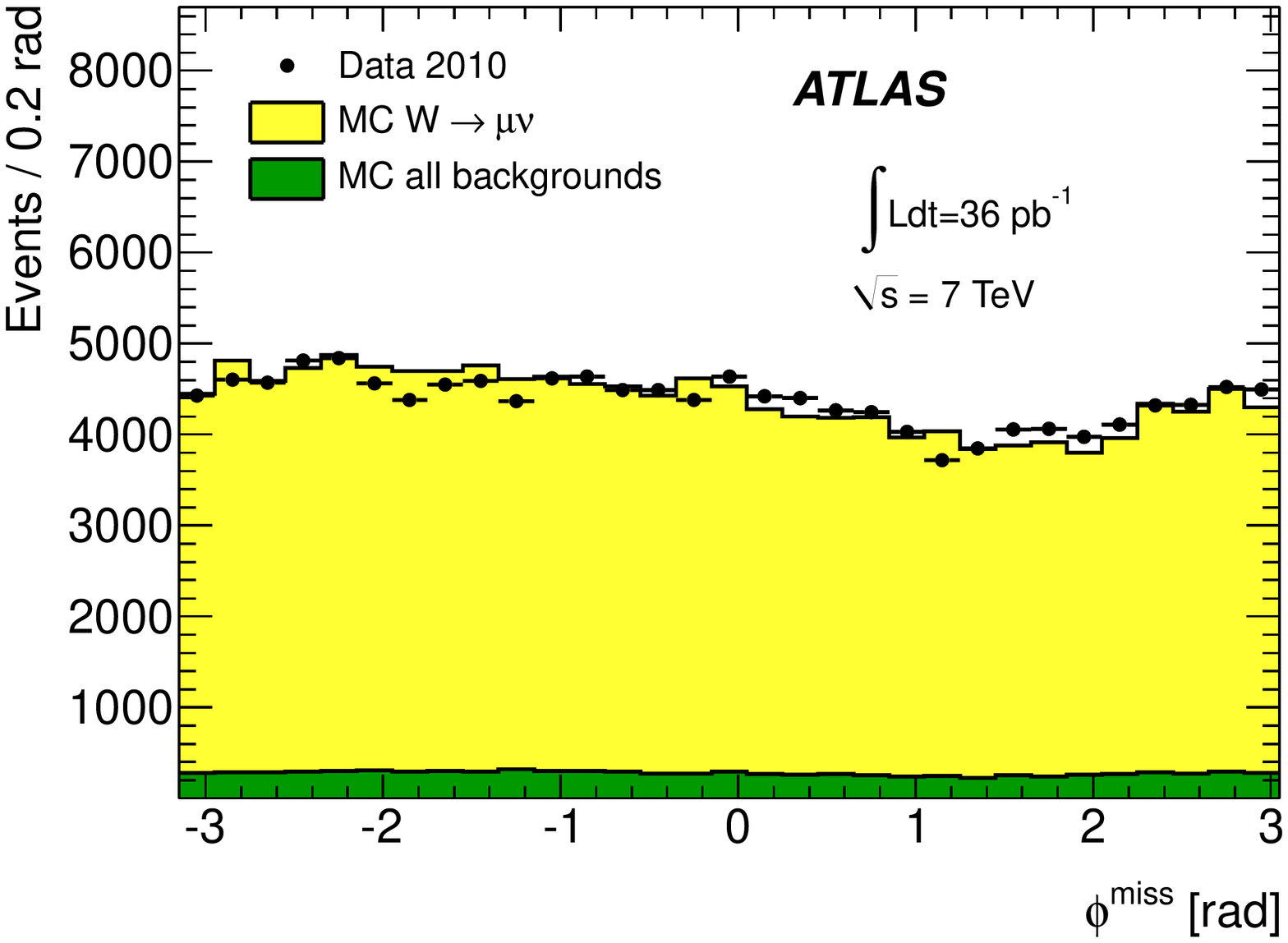} 
\end{tabular}
\end{center}
\caption{\it Distribution of \etmissmag~(top) and  \phimiss~(bottom)  as measured in a data sample of \Wen~(left) and \Wmun~(right) events. 
The expectation from Monte Carlo simulation 
is superimposed and
 normalized to data, after each MC sample is 
weighted with its corresponding cross-section. The sum of all backgrounds is shown in the lower plots.
}
\label{fig:METW_basic}
\end{figure*}

\begin{figure*}[htbp]
\begin{center}
\begin{tabular}{lr}
\includegraphics[width=.49\linewidth,height=\myFigSize]{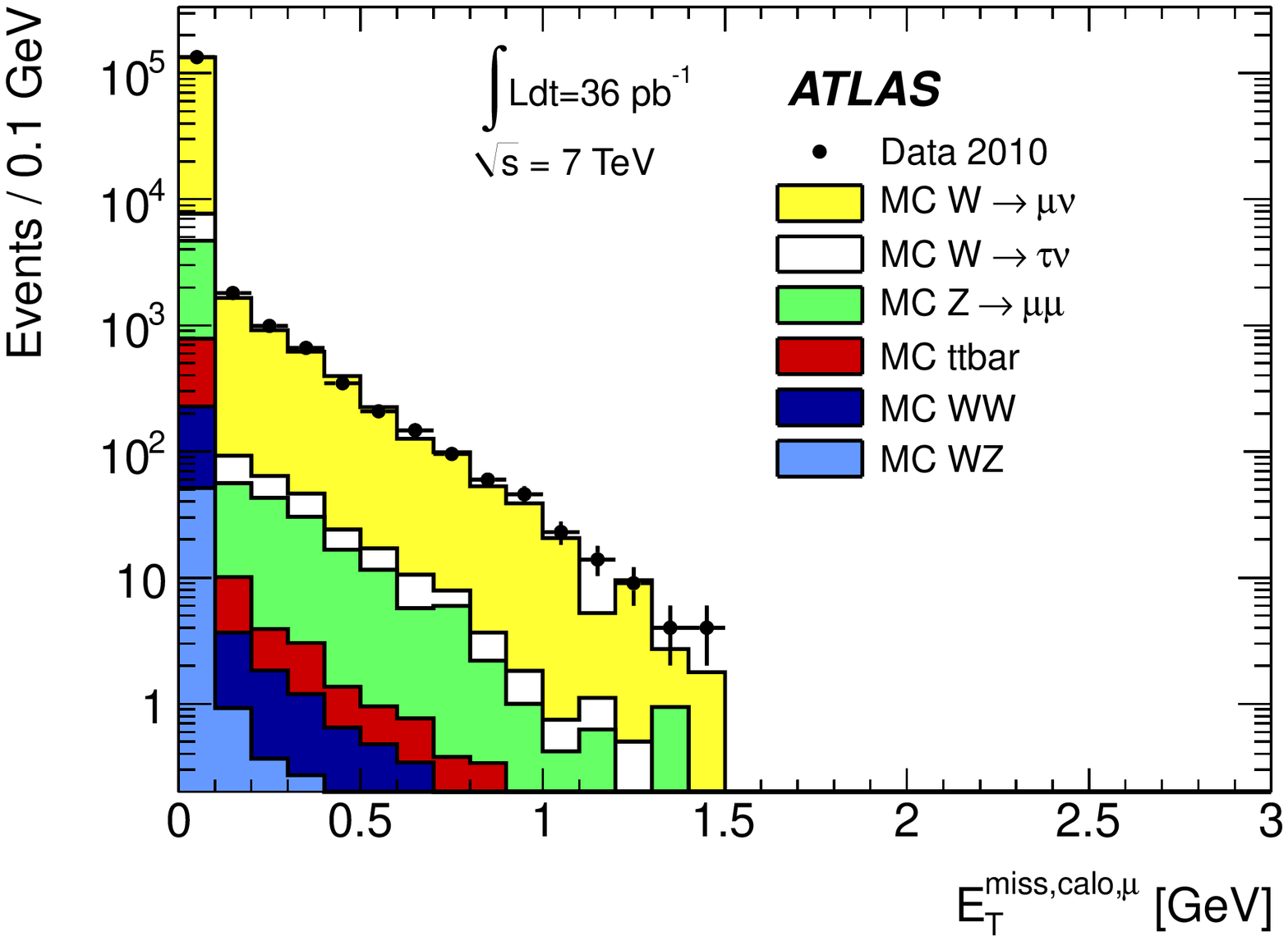}
\includegraphics[width=.49\linewidth,height=\myFigSize]{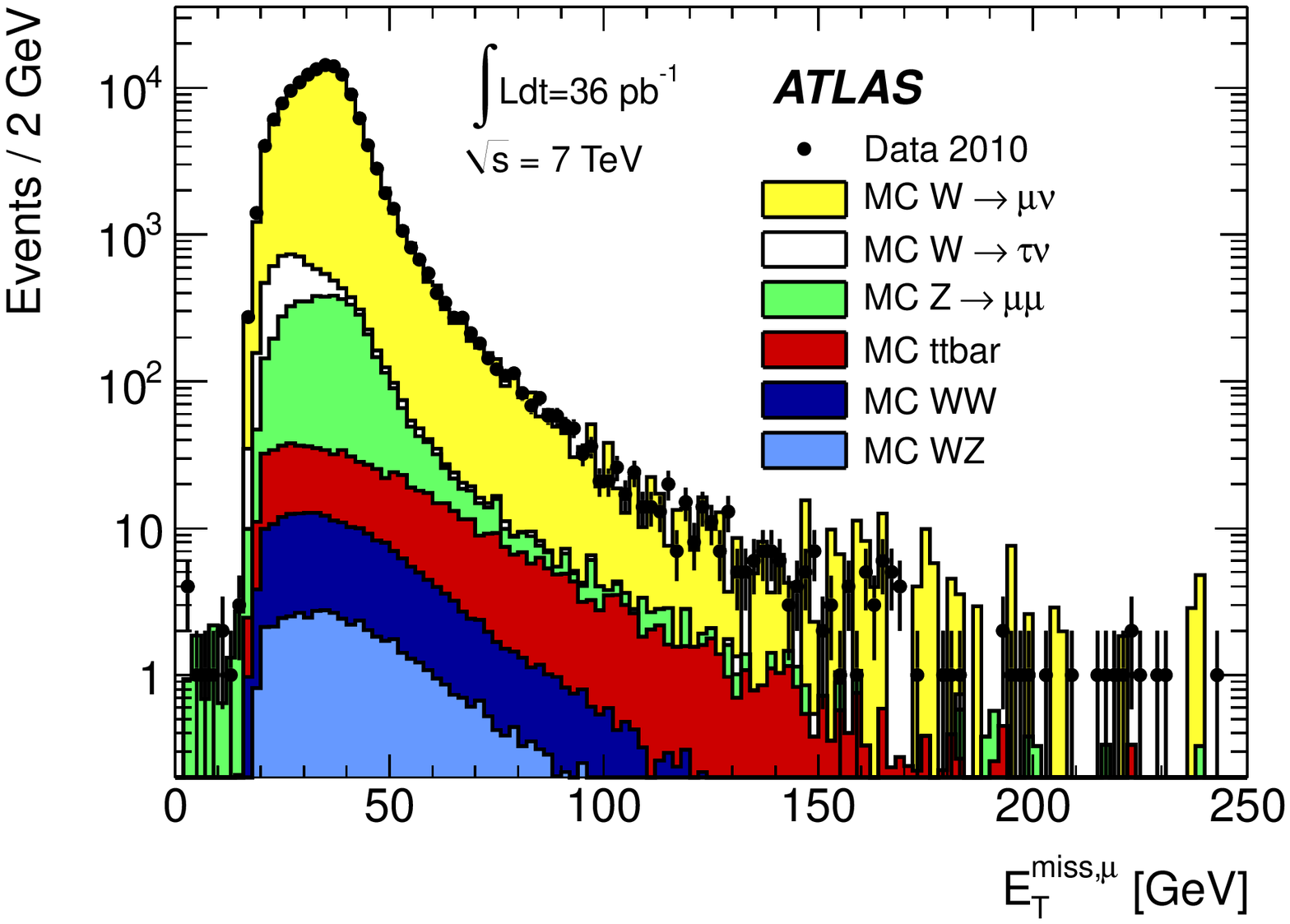}
\end{tabular}
\end{center}
\caption{\it Distribution of \etmissmag~ computed with cells from
muons  ($E_{\mathrm{T}}^{\mathrm{miss,calo},\mu}$) (left) and reconstructed muons
 ($E_{\mathrm{T}}^{\mathrm{miss,muon}}$) (right) for \Wmun~ data.
The expectation from Monte Carlo simulation 
is superimposed  and normalized to data, after each MC sample is 
weighted with its corresponding cross-section.
}  
\label{fig:METW_terms2}
\end{figure*}

\begin{figure*}[htbp]
\begin{center}
\begin{tabular}{lr}
\includegraphics[width=.49\linewidth,height=\myFigSize]{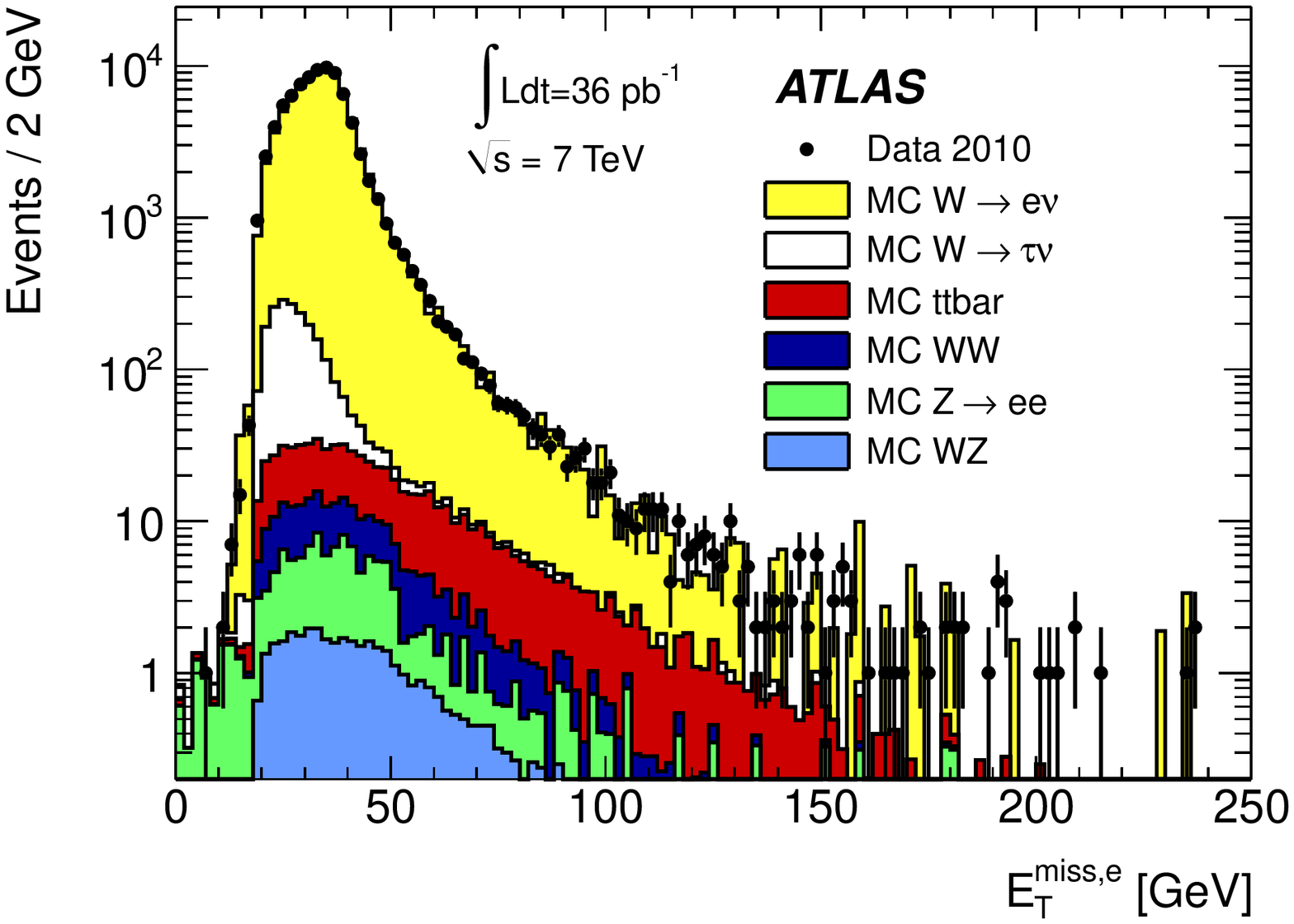}
\includegraphics[width=.49\linewidth,height=\myFigSize]{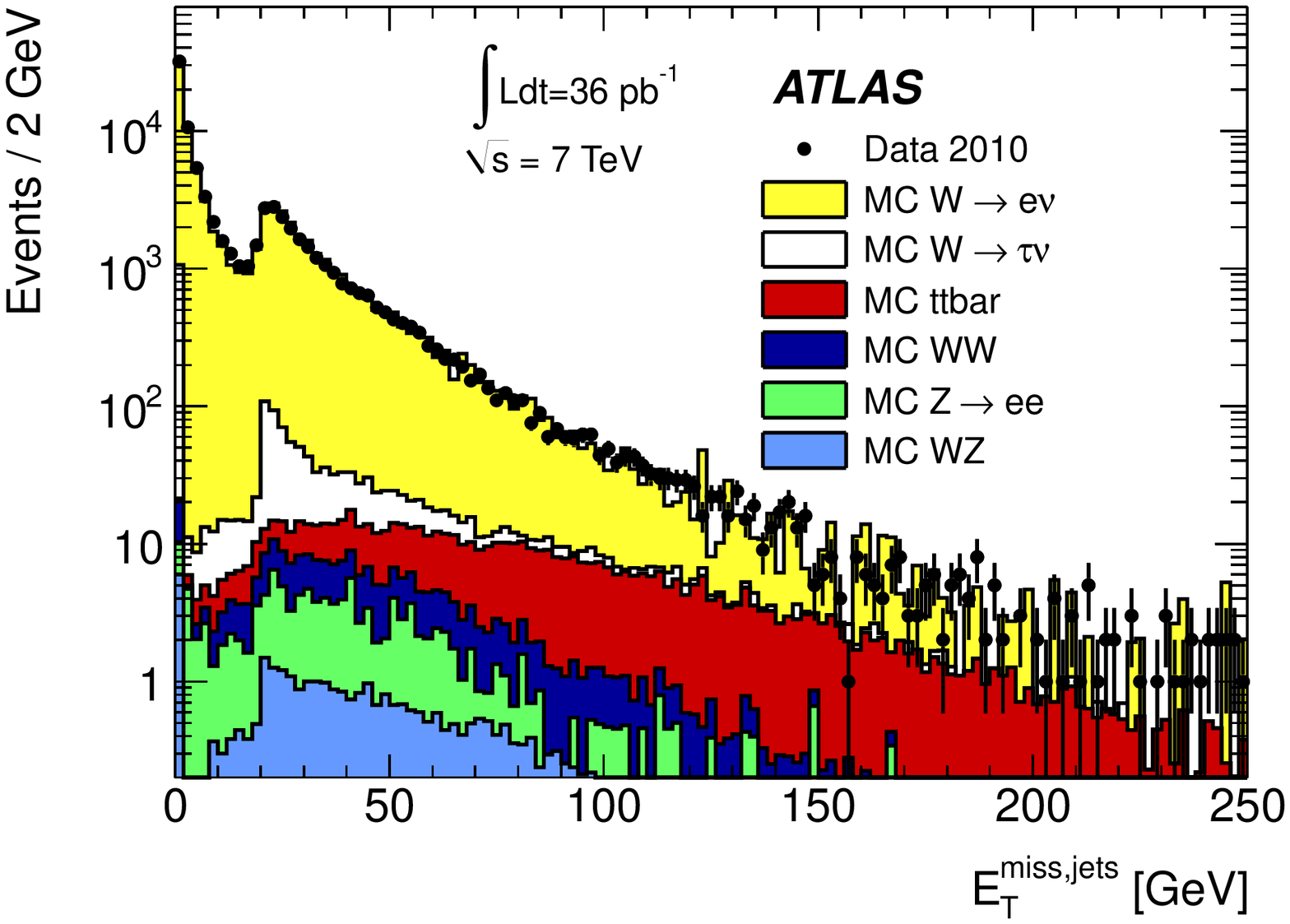}\\
\includegraphics[width=.49\linewidth,height=\myFigSize]{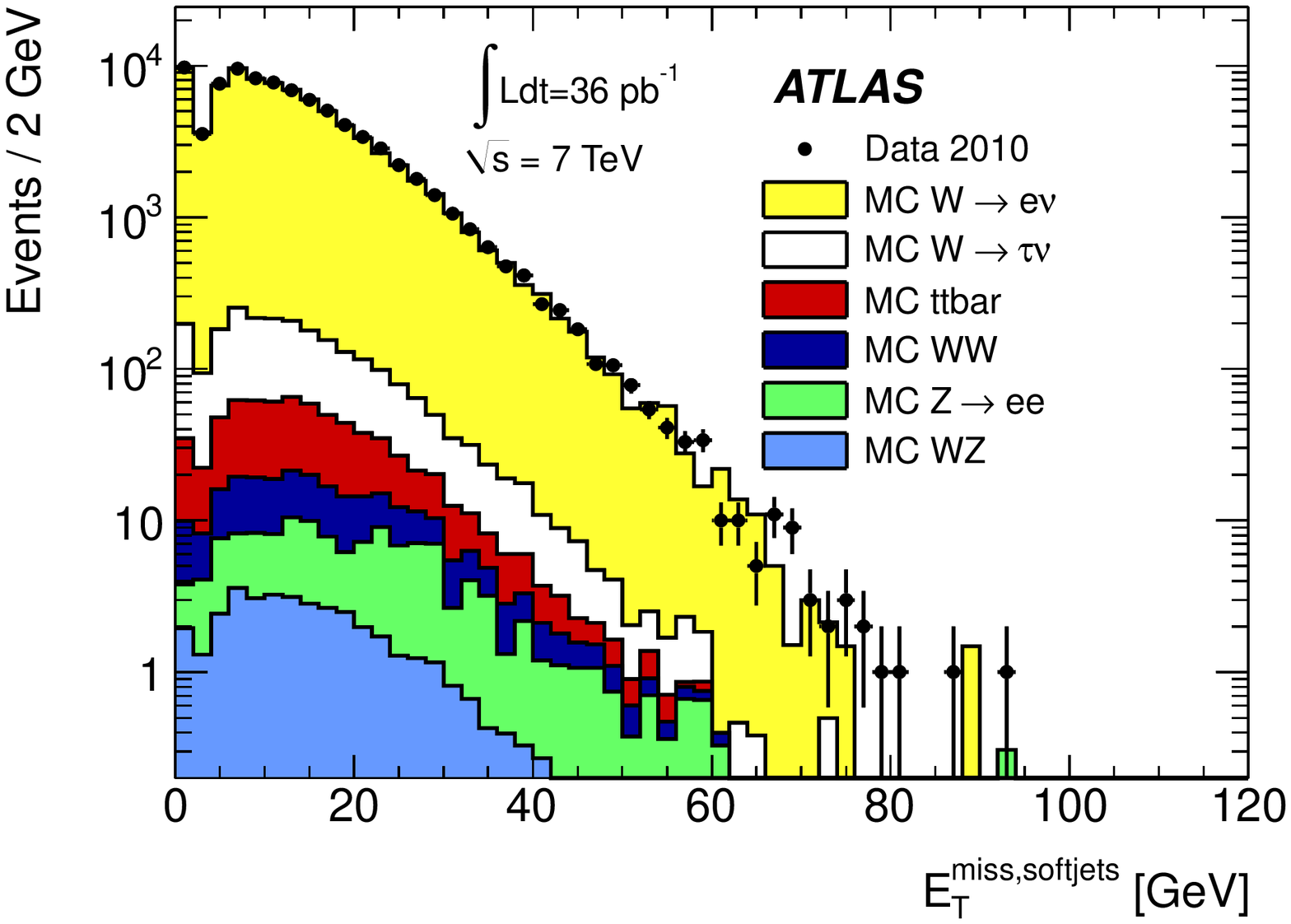}
\includegraphics[width=.49\linewidth,height=\myFigSize]{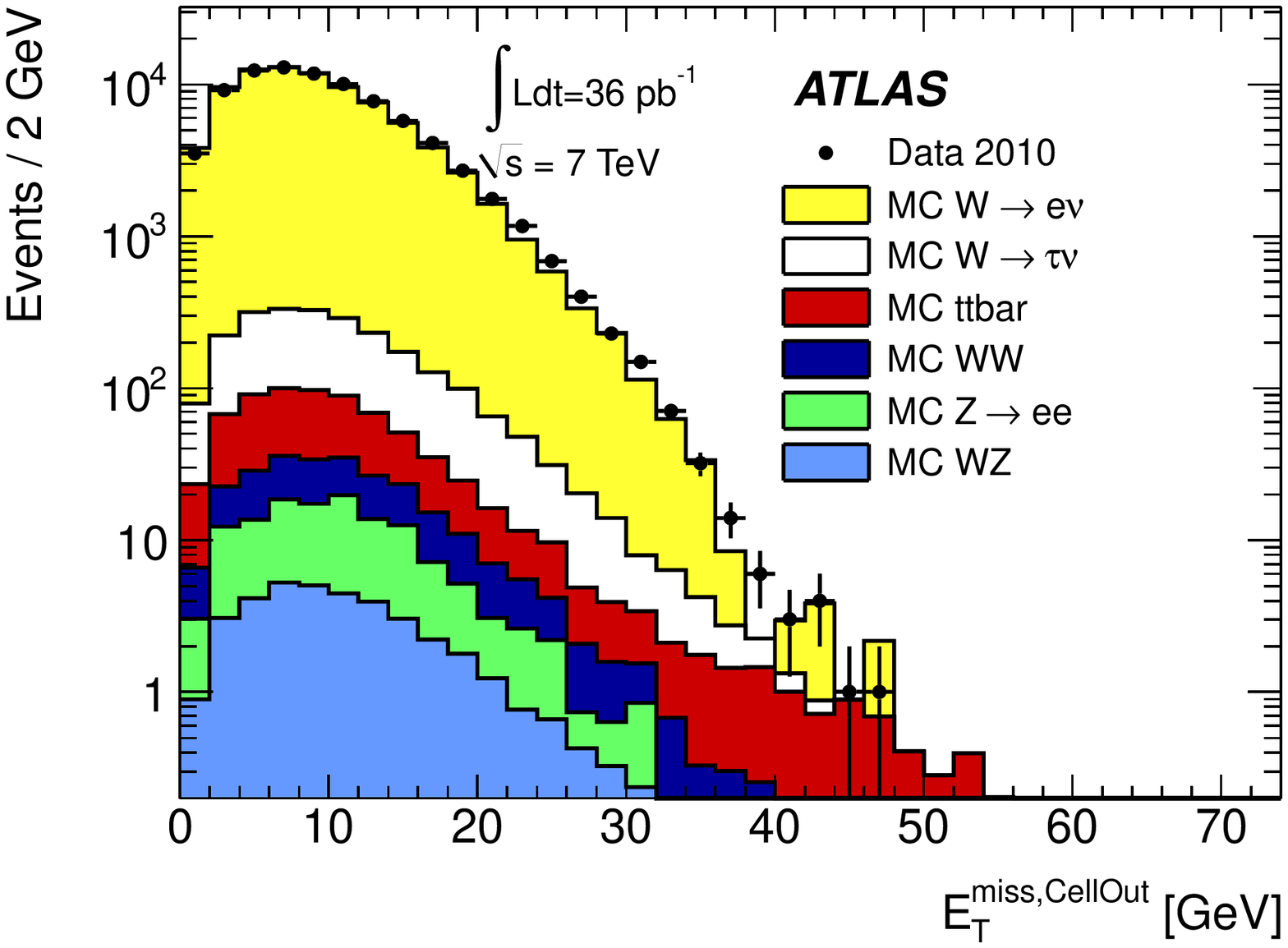}
\end{tabular}
\end{center}
\caption{\it Distribution of \etmissmag~computed with cells associated to 
electrons ($E_{\mathrm{T}}^{\mathrm{miss},e}$) (top left), 
jets with \pT~ $> 20$ GeV ($E_{\mathrm{T}}^{\mathrm{miss,jets}}$) (top right), jets with \pT~ $< 20$
GeV ($E_{\mathrm{T}}^{\mathrm{miss,softjets}}$) (bottom left) 
  and from topoclusters outside reconstructed objects
  ($E_{\mathrm{T}}^{\mathrm{miss,CellOut}}$) (bottom right) for data.
The expectation from Monte Carlo simulation 
is superimposed  and normalized to data, after each MC sample is 
weighted with its corresponding cross-section.
 }  
\label{fig:METW_terms}
\end{figure*}

The distributions of \etmissmag~ and \phimiss~ in data and in MC simulation are shown in
Figure~\ref{fig:METW_basic} for \Wen~ and \Wmun~ events.
The contributions  due to muons 
are shown  for \Wmun~ events in Figure~\ref{fig:METW_terms2}.
 Both, the \etmissmag~ contribution  from
energy deposited in calorimeter cells associated to muons, taken at the EM scale, and the \etmissmag~  
contribution from reconstructed muons are shown. 
The contributions given by the electrons, jets, soft jets and topoclusters
outside reconstructed objects  are shown in Figure~\ref{fig:METW_terms} for \Wen~ events.
The MC expectations are also shown, both from \Wln~ events, and from the dominant SM backgrounds.
The MC simulation describes all of the quantities well, with the exception that very small data-MC discrepancies are observed in the distribution 
of the $E_{\mathrm{T}}^{\mathrm{miss,e}}$ at  low \etmissmag~ values.
This can be attributed to the QCD jet background, which would predominantly populate the region of low \etmissmag~ \cite{WZXS}, 
but which is not included in the MC expectation shown.

\subsubsection{\etmiss ~ linearity in \Wln~ MC events}
\label{sec:Linearity}
 
\begin{figure*}
\begin{center}
\begin{tabular}{lr}
\hfill \includegraphics[width=.49\linewidth,height=\myFigSize]{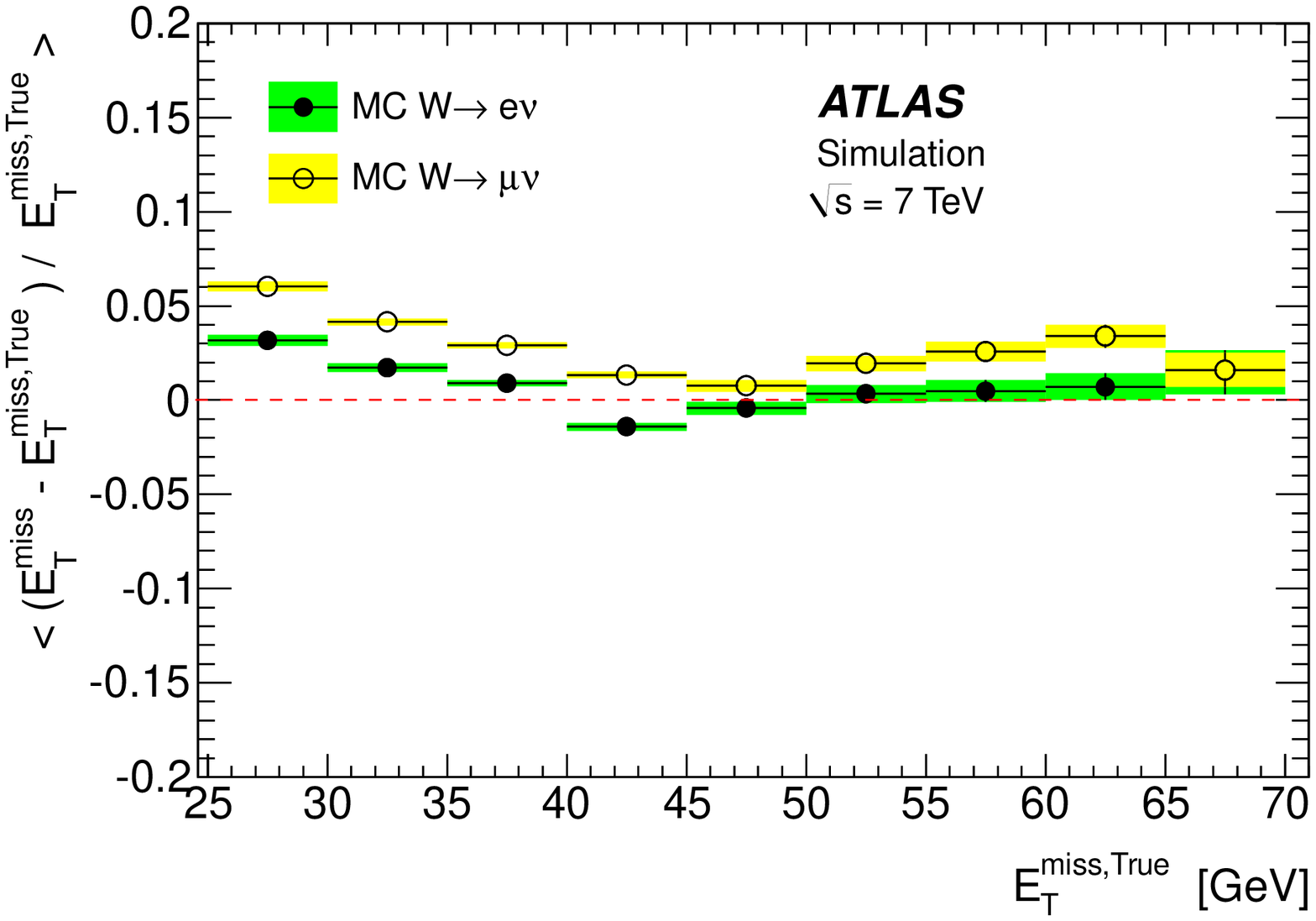}
\end{tabular}
\end{center}
\caption{\it \etmissmag~ linearity in \Wen~  and \Wmun~ MC events as
a function of $E_{\rm T}^{\rm miss,True}$}
\label{fig:MET_Lin_all}
\end{figure*}

The expected \etmissmag~ linearity,  which is defined as the mean value of  the ratio: $(E_{\mathrm{T}}^{\rm miss}-E_{\mathrm{T}}^{\rm miss,True})/E_{\mathrm{T}}^{\rm miss,True}$,   is shown as a function of $E_{\mathrm{T}}^{\rm miss,True}$ in Figure \ref{fig:MET_Lin_all} for \Wen~ and \Wmun~ MC events.
The mean value of this ratio
  is expected to be zero if the reconstructed \etmissmag~ has the correct 
scale. In Figure \ref{fig:MET_Lin_all}, it can be seen that there is a displacement
from zero which varies with the true \etmissmag. 
The bias at  low $E_{\mathrm{T}}^{\rm miss,True}$~ values  is about 5$\%$ and is due to the finite resolution of the \etmiss~ measurement. The reconstructed \etmissmag~ is positive by definition, so the relative difference is positive when the  $E_{\mathrm{T}}^{\rm miss,True}$ is small. The effect extends up to 40 GeV.
The bias is in general larger for \Wmn~ events  than for \Wen~ events. 
Considering only events with  $E_{\mathrm{T}}^{\rm miss,True}>40$ GeV, the \etmissmag~ linearity is better than 1$\%$  in \Wen~ events, while there is a non-linearity up to about 3$\%$ in \Wmun~ events.  This may be explained by an underestimation of the $E_{\rm{T}}^{\mathrm{miss,calo},\mu}$ term, in which too few calorimeter cells are associated to the reconstructed muon.

\subsection{\etmiss\ resolution}
\label{sec:resol_Z}

A more quantitative evaluation of the \etmiss~ performance can be obtained 
from a study of the $(\emiss{x},\emiss{y})$ resolutions as a function of 
\sumet.
In \Zll~ events, as well as in minimum bias and QCD jet events, no genuine \etmissmag~is expected,
so the resolution of the two \etmiss~components 
is measured directly from reconstructed quantities, 
 assuming that the true values of \px~ and
\py~ are equal to zero. 
The resolution is estimated from the width of the combined distribution of
$\emiss{x}$ and $\emiss{y}$ (denoted  $(\emiss{x},\emiss{y})$  distribution) in bins of \sumet.
The core of the distribution is fitted, for each \sumet~bin,  with a Gaussian  over twice 
the expected resolution obtained from previous studies \cite{JET_ETMISS}
and the fitted width, $\sigma$, is examined as a function of \sumet.
 The \etmiss\ resolution follows an approximately stochastic
behaviour as a function of \sumet, which can be described with the function 
$\sigma= k \cdot \sqrt{\Sigma E_{\mathrm{T}}}$,
but  deviations from this simple law are expected in the low
\sumet\ region due to noise and in the very
large \sumet\ region due to the constant term.
 
\begin{figure*}
\begin{center}
\begin{tabular}{lr}
\includegraphics[width=.49\linewidth, height=\myFigSize]{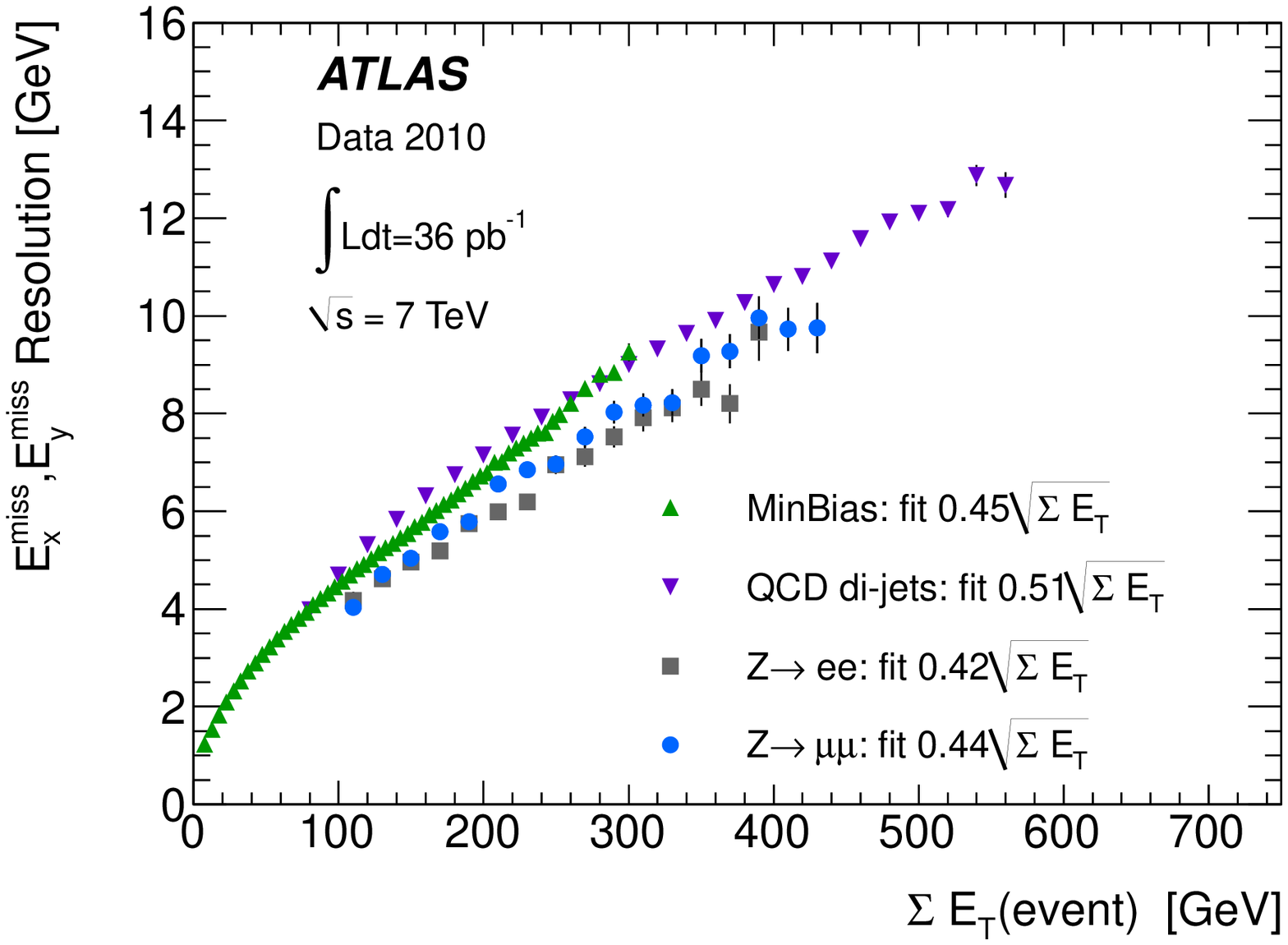}
\includegraphics[width=.49\linewidth, height=\myFigSize]{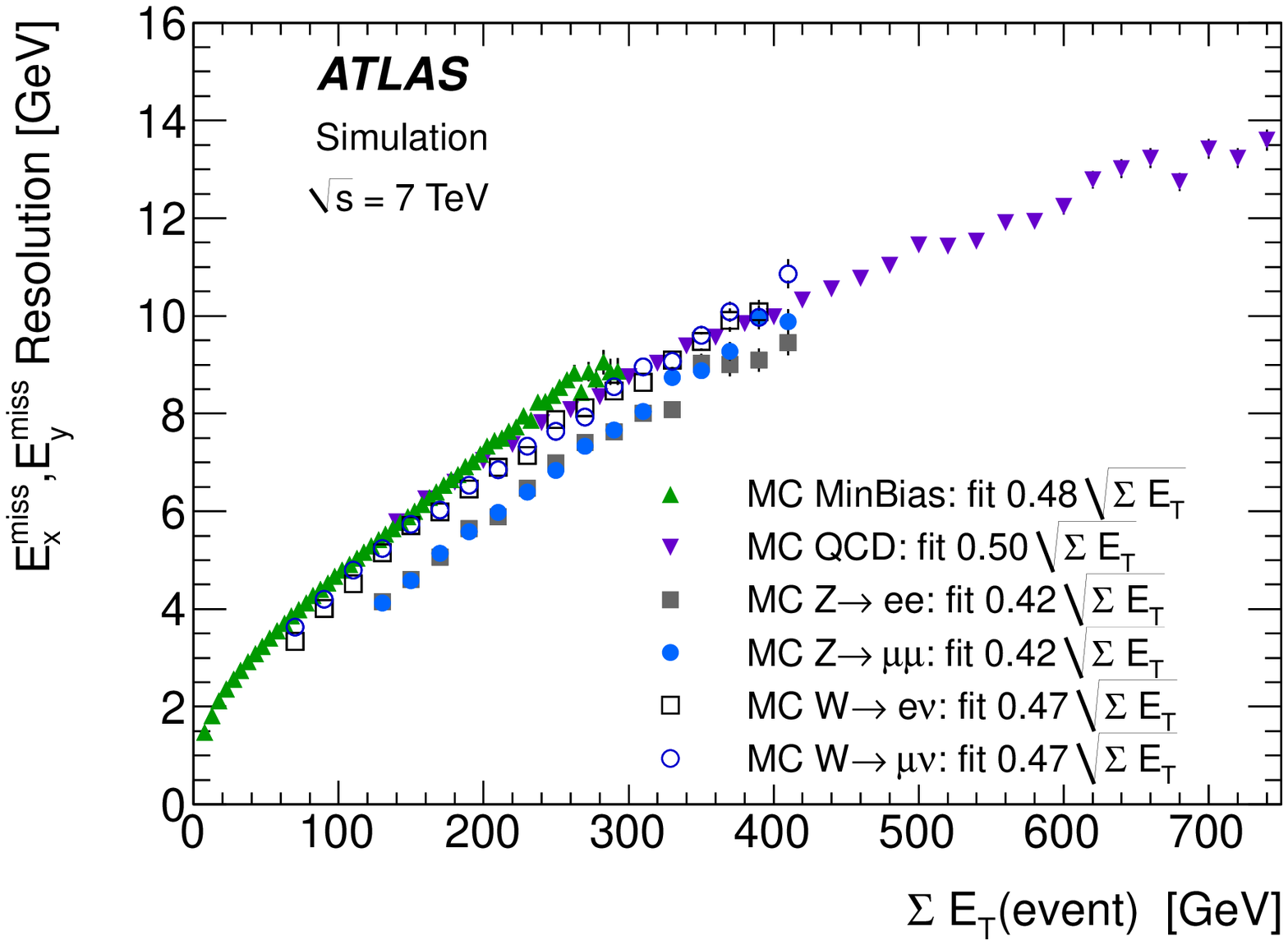}
\end{tabular}
\end{center}
\caption{\it \px~and \py~ resolution  as a function of 
the total transverse energy in the event calculated by summing the \pT~ of muons and the total transverse energy in the calorimeter in data  at  $\sqrt{s}$ = 7 TeV (left) and MC (right). 
The resolution of the two \etmiss~ components is fitted with a function $\sigma= k \cdot \sqrt{\Sigma E_{\mathrm{T}}}$ and the fitted values of the parameter  $k$, expressed in GeV$^{1/2}$,  are reported in the figure.  
}
\label{fig:resol_tutti}
\end{figure*}

 Figure \ref{fig:resol_tutti} (left) shows the resolution from data at $\sqrt{s}=7$ TeV for  \Zll~ events, minimum bias and  di-jet events
as a function of the total transverse energy in the event, obtained by summing the \pT~ of muons and the \sumet~ in calorimeters, calculated as described in Section \ref{sec:met_rec}.
 If the resolution is shown as a function of the \sumet~ in calorimeters, a difference between \Zee~ and \Zmm~ events is observed due to the fact that
 \sumet~ includes electron momenta in \Zee~ events while muon momenta are not included in
 \Zmm~ events. 

The resolution of the two \etmiss\ components is fitted with the simple
function given above. The fits  are acceptable
and are of similar quality for all different channels studied.
This allows to use the parameter $k$ as an estimator for the
resolution and to compare it in various physics channels in data 
and MC simulation.
 There is a reasonable agreement
in the \etmiss~  resolution in the different physics channels, as can be seen from the fit parameters $k$ reported in the figure.  The $k$ parameter has fit values ranging from 0.42 GeV$^{1/2}$ for \Zll~ events to 0.51 GeV$^{1/2}$ for di-jet events.  
 The \etmiss~ resolution is better in \Zll~ events because the lepton momenta are measured with better precision than jets.

In Figure  \ref{fig:resol_tutti} (right) the \etmiss~ resolution is shown for MC events. In addition to the \Zll, minimum bias and di-jet events, the resolution is also shown  for \Wln~ MC events. 
In $W$ events the resolution of the two \etmiss~ components is estimated
 from the width of  $(E_{x}^{\rm miss}-E_{x}^{\rm miss,True}, E_{y}^{\rm miss}-E_{y}^{\rm miss,True})$  in bins of \sumet, fitted with a Gaussian  as explained above.
There is a reasonable agreement in the \etmiss~ resolution in the different MC channels studied with the fitted value of $k$ ranging from 0.42 GeV$^{1/2}$ for \Zll~ events to 0.50 GeV$^{1/2}$ for di-jet events.  As observed for data, the \etmiss~ resolution  is better in \Zll~ events and slightly better in \Wln~ events, due to the presence of the leptons which are more precisely  measured. 

The resolution in MC minimum bias events is slightly worse than in data. This is  probably due to imperfections of the modelling of soft particle activity in MC simulation,
  while there is a good data-MC agreement in the resolution for other channels.

\section{Evaluation of the systematic uncertainty on the \etmiss~ scale}
\label{sec:syst}

For any analysis using \etmiss, it is necessary to be able to evaluate the 
systematic uncertainty on the \etmissmag~ scale. 
The \etmiss, as defined in Section \ref{sec:RefFinalConfig}, is the sum of several terms 
corresponding to different types of reconstructed objects.  
The uncertainty on each individual term can be evaluated given 
the knowledge of the reconstructed objects \cite{WZXS,atlasjet2010} that are used to build it and 
this uncertainty can be propagated to
 \etmissmag. 
The overall systematic uncertainty on the \etmissmag~ scale is then calculated by combining the uncertainties on each term.

The relative impact of the uncertainty of the constituent terms on \etmissmag~
differs from one analysis to another depending on the final state being studied.  
In particular, in events containing $W$ and $Z$ bosons decaying to leptons,  uncertainties on the scale and resolution in the measurements of the charged leptons, together with uncertainties on the jet energy scale,  need to be propagated to the systematic uncertainty estimate of \etmissmag.  
Another significant contribution to the \etmissmag~ scale uncertainty in $W$ and $Z$ boson final 
states comes from the contribution of topoclusters outside reconstructed objects and from soft jets.
In the next three subsections, two complementary methods for the evaluation of the systematic uncertainty on the \etmissmag$^{,\mathrm{CellOut}}$ and the \etmissmag$^{\mathrm{,softjets}}$ terms are described.
 Finally the overall \etmissmag~ uncertainty for \Wln~ events is calculated.

\subsection{Evaluation of the systematic uncertainty on the \etmiss$^{\mathrm{,CellOut}}$ scale using Monte Carlo simulation}
\label{sec:syst_CellOut_MC}

There are several possible sources of systematic uncertainty in the calculation 
of \etmissmag$^{\mathrm{,CellOut}}$.
These sources include inaccuracies in the description
 of the detector material,
 the choice of shower model and the model for the underlying event in the simulation.  
The systematic uncertainty due to each of these sources is estimated 
with dedicated MC simulations. The MC jet samples,  generated with  {\sc Pythia}, are those used to assess the systematic uncertainty on the jet energy scale \cite{JES}. Table~\ref{tab:UncertaintySource} lists the simulation samples considered, referred to in the following as ``variations'' with respect to the nominal sample.

\begin{table*}[htp]
 \centering
 \begin{tabular}{c|c}
   \hline
   \hline
   Variation & Description\\
   \hline
   \hline
   Dead Material & 5\% increase in the inner detector material\\
   & 0.1 $X_0 $  in front of the cryostat of the EM barrel calorimeter \\
 & 0.05 $X_0 $  between presampler and EM barrel calorimeter\\
& 0.1 $X_0 $  in the cryostat after the EM barrel calorimeter\\
& density of material in  barrel-endcap transition of the EM calorimeter $ \times $1.5 \\
   FTFP\_BERT & An alternative shower model for hadronic interaction in GEANT4 \\ 
   QGSP & An alternative shower model for hadronic interaction in GEANT4\\ 
   {\sc Pythia} Perugia 2010 tune  & An alternative setting of the  {\sc Pythia}  parameters\\
  & with increased final state radiation and more soft particles\\
   \hline
   \hline
   \end{tabular}
   \caption{Variations of the default simulation settings used for the estimate of the \etmissmag$^{\mathrm{,CellOut}}$ term systematic uncertainty. See Ref.~\cite{JES} for details of the parameters.}
\label{tab:UncertaintySource}
\end{table*}

 The estimate of the uncertainty on  \etmissmag$^{\mathrm{,CellOut}}$ for a
variation {\it i} is determined by calculating the percentage difference between the mean
value of this term for the nominal sample, labelled $\mu_0$, and that for the variation
sample, labelled $\mu_i$. This approach assumes that the variations affect the total scale and none of the variations introduces a shape dependence in the \etmissmag$^{\mathrm{,CellOut}}$ term, as verified in Ref. \cite{etmissCONF}.
In order to cross-check for a possible dependence on the event total transverse energy,
the relative difference $R_i = (\mu_i - \mu_0)/\mu_0$ between different variations
is computed in bins of \sumet\ for the 
jet samples. No significant dependence of $R_i$ on \sumet\  is observed.
A cross-check on the topology dependence is done using  \Wln~  samples simulated by introducing the variations $i$. 
Table~\ref{tab:syst_finalNumbers} shows the $R_i$ values as computed in both the 
QCD 
jet samples and the \Wln~  samples. The results are consistent, showing that the estimated uncertainty does not have a large dependence on the event topology.

A symmetric systematic uncertainty on the \etmissmag$^{\mathrm{,CellOut}}$ scale is obtained by summing in quadrature the estimated uncertainties averaged between 
simulated jet and $W$ events.
The total estimated uncertainty\footnote{In this uncertainty evaluation using MC simulation, the uncertainty on the absolute electromagnetic energy scale in the calorimeters should also be taken into account. 
For the bulk of the LAr barrel electromagnetic calorimeter a 1.5$\%$ uncertainty is found on the cell energy measurement, increasing to 5$\%$ for the presampler and 3$\%$ for the tile calorimeter \cite{eoverp2011}.}
on the \etmissmag$^{\mathrm{,CellOut}}$ term is 2.6\%. 

\begin{table}
 \centering
 \begin{tabular}{c|c|c}
   \hline
   \hline
   Variation &  jet events& $W$ production \\
   \hline
   \hline
   Dead Material & $(-0.5 \pm 0.1)$\% & $(-0.6\pm 0.2)$\%\\ 
   FTFP\_BERT & $(0.1 \pm 0.4)$\% & $(0.5\pm 0.2)$\%\\ 
   QGSP & $(-1.6 \pm 0.4)$\% & $(-2.2\pm 0.2)$\%\\ 
    {\sc Pythia} Perugia 2010 tune& $(-1.7 \pm 0.1)$\% & $(-1.5\pm 0.2)$\% \\ 
   \hline
   \hline
   \end{tabular}
   \caption{Systematic uncertainties ($R_i$) on $E_{\mathrm{T}}^{\mathrm{miss,CellOut}}$ associated with variations in the dead material (all the variations listed in Table 
   \ref{tab:UncertaintySource} are applied at the same time), in the calorimeter shower modelling (FTFP\_BERT, QGSP) and in the event generator settings ({\sc Pythia} Perugia 2010 tune). }
\label{tab:syst_finalNumbers}
\end{table}

\subsection{Evaluation of the systematic uncertainty on  the \etmiss$^{\mathrm{,CellOut}}$ scale from the topocluster energy scale uncertainty}
\label{sec:syst_CellOut_eoverp}

The uncertainty on the scale of the \etmissmag$^{\mathrm{,CellOut}}$ term, which
is built from topoclusters with a correction based on tracks (see Section \ref{sec:eflow}), can also be calculated from the topocluster 
energy scale uncertainties.
 These uncertainties can be estimated from comparisons between 
data and MC simulation using  the $E/p$ response from single tracks,
measured by summing the energies of all calorimeter clusters 
  around a single isolated track
\cite{eoverp2011}.
The effects of these uncertainties on the
\etmissmag$^{\mathrm{,CellOut}}$
 term can be evaluated by varying the energy scale  of 
topoclusters that contribute to  the \etmissmag$^{\mathrm{,CellOut}}$
 term in \Wen~ MC samples, as was done in Ref. \cite{WZXS}.
 
The shift in the topocluster energy scale is applied by multiplying the topocluster energy  by the function:
\begin{eqnarray}
       1\pm a \times(1+b/p_T)
        \label{eq:sys} 
\end{eqnarray}
with $a=3 (10)\%$~for $|\eta|<(>)$3.2 and  $b=1.2$ GeV. 

The $a$ parameter in Equation \ref{eq:sys} addresses the uncertainty on the cluster energy scale,  obtained by comparing the ratio of the cluster energy and the measured track momentum, $E/p$, 
in data and MC  simulation \cite{eoverp2011}.
The value in the forward region, 
 where tracks cannot be used to validate the energy scale, is estimated from the transverse momentum balance of one jet in the central region and one jet in the forward region in events with only two jets at high transverse momenta.
            
The $b$ parameter in Equation \ref{eq:sys}  addresses the possible change in the clustering efficiency and scale in a non-isolated environment. 
To go from the response for single isolated particles to the cluster energy scale,  possible effects from the
noise thresholds in the configuration with nearby particles are taken into account .

Because of threshold effects, more energy is clustered for nearby particles than for isolated ones.
In an hypothetic worst case scenario, the environment is so busy that the clustering algorithm is forced to cluster all the deposited energy, with no bias due to the noise thresholds.
Therefore, the maximal size of the noise threshold effect can be evaluated by comparing the ratio  $E_{\mathrm{cell}}/p$ of the total energy $E_{\mathrm{cell}}$ deposited into all cells around an isolated track to the track momentum, to the ratio $E/p$ of the clustered energy $E$ to the track momentum, in data and MC simulation.

 The fractional $E_{\rm T}^{{\rm miss},\mathrm{CellOut }}$ uncertainty is evaluated from:
 \begin{eqnarray}
(\Delta^{\mathrm{CellOut +}}  +  \Delta^{\mathrm{CellOut -}} )  /(2 \times E_{\rm T}^{{\rm miss}\mathrm{,CellOut}})
 \label{eq:syscalc1} 
\end{eqnarray}
   where
\begin{eqnarray}
\Delta^{\mathrm{CellOut +}} = |E_{\rm T}^{{\rm miss}\mathrm{,CellOut +}} - E_{\rm T}^{{\rm miss}\mathrm{,CellOut}}| \nonumber\\ 
 \Delta^{\mathrm{CellOut -}}  =  |E_{\rm T}^{{\rm miss},\mathrm{CellOut -}} - E_{\rm T}^{{\rm miss},\mathrm{CellOut}}|  
 \label{eq:syscalc} 
\end{eqnarray}
with  \etmissmag$^{\mathrm{,CellOut +}}$ and 
  \etmissmag$^{\mathrm{,CellOut -}}$ 
 obtained by shifting the  topocluster energies up and down,   respectively, using  Equation \ref{eq:sys}.
     The value of the fractional  \etmissmag$^{\mathrm{,CellOut}}$ uncertainty is found
   to be approximately 13\%, decreasing slightly
   with increasing \sumet$^{\mathrm{CellOut}}$.
 This uncertainty  is much larger than the uncertainty due to the detector description estimated from the first three lines of Table \ref{tab:syst_finalNumbers}.
   The main reason is that the values of $a$ and $b$ which enter into Equation \ref{eq:sys} are conservative, to include the effects described above.
    In particular the cluster energy uncertainty in the forward region is
     conservatively estimated, since the uncertainty cannot be evaluated
     using tracks.
Moreover, the procedure does not take into account the fact that  when the clusters are shifted up in \pT, some of them can form jets above threshold and they are therefore included in the soft jet term in \etmissmag. These clusters should be removed from the  \etmissmag$^{\mathrm{,CellOut}}$, they are in fact kept and this increases the uncertainty.
It should also be noted that in the calculation of \etmissmag$^{\mathrm{,CellOut}}$ the track momentum is used instead of the topocluster energy when there is a track-topocluster matching (see Section \ref{sec:eflow}). This would result  in a reduced uncertainty due to the more precise measurement of the track momentum, which is not taken into account here.
Further study is
expected to provide a reduction in this uncertainty in future,
by considering the described  effects in detail.

\vskip 0.5cm
To give an 
estimate of the \etmissmag$^{\mathrm{,CellOut}}$ systematic uncertainty, the  calorimeter contribution can be taken from Section \ref{sec:syst_CellOut_eoverp}, and the uncertainty from the event generator settings from Section \ref{sec:syst_CellOut_MC} ({\sc Pythia} Perugia 2010 tune). 
This results in a total systematic uncertainty on the scale of \etmissmag$^{\mathrm{,CellOut}}$ of about 13$\%$, which slightly decreases when  \sumet$^{\mathrm{CellOut}}$ increases.

\subsection{Evaluation of the systematic uncertainty on the  \etmiss$^{\mathrm{,softjets}}$ scale}
\label{sec:syst_SoftJet}

The same procedure described in the previous sections is used to assess the systematic uncertainty on the \etmissmag~ term calculated from soft jets (see Section \ref{calorec}). 

Using the MC approach described in Section \ref{sec:syst_CellOut_MC}, it is found that 
the uncertainty on   \etmissmag$^{\mathrm{,softjets}}$ does not exhibit a large dependence on the event \sumet, as was also found  for the uncertainty on the $E_{\rm T}^{\mathrm{miss,CellOut}}$ scale. The results 
are consistent between  the 
QCD 
jet samples and the $W$  samples, as can be seen from Table~\ref{tab:syst_softJetsNumbers} which gives the systematic uncertainties $R_i$ as computed in jet samples and in \Wln~ samples.
\begin{table}[htp]
 \centering
 \begin{tabular}{c|c|c}
   \hline
   \hline
   Variation & jet events& $W$ production \\
   \hline
   \hline
   Dead Material & $(-1.5 \pm 0.1)$\% & $(-1.5\pm 0.2)$\%\\ 
   FTFP\_BERT & $(0.3 \pm 0.4)$\% & $(0.8\pm 0.2)$\%\\ 
   QGSP & $(-2.6 \pm 0.4)$\% & $(-2.5\pm 0.2)$\%\\ 
     {\sc Pythia} Perugia 2010 tune & $(-1.4 \pm 0.1)$\% &$(-1.0\pm 0.2)$\% \\ 
   \hline
   \hline
   \end{tabular}
   \caption{Systematic uncertainties ($R_i$) on  \etmissmag$^{\mathrm{,softjets}}$ associated with variations in the dead material (all the variations listed in Table 
   \ref{tab:UncertaintySource} are applied at the same time), in the calorimeter shower modelling (FTFP\_BERT, QGSP) and in the event generator settings ({\sc Pythia} Perugia 2010 tune).}
\label{tab:syst_softJetsNumbers}
\end{table}

A total, symmetric, systematic uncertainty of about 3.3\% on the  \etmissmag$^{\mathrm{,softjets}}$  term is obtained by combining the results in Table \ref{tab:syst_softJetsNumbers}, as was done in Section \ref{sec:syst_CellOut_MC}.
With the same data-driven approach  utilising the uncertainty on the topocluster energy scale described  in Section \ref{sec:syst_CellOut_eoverp}, the systematic uncertainty on \etmissmag$^{\mathrm{,softjets}}$ is evaluated to be about 10$\%$.

\vskip 0.5cm
As for \etmissmag$^{\mathrm{,CellOut}}$, the uncertainty on the \etmissmag$^{\mathrm{,softjets}}$ scale found 
by shifting the topocluster energies is 
larger than the uncertainty estimated from MC simulation.
To give an 
estimate of the systematic uncertainty on  \etmissmag$^{\mathrm{,softjets}}$, the  contribution from the calorimeter response can be taken from the data-driven evaluation and the contribution from the event generator settings from Table \ref{tab:syst_softJetsNumbers}. 
This results in an overall systematic uncertainty of about 10$\%$ on \etmissmag$^{\mathrm{,softjets}}$, slightly increasing as \sumet~ increases.

\subsection{Evaluation of the overall systematic uncertainty on the \etmiss~ scale in 
$\textit{{\textbf{W}}} \  \rightarrow e \nu$ and $\textit{{\textbf{W}}} \  \rightarrow \mu \nu$ 
events}

\begin{figure*}[htbp]
\begin{center}
\begin{tabular}{lr}
\includegraphics[width=.49\linewidth,height=\myFigSize]{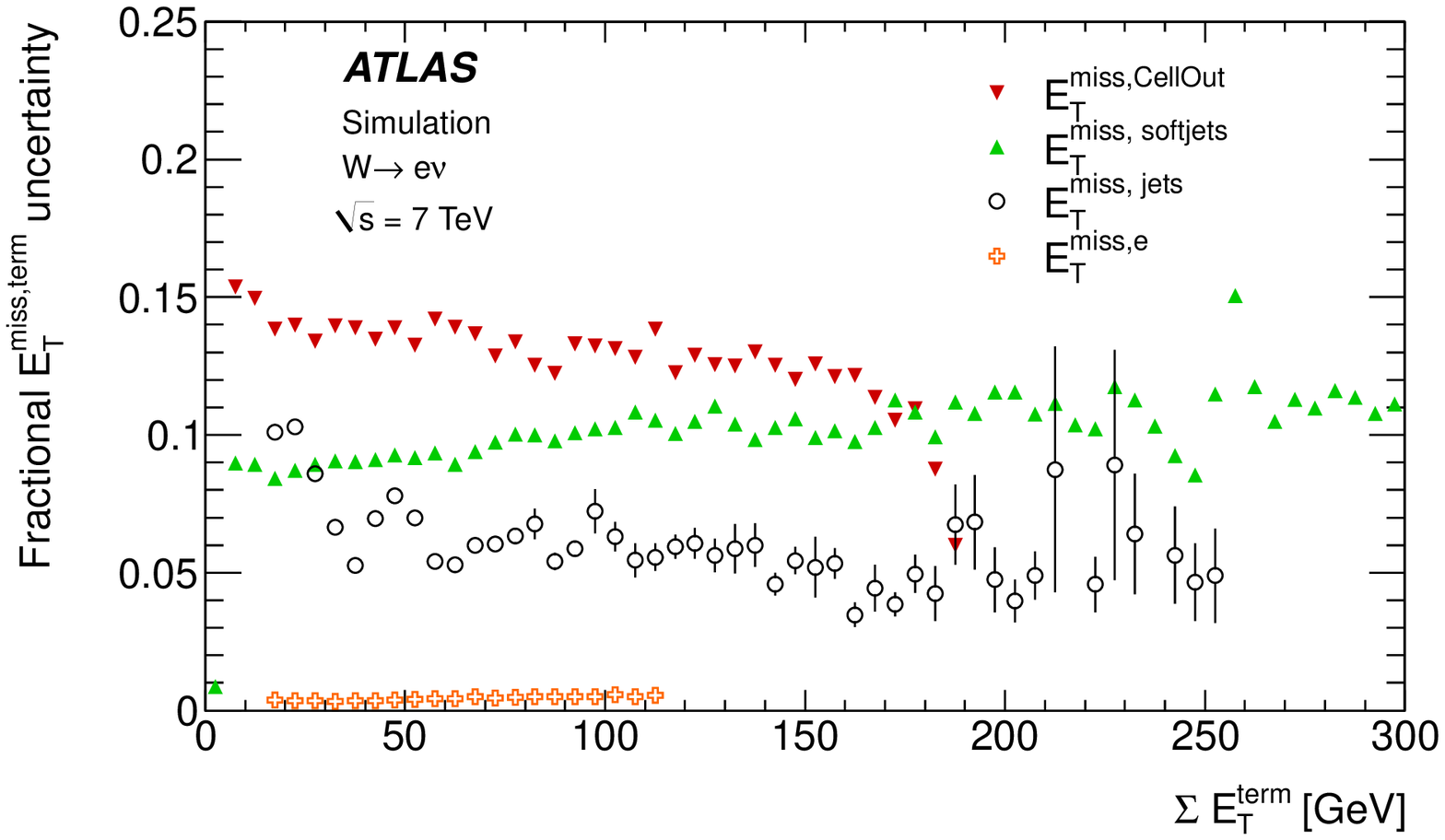} 
\includegraphics[width=.49\linewidth,height=\myFigSize]{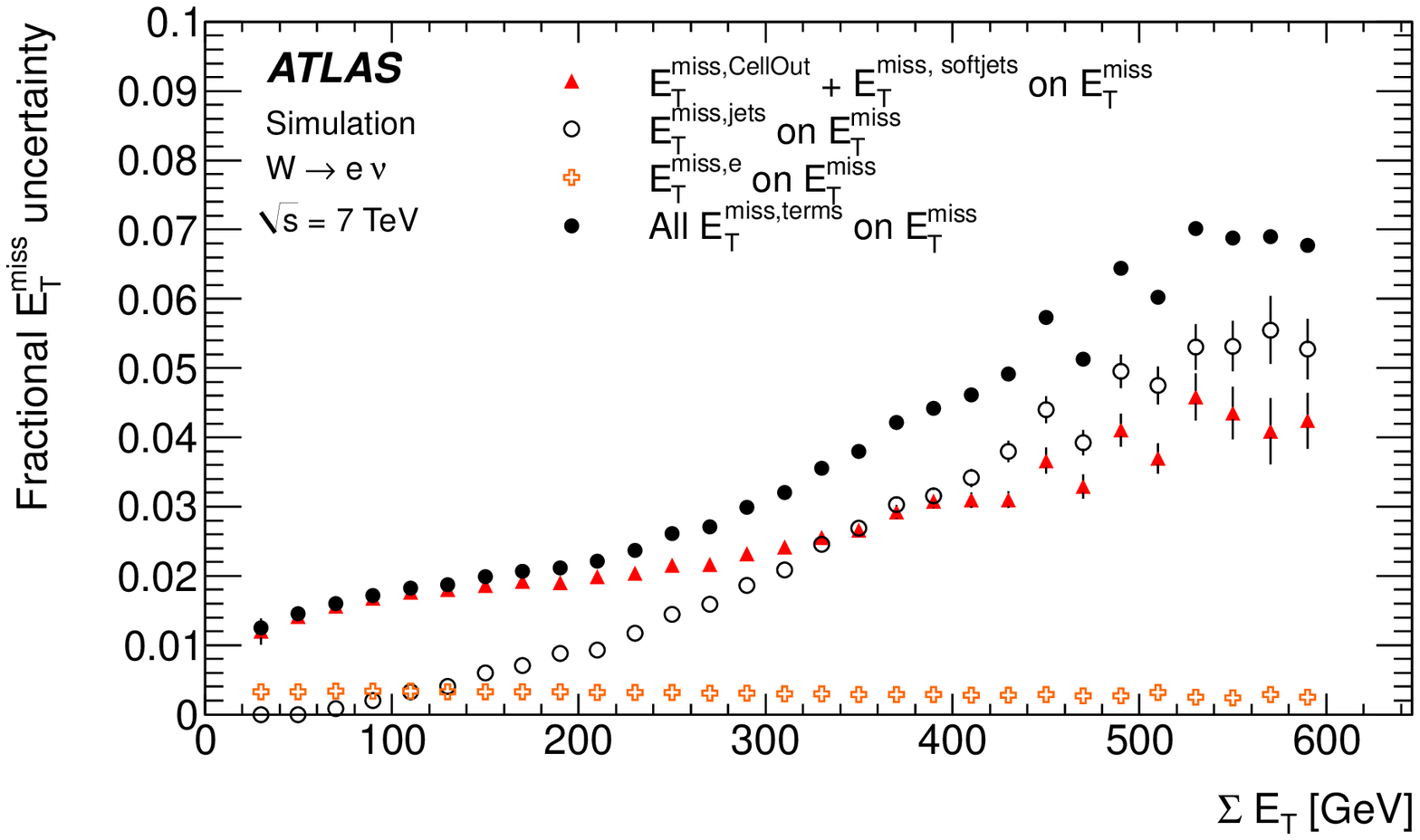} \\
\includegraphics[width=.49\linewidth,height=\myFigSize]{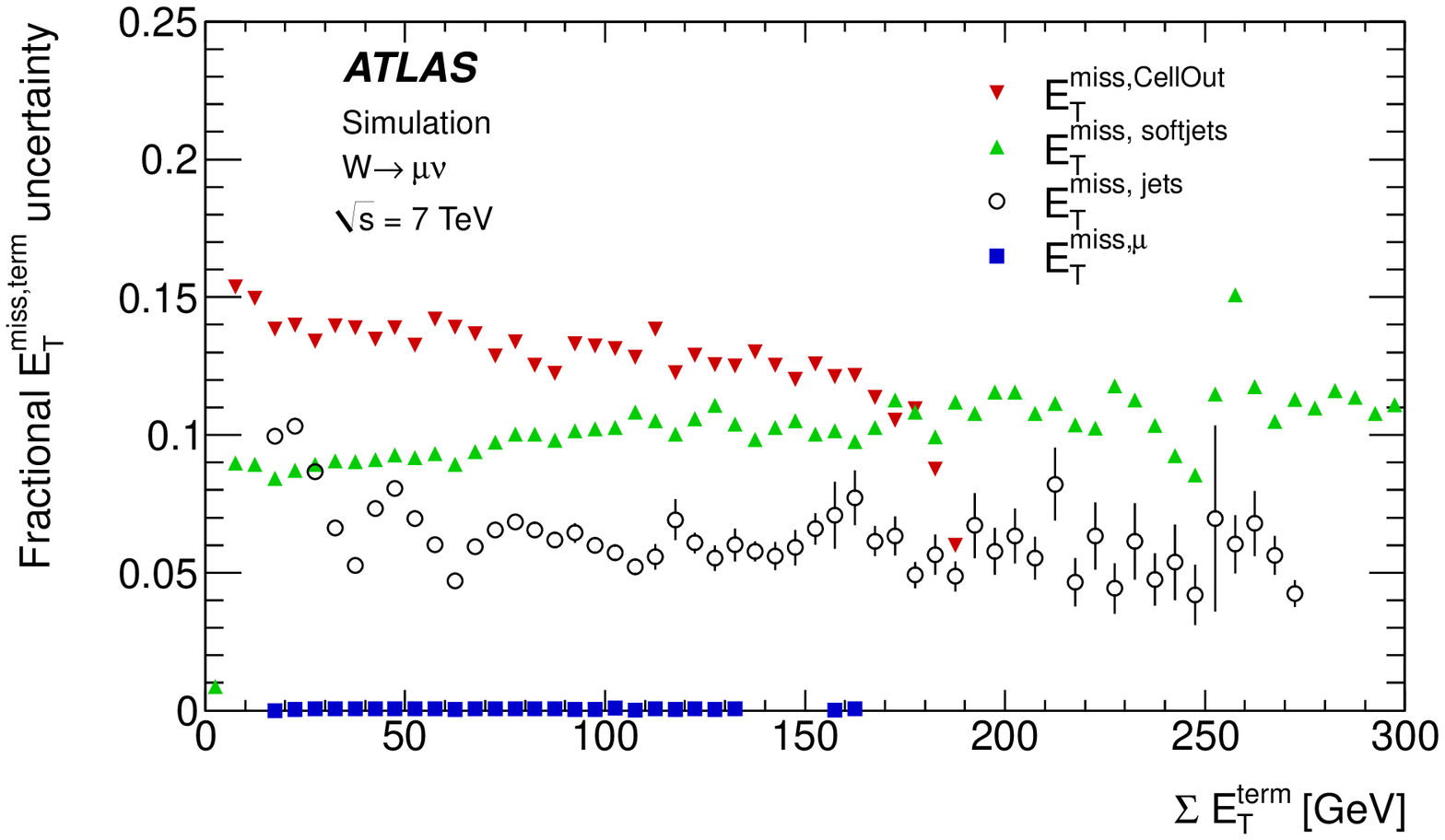} 
\includegraphics[width=.49\linewidth,height=\myFigSize]{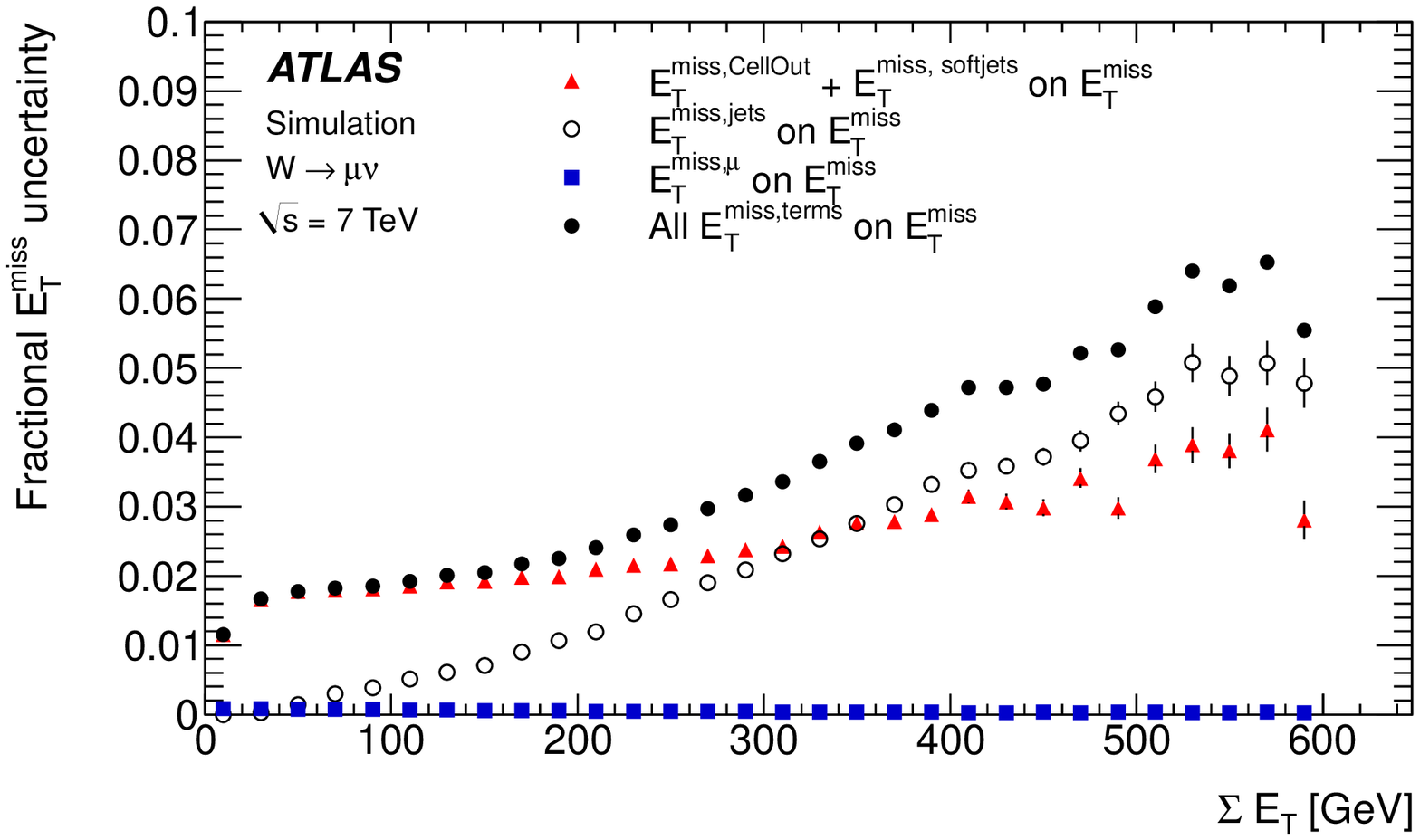} 
\end{tabular}
\end{center}
\caption{\it  Fractional systematic uncertainty (calculated as in 
 Equations  \ref{eq:syscalc1} and \ref{eq:syscalc})
 on different \etmissmag~ terms  as a function of  respective \sumet$^{\mathrm{term}}$~ (left) 
and contributions of  different term uncertainties on \etmissmag~ uncertainty 
as a function of   \sumet~ (right) in  MC \Wen~ events (top) and \Wmn~ events (bottom).
The overall systematic uncertainty on the \etmissmag~ scale, obtained combining the various contributions is
 shown in the right plots (filled circles). The uncertainties on \etmissmag$^{\mathrm{,softjets}}$   and   \etmissmag$^{\mathrm{,CellOut}}$  are considered to be fully correlated. 
}
\label{fig:Totsys_Wln}
\end{figure*}

Using as inputs the systematic uncertainties on the different reconstructed objects 
 \cite{WZXS,JES} 
and on \etmissmag$^{,\mathrm{CellOut}}$ and \etmissmag$^{\mathrm{,softjets}}$ evaluated in 
the previous sections,   the overall \etmissmag~ systematic
uncertainty  in \Wen~ and \Wmn~  events is estimated.
The same method  can be applied to any final state event topology.
Figure~ \ref{fig:Totsys_Wln} shows, for both  \Wen~ and \Wmn~ events, the systematic uncertainties 
on   each of the terms \etmissmag$^{,e} $ (\etmissmag$^{,\mu}$),   \etmissmag$^{\mathrm{,jets}}$,               
 \etmissmag$^{\mathrm{,softjets}}$   and   \etmissmag$^{\mathrm{,CellOut}}$ as a function of their
individual contribution to \sumet\, labelled  \sumet$^{\mathrm{term}}$.
All the uncertainties are calculated with the formulae in Equations  \ref{eq:syscalc1} and \ref{eq:syscalc}.
In the same figure the uncertainty on \etmissmag~ due to the uncertainties on the different terms is also shown as a function of the total \sumet, together with the overall uncertainty on \etmiss, obtained by combining the partial terms. 
The uncertainties on \etmissmag$^{\mathrm{,softjets}}$   and   \etmissmag$^{\mathrm{,CellOut}}$  are considered 
to be fully correlated. 
 In \Wen~ and \Wmn~ events, selected as described in Section \ref{sec:wl_event_selection}, 
 the overall uncertainty on the \etmissmag~ scale increases with \sumet~ from $\sim1 \%$ to 
$\sim7 \%$. It is estimated  to be, on average, about 2.6\% for both channels.

 The \etmissmag~ scale uncertainty depends on the event topology because the contribution of a given \etmissmag~ term can vary for different final states.

\section{Determination of the \etmiss~ scale from $\textit{{\textbf{W}}} \  \rightarrow \ell \nu$  events}
\label{sec:W_in-situ}

   The  determination of the  absolute 
\etmissmag~ scale  is important 
in a range of  analyses  involving \etmissmag\ measurements,
ranging from precision measurements to searches  for  new  physics.

In this section two complementary methods to determine the absolute scale
of \etmissmag\  using \Wln~ events are described.  The  first method uses a  fit to the
distribution of the transverse mass, \mT, of the lepton-\met\ system, 
and is sensitive both to the scale and the resolution of \etmissmag.  The
second method  uses
the interdependence of the neutrino and lepton momenta 
in the  $W\to e\nu$ channel, and the \etmissmag~ scale  is determined as a
function of  the reconstructed electron  transverse momentum.   
Both  methods  allow checks  on  the
agreement   between   data  and   MC simulation  for   the
\etmissmag\ scale.

\subsection{Reconstructed transverse mass method}
\label{sec:Wscale_fromMT}
The method described in this section
uses the shape of  the \mT~distribution and is sensitive to both the \met\ resolution and scale.
The lepton transverse momentum, \pTell, and the \etmissmag~ are used to calculate \mT~as:
\begin{eqnarray}
 m_T=\sqrt{2p_{\rm T}^{\ell} E_{\mathrm{T}}^{\mathrm{miss}}(1-\cos\phi)}
\label{eq:trasv_mass}
\end{eqnarray}
where $\phi$ is the azimuthal angle between the lepton momentum and \met\ directions.
The true \mT~is reconstructed from the simulation  under the
hypothesis that \met\ is entirely due to the neutrino momentum, \pTn. 
Template histograms of the
\mT~distributions are generated by convoluting the true transverse mass distribution 
with a Gaussian function: 
\begin{equation}
 E_{x(y)}^{\mathrm{miss,smeared}} = \alpha \, E_{x(y)}^{\rm miss,True} * {Gauss}(0,k \cdot  \sqrt{\Sigma E_{\rm T}})
\label{eq:smeared_met}
\end{equation}
where the parameters $\alpha$ and $k$ 
 are the \met\ scale and resolution respectively.

The  $\alpha$ and $k$ parameters are determined through a  fit of the \mT~distribution to data  using a linear combination of signal and background  \mT~distributions obtained from simulation. 
All the backgrounds, with the exception of the jet background, are evaluated from the same MC samples used in Section 
\ref{sec:Perf_W} and the normalization is fixed according to their cross-sections. The shape of the 
jet background is also evaluated from MC simulation and its normalization is obtained from the fit, in addition to  $\alpha$ and $k$.

\begin{figure*}[!htbp]
\hfill \includegraphics[width=.49\linewidth,height=\myFigSize]{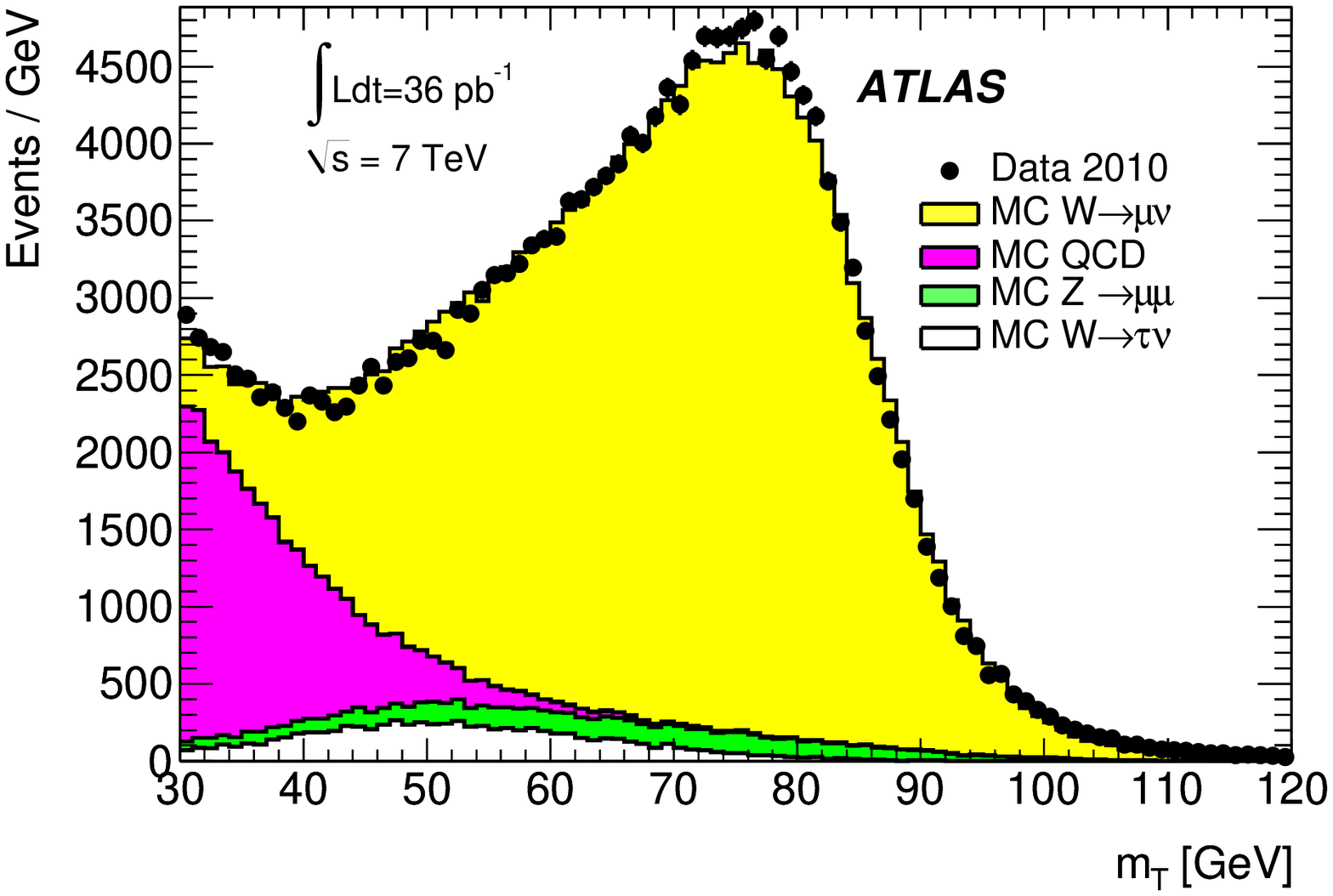}
\hfill \includegraphics[width=.49\linewidth,height=\myFigSize]{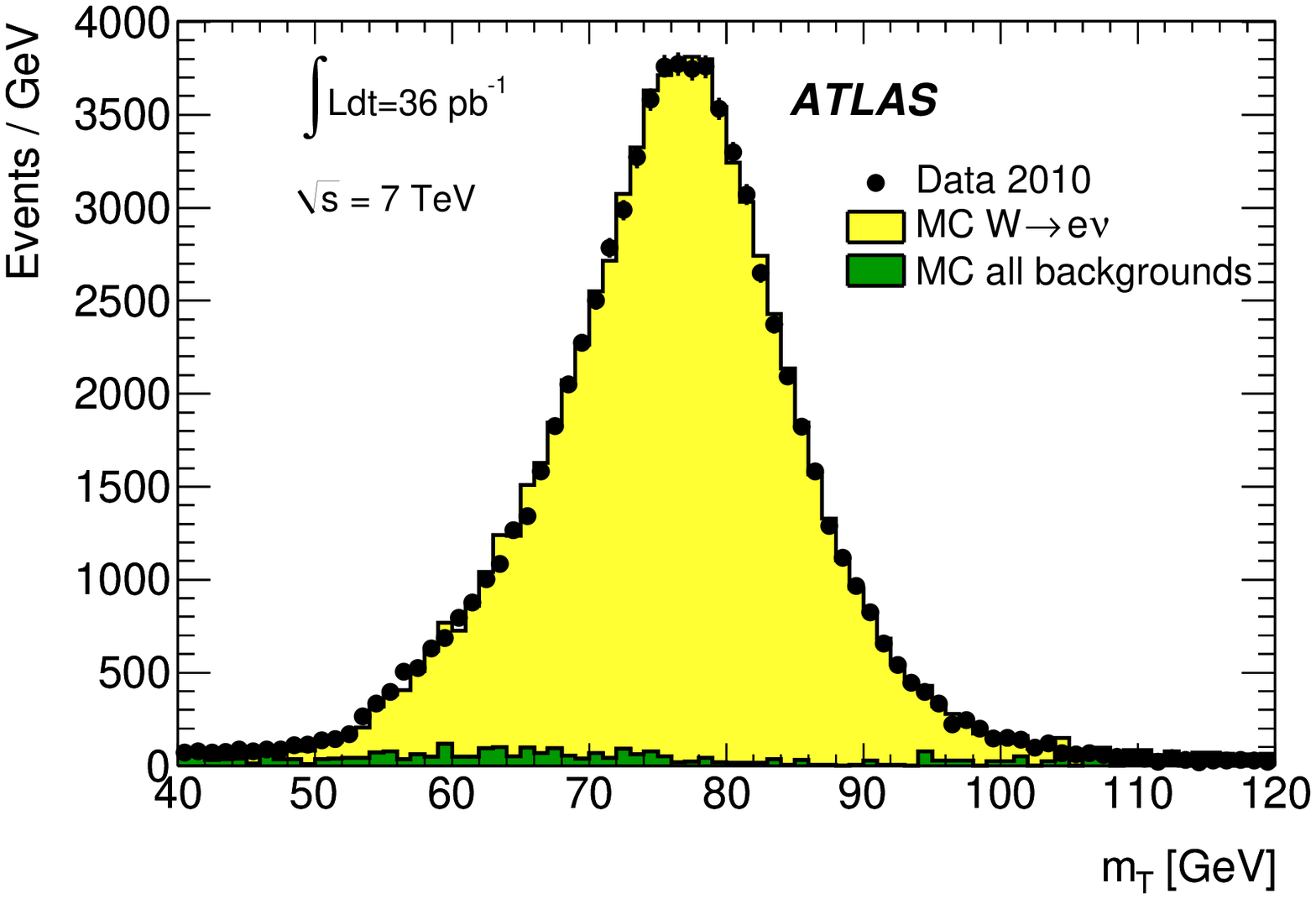}
\caption[$m_{T(Reco\met-muon}$ distribution in data and MC simulation.]{\it  Distributions of the transverse mass,
 \mT, of the muon-\met\ system (left) and of the electron-\met\ system (right) for data.
 The \mT~distributions from Monte Carlo simulation are superimposed,  after each background sample is weighted  as explained in the text. The main backgrounds are shown for \Wmn, the sum of all backgrounds is shown for \Wen.
  The \Wln~ MC signal histogram is obtained  using  the true \met  smeared as in Equation (\ref{eq:smeared_met}) with the scale and resolution parameters obtained from the fit. 
}  \label{fig:fit_muons_ele}
\end{figure*}

To select  $W\to\mu\nu$ events, the same criteria as described in
Section \ref{sec:wl_event_selection} are used, with the exception that no cut on \met\  is applied and a looser cut, \mT~$>30$ GeV, is applied in order that the background normalization can be fitted.
The $\alpha$ and $k$ parameters  obtained from the fit are shown in Table \ref{tab:Wscale1},
together with the  numbers of events for the signal and backgrounds and 
the $\chi^2$/ndof of the fit. 
In the table, instead of the values of $\alpha$,  the  values of $\alpha - 1 = \langle(E^{\rm miss}_{x(y)} -   E_{x(y)}^{\rm miss,True}) / E_{x(y)}^{\rm miss,True}\rangle$ are reported, in order to compare with the result in Sections \ref {sec:Linearity} and \ref{sec:Wscale_fromPT}. 
The results for the $\alpha$ and $k$ parameters using the \mT~distribution of the simulated signal 
are also shown in Table \ref{tab:Wscale1},
 and they are  in good agreement with the results from data. 
 The result of the fit to data and MC simulation is shown in Figure \ref{fig:fit_muons_ele}.
\begin{table*}[htbp]
\begin{center}
\begin{tabular}{|c|c|c|c|c|c|c|}
\hline
Channel  & $\alpha-1$ (\%)  & $k$ &  Signal &  EW(fixed) & QCD & $\chi^2/ndof$\\
\hline
\hline
$W\to\mu\nu$ data  &  $5.1 \pm 0.8$  &  $0.52 \pm 0.01$ & $164920\pm 840$ & 14760&   $24870 \pm  840$ &  68/87\\ 
\hline
 $W\to\mu\nu$ MC  & $5.5 \pm 0.8$ &  $ 0.50 \pm 0.01$    &\multicolumn{2}{r}{}&  & 70/78\\
 \hline
 $W\to e\nu$ data  &  $ -0.8 \pm 1.6$  &$ 0.49 \pm  0.01 $ & $75660 \pm 180$ &  1210 & $980 \pm 180$  &  54/75\\ 
\hline
 $W\to e\nu$ MC  & $ 1.8 \pm 1.7$ &  $0.50 \pm  0.01$  & \multicolumn{2}{r}{} && 38/54\\
\hline
\end{tabular} 
\caption{Results of \mT~fit in \Wln~ events. The second and third columns show the scale and resolution parameters obtained. The numbers of events for the signal, the electroweak and  QCD 
backgrounds obtained from the fit are shown in the fourth, fifth  and sixth columns for data. In the last column the $\chi^2/$ndof of the fit is reported. 
The errors are statistical and take into account background subtraction uncertainties and correlations.}
\label{tab:Wscale1}
 \end{center}
 \end{table*}
 
To select  $W\to e\nu$ events, the selection described in Section \ref{sec:wl_event_selection} 
is used with the addition of tighter cuts.
 A cut \met~$>$ 36 GeV  is applied
 to exclude the region where the \etmissmag~  response is not linear 
  (see Figure \ref{fig:MET_Lin_all}).
 A  cut  \mT~$>$ 40 GeV is also applied.
 The $\alpha$ and $k$ parameters  obtained from the fit are shown in Table \ref{tab:Wscale1}, together with the results obtained from the MC, which are in good agreement with data. 
 The result of the fit  to data and MC simulation is shown in Figure \ref{fig:fit_muons_ele}.
 
The results obtained with this method are compatible, at the few percent level,  with the results
 shown in Figure \ref{fig:MET_Lin_all} and Figure \ref{fig:resol_tutti},  which were derived using only simulation.
From those figures, for the \Wmn~ channel 
 $\alpha-1$ has values up to 3\%\ and the resolution is 0.47$\sqrt{\sum E_{\mathrm{T}}}$;
 for the \Wen~ channel  $\alpha-1$ is close to zero for high \met\ values and the resolution is 0.47$\sqrt{\sum E_{\mathrm{T}}}$.

The uncertainty due to background subtraction is already included in the uncertainty reported in  Table~\ref{tab:Wscale1}. 
The systematic uncertainty on  $\alpha-1$ is  determined to be about 1\% for  each channel, 
by checking the stability of the results using  different cuts on  \etmissmag~  and  using a different 
  generator,  {\sc MC@NLO}.
In summary, with this method the \etmissmag\ absolute scale is determined from \Wln~ events, in a data sample corresponding to an integrated luminosity of about 36 \ipb, with an uncertainty (adding the uncertainties reported in Table ~\ref{tab:Wscale1} with the systematic uncertainty) of about 1.5\% and about 2\%  for the \Wmn~ and \Wen~  decay channels, respectively.

\subsection{Method based on the correlation between electron and neutrino transverse momenta in $\textit{{\textbf{W}}} \  \rightarrow e \nu$} 
\label{sec:Wscale_fromPT}

In this section the correlation between the transverse momenta of charged
and neutral leptons from $W$
boson decays is used to determine the \etmissmag~  scale.
The  mean measured  \etmissmag is compared to the mean true  \etmissmag from signal MC events.  
The  relative bias  in the reconstructed  \etmissmag,
 ($\langle$\etmissmag$\rangle - \langle$\etmissmag$^{\rm ,True}\rangle$)/$\langle$\etmissmag$^{\rm ,True}\rangle$, is
studied as a function of \pTe~ because the MC simulation of the
electron response is more accurate than that for hadrons.

This method is shown for  \Wen~ events by applying selection criteria similar to the ones described in Section \ref{sec:wl_event_selection}, but with isolation requirements both on the
electron track and calorimeter signal.
The \etmissmag~ is required to be greater than 20 GeV and no cut is applied on \mT.

MC  samples  are generated with  {\sc MC@NLO}  \cite{MCAtNLO}.  
A next-to-leading-order (NLO) generator is used 
 for this study because in  this  approach the  \etmissmag~  scale is validated on the basis of the  known decay properties of the $W$ boson.
The correlation between \pTn~ and \pTe~ is important for this study, and is
  poorly described by leading-order generators such as {\sc PYTHIA}, whereas it
  is much improved in {\sc MC@NLO}.
The MC events are 
weighted  such that the  true $W$ boson transverse momentum,  \pTW,  
 and pseudorapidity $\eta^W$  agree  with that
  generated using the {\sc RESBOS} ~\cite{ResBos}
generator  which is  more accurate  in describing \pTW~  at low  values.
Finally, an additional
  smearing is applied to the reconstructed electron momentum in the MC samples,
  to match the electron resolution measured in data, and the correction is
  propagated to \etmissmag.

\begin{figure*}
   \centering \includegraphics[width=.49\linewidth,height=\myFigSize]{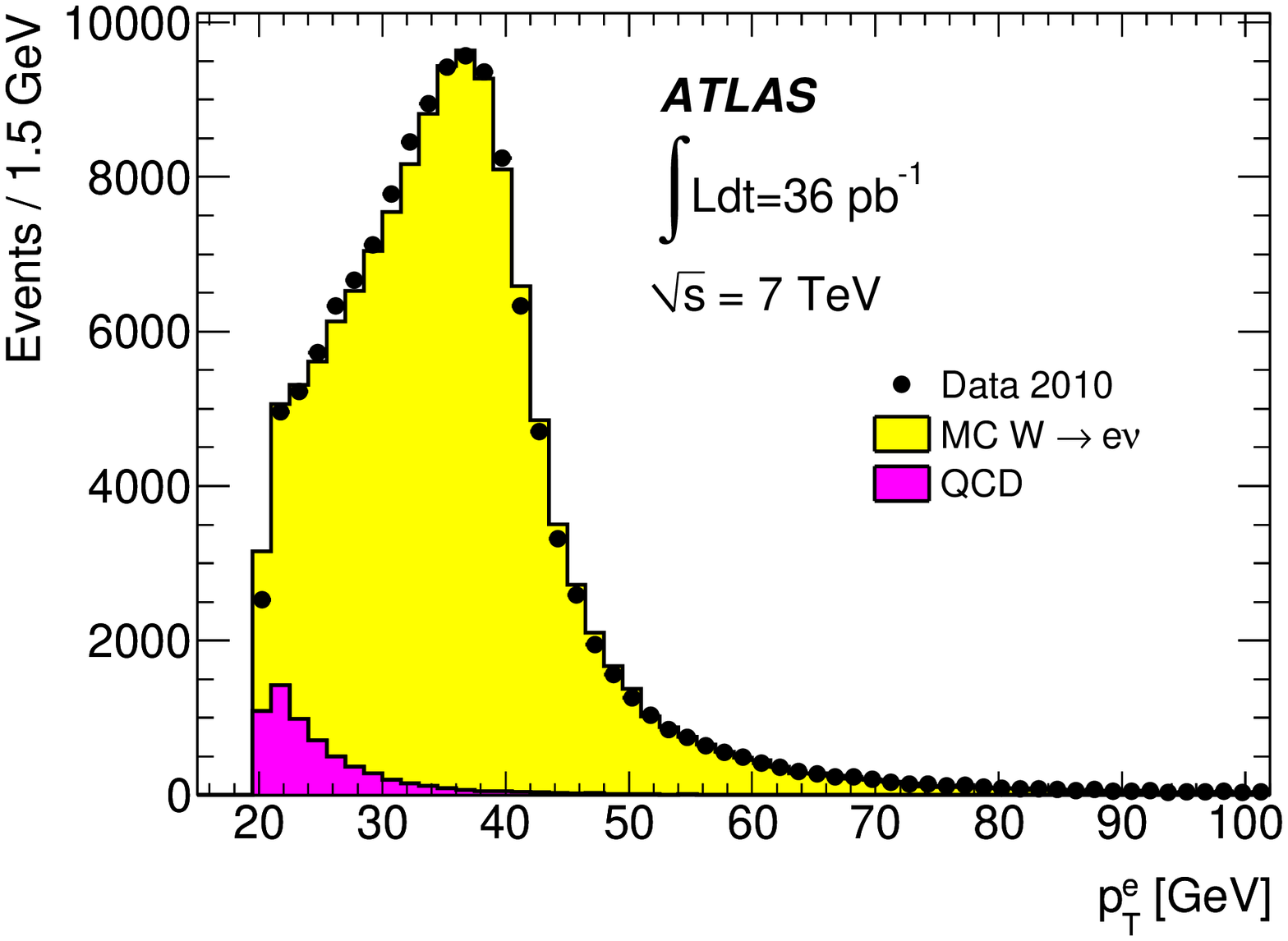}
      \centering \includegraphics[width=.49\linewidth,height=\myFigSize]{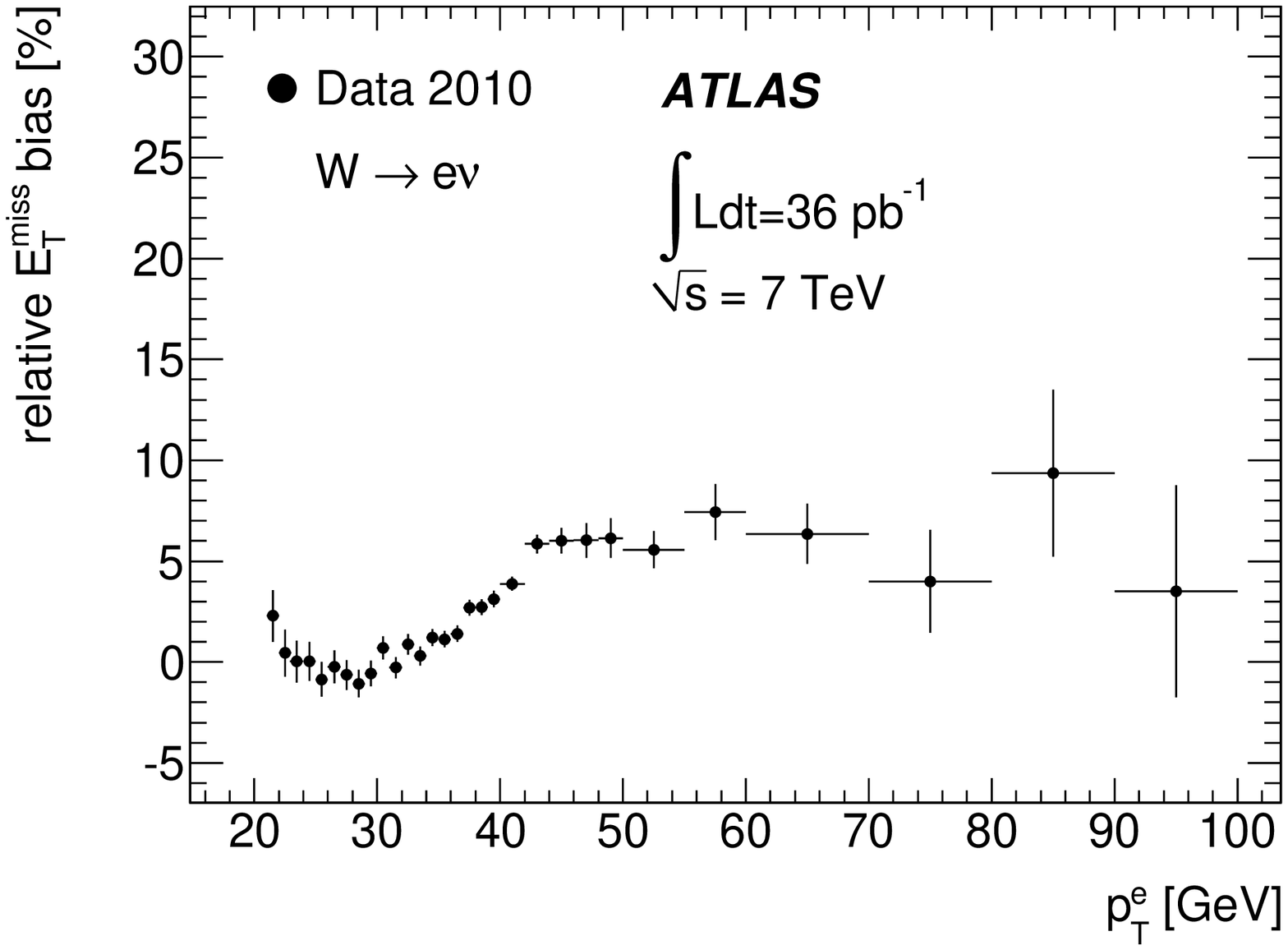} 
   \caption{\it Transverse
   momentum  distribution of electron  candidates in  data,
   in  signal MC   with nominal  event  selection and  with
   reversed cuts  for background
   (QCD)     from data (left).
    Relative  bias  in   the
   reconstructed \etmissmag (right). Only statistical uncertainties are shown.
\label{fig:Wscale2}}
\end{figure*}

A data-driven technique is used  to estimate
the  impact of  jet background,
which is small (see Figure~\ref{fig:Wscale2} left) and concentrated at low \pTe.
\Wtau~ events, where the $\tau$  decays to
an electron, are  the second  largest background, but  the impact  on the mean
value of \etmissmag~ is found to be negligible.

The distribution of \pTe~ is shown in Figure~\ref{fig:Wscale2}.  The  distribution from data after  event selection is fitted by  varying the normalization of signal MC and QCD 
background distributions.  A satisfactory  description of data is achieved
except  for the first  bin, which  is excluded  from the  fit.  
For each \pTe~ bin, the corrected distribution
  of \etmissmag~ is obtained by subtracting that of the background sample (after
  normalizing it according to the fit) from the data distribution.
The largest impact  of background  corresponds to  \pTe ~= 20 GeV, with an effect 
of about 2  GeV on the mean value of \etmissmag; the effect decreases quickly to 0.2 GeV at \pTe~ = 30 GeV. 

Since a cut  on  \etmissmag~ is used for the event  selection and
the \etmiss~ resolution is finite,  the results are biased. 
To correct for the bias in signal MC events  the requirement
of reconstructed \etmissmag~ $>$~20 GeV is replaced by a cut on true \etmissmag~   $>$~20 GeV.

The  mean measured  \etmissmag, corrected  for  background and for the event selection bias,  is used to calculate the  relative bias  in the reconstructed  \etmissmag,
 ($\langle$\etmissmag$\rangle - \langle$\etmissmag$^{\rm ,True}\rangle$)/$\langle$\etmissmag$^{\rm ,True}\rangle$, which  is  shown   in 
Figure~\ref{fig:Wscale2} as a function of  \pTe.  
The figure shows that the 
\etmissmag~  scale is correct at low values of  \pTe~  while it is overestimated 
 at high values  of  \pTe.    

The bias on \etmissmag~ is on the percent level between 25 and 35 GeV, 
then it rises up to  $7\%$  and it is  2 +- 0.1$\%$ on average.
For comparison, if the entire calculation is performed on signal MC events alone, 
the resulting  average bias in \etmissmag~ is  $2.9  \pm  0.1\%$.
The  method relies  on simulation to derive the correlation between
\etmissmag$^{\rm ,True}$~ and  \pTe, so it can be sensitive to details of the simulation.
  In particular, the 
jet factorization   and   renormalization   scales,  as well as the choice of  {\sc PDF},  
 can  affect the  results, but  all these also
change the  \pTW~  distribution.  Therefore the
shape of the \pTW~ distribution was distorted by $\pm 10\%$, justified by 
the comparison of a recent  measurement of  the  \pTZ~  distribution~\cite{ZPtPaper}
with  {\sc  RESBOS} predictions, and the relative bias was calculated again.
A systematic uncertainty on the relative \etmissmag~ scale bias 
of  $\pm 2\%$ is evaluated.
The results  for the average \etmissmag~ scale  are  summarized   in Table~\ref{tab:Wscale2}. 
These results agree within errors 
with the values of $\alpha-1$ shown in Table~\ref{tab:Wscale1}.

 \begin{table}[htbp]
\begin{center}
\begin{tabular}{c|c}
\hline
\hline
source  & scale bias (\%)  \\
\hline
data  &  $2.0 \pm 0.1 \pm 2.0$  \\ 
\hline
MC  &  $ 2.9 \pm 0.1 $  \\ 
\hline
\hline
\end{tabular} 
\caption{Average relative \etmissmag~  scale bias obtained from data and MC simulation from the electron-neutrino correlation method.
The statistical and the systematic uncertainties are given for data.}
\label{tab:Wscale2}
 \end{center}
 \end{table}

\section{Conclusion}
\label{sec:conclu}

The missing transverse momentum (\etmiss) has been measured
in minimum bias, di-jet, \Zll~ and \Wln~ events in 
7 TeV $pp$ collisions recorded with the ATLAS detector in 2010.

The value of \etmiss~is reconstructed from calorimeter cells in topological clusters, with the exception of electrons and photons for which a different clustering algorithm is used, and from reconstructed muons.
 The cells are calibrated according to their parent particle type.
The scheme yielding the best performance is evaluated to be that in which electrons are
calibrated with the default electron calibration and
 photons are used at the EM scale,
  the $\tau$-jets and  jets are calibrated with the  local hadronic calibration (LCW), 
  the jets with \pT~ greater than 20 GeV are scaled to the jet energy scale,
and the contribution from topoclusters not associated to high-\pT~ objects is calculated 
with LCW calibration combined with tracking information.

Monte Carlo simulation is found to describe the data in general rather well.
No large tails are observed in the \etmissmag~distribution in minimum bias,  di-jet and \Zll~ events, where no significant \etmissmag~ is expected. 
The tails are not completely well described by MC 
simulation especially in di-jets events, where there are more
events in the tail in data. 

There is some difference observed
between data and MC simulation
for the reconstructed total transverse energy.
The precise difference is dependent on the model used to
simulate soft-physics processes.

The \etmiss~resolution 
is similar in the different channels studied and in agreement with 
the resolution in the MC simulation. The resolution follows a function $\sigma= k \cdot \sqrt{\Sigma E_{\mathrm{T}}}$, where the parameter $k$ is about 0.5 GeV$^{1/2}$.

The linearity of the \etmiss~  measurement in \Wln~ events is studied in MC simulation as a function of the true \etmissmag. Except for the bias observed at  small true \etmissmag~ values
 (visible up to 40 GeV), due to the finite \etmiss~ resolution, the linearity is better than 1$\%$ in \Wen~ events, while a small non-linearity up to about 3$\%$ is observed in \Wmn~ events.

The \etmiss~ projected along the $Z$ direction in \Zll~ events is observed to have a 
 bias up to 6 GeV at large values of $p_T^Z$ in events with no jets, 
suggesting that some improvements are still needed in the calibration of low-\pT~ objects.

The overall systematic uncertainty on \etmissmag scale, calculated by combining the uncertainties on the various terms entering the full \etmiss~ calculation,  is estimated to be,
on average, 
2.6\% in events with a $W$ decaying to a lepton (electron or muon) and neutrino.
 The uncertainty is larger at large \sumet.

Two methods are used for determining the \etmissmag~ scale from  \Wln~ events in data,
giving results in agreement with that evaluated using MC simulation. The 
resulting uncertainty on the \etmissmag~ scale determined  in-situ with 36 \ipb~ of data is, on average, about 2\%.

\section{Acknowledgements}

We thank CERN for the very successful operation of the LHC, as well as the
support staff from our institutions without whom ATLAS could not be
operated efficiently.

We acknowledge the support of ANPCyT, Argentina; YerPhI, Armenia; ARC,
Australia; BMWF, Austria; ANAS, Azerbaijan; SSTC, Belarus; CNPq and FAPESP,
Brazil; NSERC, NRC and CFI, Canada; CERN; CONICYT, Chile; CAS, MOST and
NSFC, China; COLCIENCIAS, Colombia; MSMT CR, MPO CR and VSC CR, Czech
Republic; DNRF, DNSRC and Lundbeck Foundation, Denmark; ARTEMIS, European
Union; IN2P3-CNRS, CEA-DSM/IRFU, France; GNAS, Georgia; 
\par\noindent
BMBF, DFG, HGF, MPG
and AvH Foundation, Germany; 
\par\noindent
GSRT, Greece; ISF, MINERVA, GIF, DIP and
Benoziyo Center, Israel; INFN, Italy; MEXT and JSPS, Japan; CNRST, Morocco;
FOM and NWO, Netherlands; RCN, Norway; MNiSW, Poland; GRICES and FCT,
Portugal; MERYS (MECTS), Romania; MES of Russia and ROSATOM, Russian
Federation; JINR; MSTD, Serbia; MSSR, Slovakia; ARRS and MVZT, Slovenia;
DST/NRF, South Africa; MICINN, Spain; SRC and Wallenberg Foundation,
Sweden; SER, SNSF and Cantons of Bern and Geneva, Switzerland; NSC, Taiwan;
TAEK, Turkey; STFC, the Royal Society and Leverhulme Trust, United Kingdom;
DOE and NSF, United States of America.

The crucial computing support from all WLCG partners is acknowledged
gratefully, in particular from CERN and the ATLAS Tier-1 facilities at
TRIUMF (Canada), NDGF (Denmark, Norway, Sweden), CC-IN2P3 (France),
KIT/GridKA (Germa-
\par\noindent
ny), INFN-CNAF (Italy), NL-T1 (Netherlands), PIC (Spain),
ASGC (Taiwan), RAL (UK) and BNL (USA) and in the Tier-2 facilities
worldwide.

\bibliographystyle{atlasnote}
\bibliography{Biblio1}

\providecommand{\href}[2]{#2}\begingroup\raggedright\begin{thebibliography}{10}

\bibitem{ATLAS_PHYS_PAP}
{The ATLAS} Collaboration, {\em {The ATLAS Experiment at the CERN Large Hadron
  Collider}\/},  JINST {\bf 3} (2008)  S08003.

\bibitem{ATLAS_PERF_first}
{The ATLAS} Collaboration, {\em {Performance of the ATLAS detector using first
  collision data}\/},  JHEP {\bf 09} (2010)  056.

\bibitem{SUSY_PAPER}
{The ATLAS} Collaboration, {\em Search for squarks and gluinos using final
  states with jets and missing transverse momentum with the ATLAS detector in 7
  TeV proton-proton collisions\/},  Phys. Lett. B {\bf 701} (2011)  186.

\bibitem{CONF_NOTE_JET}
{The ATLAS} Collaboration, {\em Properties of jets and inputs to jets
  reconstruction and calibration with the ATLAS detector using proton-proton
  collisions at $\sqrt{s}=7$ TeV\/},  {
  \href{http://cdsweb.cern.ch/record/1281310/}{ATLAS-CONF-2010-053}}.

\bibitem{ANTI_KT}
M.~Cacciari, G.~P. Salam, and G.~Soyez, {\em The anti-k$_{t}$ jet clustering
  algorithm\/},  JHEP {\bf 05} (2008)  063,
  \href{http://arxiv.org/abs/0802.1189}{{\tt arXiv:0802.1189}}.

\bibitem{Lumi}
{The ATLAS} Collaboration, {\em Luminosity Determination in pp Collisions at 7
  TeV using the ATLAS Detector at the LHC\/},  EPJC {\bf 071} (2011)  1630,
  \href{http://arxiv.org/abs/1101.2185}{{\tt arXiv:1101.2185}}.

\bibitem{Lumi1}
{The ATLAS} Collaboration, {\em Updated Luminosity Determination in pp
  Collisions at sqrt(s)=7 TeV using the ATLAS Detector\/},
  {\href{HTTP://CDSWEB.CERN.CH/RECORD/1334563/}{ATLAS-CONF-2011-011}}.

\bibitem{WZXS}
{The ATLAS} Collaboration, {\em Measurement of the $W\rightarrow l\nu$ and
  $Z\rightarrow ll$ production cross sections in proton-proton collisions at
  $\sqrt{s}=7$ TeV with the ATLAS detector\/},  JHEP {\bf 12} (2010)  060,
  \href{http://arxiv.org/abs/1010.2130}{{\tt arXiv:1010.2130}}.

\bibitem{MINBIAS_PAPER}
{The ATLAS} Collaboration, {\em Charged-particle multiplicities in pp
  interactions measured with the ATLAS detector at the LHC\/},  New J. Phys.
  {\bf 13} (2011)  053033.

\bibitem{Muon_Perf}
{The ATLAS} Collaboration, {\em Muon Reconstruction Performance\/},
  {\href{HTTP://CDSWEB.CERN.CH/RECORD/1281339/}{ATLAS-CONF-2010-064}}.

\bibitem{egamma}
{The ATLAS} Collaboration, {\em Electron and photon reconstruction and
  identification in ATLAS: expected performance at high energy and results at
  900 GeV\/},
  {\href{HTTP://CDSWEB.CERN.CH/RECORD/1273197/}{ATLAS-CONF-2010-005}}.

\bibitem{PYTHIA}
T.~Sjostrand, S.~Mrenna, and P.~Skands, {\em PYTHIA 6.4 Physics and Manual\/},
  JHEP {\bf 05} (2006)  026.

\bibitem{MONTE-CARLO}
{The ATLAS} Collaboration, {\em Charged particle multiplicities in pp
  interactions at sqrt(s) = 0.9 and 7 TeV in a diffractive limited phase-space
  measured with the ATLAS detector at the LHC and new PYTHIA6 tune\/},
  {\href{HTTP://CDSWEB.CERN.CH/RECORD/1277665/}{ATLAS-CONF-2010-031}}.

\bibitem{GEANT4}
S.~Agostinelli et al., {\em GEANT4: A simulation toolkit\/},  NIM {\bf A 506}
  (2003)  250.

\bibitem{MCAtNLO}
S.~Frixione and B.~R. Webber, {\em Matching NLO QCD computations and parton
  shower simulations\/},  JHEP {\bf 06} (2002)  029.

\bibitem{PYTHIA8}
T.~Sjostrand, S.~Mrenna, and P.~Skands, {\em Brief Introduction to
  PYTHIA8.1\/},  Comput. Phys. Comm. {\bf 178} (2008)  852,
  \href{http://arxiv.org/abs/0710.3820}{{\tt arXiv:0710.3820}}.

\bibitem{JET_ETMISS}
{The ATLAS} Collaboration, {\em Expected Performance of the ATLAS Experiment -
  Detector, Trigger and Physics (Jet and \etmissmag~chapter)\/},
  \href{http://arxiv.org/abs/0901.0512}{{\tt arXiv:0901.0512}}.

\bibitem{clusters}
W.~Lampl et al., {\em Calorimeter clustering algorithms: Description and
  performance\/},
  {\href{HTTP://CDSWEB.CERN.CH/RECORD/1099735/}{ATL-LARG-PUB-2008-002}}.

\bibitem{TauPerf}
{The ATLAS} Collaboration, {\em Tau Reconstruction and Identification
  Performance in ATLAS\/},
  {\href{HTTP://CDSWEB.CERN.CH/RECORD/1298857/}{ATLAS-CONF-2010-086}}.

\bibitem{LCW}
T.~Barillari et al., {\em Local Hadron Calibration Properties\/},
  {\href{HTTP://CDSWEB.CERN.CH/RECORD/1112035/}{ATL-LARG-PUB-2009-001}}.

\bibitem{JES}
{The ATLAS} Collaboration, {\em Jet energy scale and its systematic uncertainty
  in proton-proton collisions at $\sqrt{s}=7$ TeV in ATLAS 2010 data\/},
  {\href{HTTP://CDSWEB.CERN.CH/RECORD/1337782/}{ATLAS-CONF-2011-032}}.

\bibitem{PYTHIA82}
R.~Corke and T.~Sjostrand, {\em Interleaved Parton Showers and Tuning
  Prospects\/},  JHEP {\bf 032} (2011)  1103,
  \href{http://arxiv.org/abs/1011.1759}{{\tt arXiv:1011.1759}}.

\bibitem{atlasjet2010}
{The ATLAS} Collaboration, {\em {Measurement of inclusive jet and dijet cross
  sections in proton-proton collisions at 7 TeV centre-of-mass energy with the
  ATLAS detector}\/},  Eur. Phys. J. C {\bf 71} (2011)  1512,
  \href{http://arxiv.org/abs/1009.5908}{{\tt arXiv:1009.5908}}.

\bibitem{etmissCONF}
{The ATLAS} Collaboration, {\em Reconstruction and Calibration of Missing
  Transverse Energy and Performance in Z and W events in ATLAS Proton-Proton
  Collisions at $\sqrt{s}=7$ TeV\/},
  {\href{HTTP://CDSWEB.CERN.CH/RECORD/1355703/}{ATLAS-CONF-2011-080}}.

\bibitem{eoverp2011}
{The ATLAS} Collaboration, {\em ATLAS Calorimeter Response to Single Isolated
  Hadrons and Estimation of the Calorimeter Jet Scale Uncertainty\/},
  {\href{HTTP://CDSWEB.CERN.CH/RECORD/1337075/}{ATLAS-CONF-2011-028}}.

\bibitem{ResBos}
C.~Balazs and C.~P. Yuan, {\em Soft gluon effects on lepton pairs at hadron
  colliders\/},  Phys. Rev. D {\bf 56} (1997)  55585583.

\bibitem{ZPtPaper}
{The ATLAS} Collaboration, {\em Measurement of the transverse momentum
  distribution of Z/$\gamma$ bosons in proton-proton collisions at $\sqrt{s}=7$
  TeV with the ATLAS detector\/},  submitted to Phys. Lett. B  ,
  \href{http://arxiv.org/abs/1107.2381}{{\tt arXiv:1107.2381}}.

\end{thebibliography}\endgroup

\clearpage
\onecolumn
\begin{flushleft}
{\Large The ATLAS Collaboration}

\bigskip

G.~Aad$^{\rm 48}$,
B.~Abbott$^{\rm 111}$,
J.~Abdallah$^{\rm 11}$,
A.A.~Abdelalim$^{\rm 49}$,
A.~Abdesselam$^{\rm 118}$,
O.~Abdinov$^{\rm 10}$,
B.~Abi$^{\rm 112}$,
M.~Abolins$^{\rm 88}$,
H.~Abramowicz$^{\rm 153}$,
H.~Abreu$^{\rm 115}$
% ...
\bigskip

% ...
$^{10}$ Institute of Physics, Azerbaijan Academy of Sciences, Baku, Azerbaijan\\
$^{11}$ Institut de F\'isica d'Altes Energies and Universitat Aut\`onoma  de Barcelona and ICREA, Barcelona, Spain\\
% ...
$^{48}$ Fakult\"{a}t f\"{u}r Mathematik und Physik, Albert-Ludwigs-Universit\"{a}t, Freiburg i.Br., Germany\\
$^{49}$ Section de Physique, Universit\'e de Gen\`eve, Geneva, Switzerland\\
% ...
$^{88}$ Department of Physics and Astronomy, Michigan State University, East Lansing MI, United States of America\\
% ...
$^{111}$ Homer L. Dodge Department of Physics and Astronomy, University of Oklahoma, Norman OK, United States of America\\
$^{112}$ Department of Physics, Oklahoma State University, Stillwater OK, United States of America\\
% ...
$^{115}$ LAL, Univ. Paris-Sud and CNRS/IN2P3, Orsay, France\\
% ...
$^{118}$ Department of Physics, Oxford University, Oxford, United Kingdom\\
% ...
$^{153}$ Raymond and Beverly Sackler School of Physics and Astronomy, Tel Aviv University, Tel Aviv, Israel\\
% ...
\end{flushleft}

\end{document}